\documentclass{aa}

\usepackage{longtable}
\usepackage{caption}

\usepackage{graphicx}
\usepackage{txfonts}
\usepackage[fleqn]{amsmath}	
\usepackage{amssymb}	
\usepackage{mathtools}
\usepackage[breaklinks=true]{hyperref}
\hypersetup{
  colorlinks=true,   
  citecolor=blue,    
  urlcolor=blue,     
  linkcolor=blue,     
}
\bibpunct{(}{)}{;}{a}{}{,} 
\usepackage{lscape}

\newcommand{\ks}{K_{\rm S}}
\newcommand{\lbol}{L_{\rm bol}}
\newcommand{\vp}{\upsilon_{\rm peak}}

\DeclareRobustCommand{\ion}[2]{%
\relax\ifmmode
\ifx\testbx\f@series
{\mathbf{#1\,\mathsc{#2}}}\else
{\mathrm{#1\,\mathsc{#2}}}\fi
\else\textup{#1\,{\mdseries\textsc{#2}}}%
\fi}

    \newcommand{\xmm}{XMM--\emph{Newton}\xspace}

   \newcommand{\emcee}{\textsc{emcee}\xspace}

\newcommand{\rev}[1]{{ #1}}

\begin{document} 

   \title{The most luminous blue quasars at 3.0\,$<$\,\textit{z}\,$<$\,3.3}
   \subtitle{III.  LBT spectra and accretion parameters}

   \author{Bartolomeo Trefoloni\inst{1,2}
   \thanks{\email{bartolomeo.trefoloni@unifi.it}},
    Elisabeta~Lusso\inst{1,2},
    Emanuele~Nardini\inst{2}, 
    Guido~Risaliti\inst{1,2},
    Giada~Bargiacchi\inst{3,4},
    Susanna~Bisogni\inst{5}, 
	Francesca~M.~Civano\inst{6},
	Martin~Elvis\inst{7},
	Giuseppina~Fabbiano\inst{7},
	Roberto~Gilli\inst{8},
	Alessandro~Marconi\inst{1,2}, 
    Gordon~T.~Richards\inst{9},
    Andrea~Sacchi\inst{7,10},
	Francesco~Salvestrini\inst{2},
    Matilde~Signorini\inst{1,2},
	Cristian~Vignali\inst{11,8}
          }
          
\institute{
$^{1}$Dipartimento di Fisica e Astronomia, Universit\`a di Firenze, via G. Sansone 1, 50019 Sesto Fiorentino, Firenze, Italy\\
$^{2}$INAF -- Osservatorio Astrofisico di Arcetri, Largo Enrico Fermi 5, I-50125 Firenze, Italy\\
$^{3}$Scuola Superiore Meridionale, Largo S. Marcellino 10, I-80138, Napoli\\
$^{4}$Istituto Nazionale di Fisica Nucleare (INFN), Sez. di Napoli, Complesso Univ. Monte S. Angelo, Via Cinthia 9, I-80126, Napoli, Italy\\
$^{5}$INAF -- Istituto di Astrofisica Spaziale e Fisica Cosmica Milano, via Corti 12, 20133 Milano, Italy\\
$^{6}$NASA Goddard Space Flight Center, Code 660, Greenbelt, MD 20771, USA\\
$^{7}$Center for Astrophysics | Harvard \& Smithsonian, 60 Garden Street, Cambridge, MA 02138, USA\\
$^{8}$INAF -- Osservatorio di Astrofisica e Scienza dello Spazio di Bologna, via Gobetti 93/3, I-40129 Bologna, Italy\\
$^{9}$Department of Physics, 32 S. 32nd Street, Drexel University, Philadelphia, PA 19104\\
$^{10}$Scuola Universitaria Superiore IUSS Pavia, Palazzo del Broletto, piazza della Vittoria 15, 27100 Pavia, Italy\\
$^{11}$Dipartimento di Fisica e Astronomia, Universit\`a degli Studi di Bologna, via Gobetti 93/2, I-40129 Bologna, Italy\\
}

\titlerunning{The most luminous blue quasars at $z\sim3$ III. LBT spectra and accretion parameters}
\authorrunning{B. Trefoloni et al.}
\date{\today}

 
   
   \abstract{We present the analysis of the rest frame ultraviolet and optical spectra of 30 bright blue quasars at $z\sim3$, selected to examine the suitability of active galactic nuclei as cosmological probes. In our previous works, based on pointed \xmm observations, we found an unexpectedly high fraction ($\approx 25 \%$) of X-ray weak quasars in the sample. The latter sources also display a flatter UV continuum and a broader and fainter \ion{C}{iv} profile in the archival UV data with respect to their X-ray normal counterparts. Here we present new observations with the Large Binocular Telescope in both the $zJ$ (covering the rest-frame $\simeq$2300--3100 \AA) and the $\ks$ ($\simeq$4750--5350 \AA) bands. We estimated black hole masses ($M_{\rm BH}$) and Eddington ratios ($\lambda_{\rm Edd}$) from the available rest-frame optical and UV emission lines (H$\beta$, \ion{Mg}{ii}), finding that our $z\sim3$ quasars are on average highly accreting ($\langle \lambda_{\rm Edd} \rangle\simeq 1.2$ and $\langle M_{\rm BH} \rangle\simeq 10^{9.7}M_\odot$), with no difference in $\lambda_{\rm Edd}$ or $M_{\rm BH}$ between X-ray weak and X-ray normal quasars. From the $zJ$ spectra, we derive the properties (e.g., flux, equivalent width) of the main emission lines (\ion{Mg}{ii}, \ion{Fe}{ii}), finding that X-ray weak quasars display higher \ion{Fe}{ii}/\ion{Mg}{ii} ratios with respect to typical quasars. \ion{Fe}{ii}/\ion{Mg}{ii} ratios of X-ray normal quasars are instead consistent with other estimates up to $z\simeq6.5$, corroborating the idea of already chemically mature broad line regions at early cosmic time. From the $\ks$ spectra, we find that all the X-ray weak quasars present generally weaker [\ion{O}{iii}] emission (EW<10 \AA) than the normal ones. The sample as a whole, however, abides by the known X-ray/[\ion{O}{iii}] luminosity correlation, hence the different [\ion{O}{iii}] properties are likely due to an intrinsically weaker [\ion{O}{iii}] emission in X-ray weak objects, associated to the shape of the spectral energy distribution. We interpret these results in the framework of accretion-disc winds.
   }
   

   \maketitle
%

\section{Introduction}
Quasars are the most luminous persistent sources in the Universe and, as such, they represent a class of objects of fundamental importance to understand the mechanisms of production of radiation and its interplay with gas and dust up to very high redshift. Quasars belong to the high luminosity ($L_{\rm bol}>10^{45}$ erg s$^{-1}$) tail of the active galactic nuclei (AGN) population and, according to the current paradigm, their emission is powered by accretion onto a supermassive black hole (SMBH, $M_{\rm BH}>10^6 M_{\odot}$). The main contribution to their broadband spectral energy distribution (SED) comes from the optical/UV emission produced by the disc (e.g., \citealt{Salpeter1964, lynden1969galactic, czerny1987constraints}) and X-ray emission from the so-called ``corona'' (e.g., \citealt{sunyaev1980comptonization, 1993ApJ...413..507H}), where the energy of the UV seed photons emitted by the disc is boosted via inverse Compton scattering. The UV emission is accompanied by a shallow bump in the infrared (due to dust reprocessing in the torus), whilst strong radio emission, if present, is generally linked to a jet.

The high luminosity observed in quasars, as well as the growing number of available observations up to high redshift (e.g., \citealt{mortlock2011luminous, banados2018}), make them valuable objects to investigate the cosmological parameters, as proposed by our group \citep[e.g.,][]{risaliti2015hubble} by making use of the non-linear relation between their UV and X-ray luminosity (the $L_{\rm X}-L_{\rm UV}$ or, equivalently, the $\alpha_{\rm OX}-L_{\rm UV}$ relation\footnote{$\alpha_{\rm OX}$ is defined as a function of the monochromatic flux densities at rest-frame 2500 \AA\ and 2 keV as $\alpha_{\rm OX}= 0.384 \log (F_{\rm 2\,keV}/F_{2500\,\AA})$.}, e.g., \citealt{avnitananbaum86}). Such a relation was found to be independent on redshift (e.g., \citealt{vignali03, steffen06, green09, lusso2016tight}); yet, the interplay between corona and disc could in principle vary with the bolometric luminosity ($L_{\rm bol}$) and/or the Eddington ratio ($\lambda_{\rm Edd}$, defined as $L_{\rm bol}/L_{\rm Edd}$ where $L_{\rm Edd}$ is the Eddington luminosity). At high $\lambda_{\rm Edd}$, for instance, the framework of geometrically thin and optically thick disc \citep{1973A&A....24..337S} could break down, as the disc is expected to thicken (\citealt{Abramowicz1988, chen2004slim, wang2014}), a behaviour also shown by simulations (\citealt{ohsuga2011global, jiang2014radiation, jiang2016iron, jiang2019global, skadowski2014numerical}). Moreover, \rev{perturbations to the standard accretion process could be associated to the presence of powerful accretion-disc winds, directly driven by the nuclear activity (e.g., \citealt{proga2005}).} A highly efficient accretion is an ideal condition for the launch of such outflows (\citealt{zubovas2013bal, nardini2015black, king2015powerful, nardini2019ton}), which could justify the observed relations between the SMBH mass and the galaxy properties (e.g., the $M_{\rm BH}-\sigma$ relation; \citealt{2000ApJ...539L...9F, gebhardt2000relationship, 2003ApJ...589L..21M, king2005agn}), although it is not yet clear whether and how AGN-driven outflows can affect their host galaxies. 

Remarkably, at high $\lambda_{\rm Edd}$ several quasar samples sharing similar UV properties have recently shown an enhanced fraction of objects ($\approx$\,25\%) \rev{whose X-ray spectra are relatively flat ($\langle \Gamma \simeq 1.6  \rangle$)} and underluminous (by factors of $>3-10$) with respect to what is expected according to the $L_{\rm X}-L_{\rm UV}$ relation (e.g., \citealt{luo2015, nardini2019, zappacosta2020, laurenti2021x}), \rev{in many cases without any clear evidence for absorption as revealed by the spectral analysis}. 

A high $\lambda_{\rm Edd}$ is also conducive to a higher prominence of the \ion{Fe}{ii} emission line complex (e.g., \citealt{bg1992, marziani2001searching, zamfir2010detailed, shenho2014}). This feature, in turn, can be employed to investigate the chemical enrichment of the broad line region (BLR) of quasars through the ratio \ion{Fe}{ii}/\ion{Mg}{ii} up to very high redshift ($z\lesssim 7$). The \ion{Fe}{ii}/\ion{Mg}{ii} ratio seems to correlate with $\lambda_{\rm Edd}$ and $M_{\rm BH}$, but does not show any clear trend with the AGN luminosity (e.g., \citealt{dong2011controls, shin2019fe,shin2021strong}). Despite plenty of studies (e.g., \citealt{kawara1996infrared, thompson1999lack, iwamuro2002fe, iwamuro2004fe, dietrich2003fe, barth2003iron, freudling2003iron, maiolino2003early, tsuzuki2006fe, jiang2007gemini, kurk2007black, sameshima2009ultraviolet, sameshima2020mg, de2011evidence, de2014black, mazzucchelli2017physical, shin2019fe}), it is still unclear whether any evolutionary trend of this ratio exists, also because of the large uncertainties on the measurements of this quantity. 

This paper is the third dedicated to the analysis of the spectral properties of 30 luminous ($L_{\rm bol}>8\times 10^{46}$ erg s$^{-1}$) quasars at redshift $z=3.0-3.3$ \citep[][Paper I]{nardini2019}. Here, we focus on the \ion{Mg}{ii} $\lambda$2798 emission line probed by dedicated observations at the Large Binocular Telescope (LBT) in the $zJ$ band, as well as on the H$\beta$--[\ion{O}{iii}] complex for a subsample observed in the $\ks$ band. Our main aim is to investigate whether any evidence of a difference in the optical/UV properties (e.g., emission line strengths, continuum) exists between X-ray normal and X-ray weak quasars. 

Throughout the paper we will refer to X-ray normal quasars as the $N$ group, and to X-ray weak and weak candidates as the \textit{W+w} group (i.e., we do not distinguish between X-ray weak, $W$, and weak candidates, $w$). For the operational definition of X-ray normal ($N$), weak ($W$) and weak candidates ($w$), we refer the interested reader to Section 2.3 of \citet[][Paper II]{lusso2021most}.
The paper is structured as follows: the quasar sample and the observations are described in Section \ref{obs_data_red}, whilst the analysis of UV and optical spectra is reported in Section \ref{analysis}. Results are presented and discussed in Sections \ref{results} and \ref{sec:discuss}, and conclusions are drawn in Section \ref{conclusions}. 

\section{Observations and data reduction}
\label{obs_data_red}

\subsection{The data set}

The quasar sample analysed here consists of 30 luminous ($\lbol$\,$>$\,10$^{46.9}$ erg s$^{-1}$) quasars at $z\sim3$, for which X-ray observations were obtained through an extensive campaign performed with \xmm. This sample, selected in the optical from the Sloan Digital Sky Survey (SDSS) Data Release 7 \citep{abazajian2009seventh} to be representative of the most luminous, intrinsically blue radio-quiet quasars, boasts by construction a remarkable degree of homogeneity in terms of optical/UV properties. The reader can find more details on the sample selection in the Supplementary Material of \citealt{risaliti2019cosmological} (see also \citealt{lusso2020} for a more general discussion on the selection criteria employed to define homogeneous samples of quasars in the optical/UV). While we refer the reader interested in the X-ray analysis to Paper I, we briefly summarize the main results below.

About two thirds of the sample show X-ray luminosities in agreement with what expected from the $L_{\rm X}-L_{\rm UV}$ relation, and  an average continuum photon index of $\Gamma_{\rm X}$\,$\sim$\,1.85, fully consistent with AGN at lower redshift, luminosity, and $M_{\rm BH}$ (e.g., \citealt{just07,piconcelli2005xmm,bianchi2009}). Their 2--10 keV band luminosities are in the range $4.5\times 10^{44}\leq L_{2-10\,\rm keV} \leq 7.2\times 10^{45}$ erg s$^{-1}$, representing one of the most X-ray luminous samples of radio-quiet quasars ever observed. 
Conversely, one third of the sources are found to be underluminous by a factor $\ga$\,3--10. X-ray absorption at the source redshift is not statistically required in general by the fits of the X-ray spectra and, despite the poor quality of the data in a handful of cases that does not allow us to definitely exclude some absorption, column densities $N_{\rm H}(z) > 3 \times 10^{22}$ cm$^{-2}$ can be confidently ruled out.

In Paper II, we analysed the \ion{C}{iv} $\lambda$1549 emission line properties (e.g., equivalent width, EW; line peak velocity, $\vp$) and UV continuum slope as a function of the X-ray photon index and 2--10 keV flux. In summary, we found that the composite spectrum of X-ray weak quasars is flatter ($\alpha_{\lambda}\sim -0.6$) than the one of X-ray normal quasars ($\alpha_{\lambda}\sim -1.5$). The \ion{C}{iv} emission line is on average fainter in the X-ray weak sample, but only a modest  blueshift (600--800 km s$^{-1}$) is reported for the \ion{C}{iv} lines of both stacks. This emission feature is found to be broader in the \textit{W+w} stacked spectrum (FWHM\,$\simeq$\,10,000 km s$^{-1}$) than in the $N$ one ($\simeq$\,7,000 km s$^{-1}$), but in agreement with previous results on the topic at similar redshifts \citep[e.g.,][]{shen2011,richards2002} and luminosities \citep[e.g.,][]{vietri2018}. When we added the sample from \citet[][filtered out according to our selection criteria]{timlin2020} in order to expand the dynamical range of the parameters of interest, we were able to confirm the statistically significant trends of \ion{C}{iv} $\vp$ and EW with UV luminosity at 2500 \AA\ for both X-ray weak and X-ray normal quasars, as well as the correlation between X-ray weakness and the EW of \ion{C}{iv}. Yet, we did not observe any clear relation between the 2--10 keV luminosity and $\vp$. We found a statistically significant correlation between the hard X-ray flux and the integrated \ion{C}{iv} flux for X-ray normal quasars, which extends across more than three (two) decades in \ion{C}{iv} (X-ray) luminosity, whilst X-ray weak quasars deviate from the main trend by more than 0.5 dex. 

To interpret these results, we argued that X-ray weakness might arise in a starved X-ray corona picture, possibly associated with an ongoing disc-wind phase. If the wind is ejected in the vicinity of the black hole, the accretion rate across the final gravitational radii will diminish, so depriving a compact, centrally-confined corona of seed UV photons and resulting in an X-ray weak quasar. Yet, at the largest UV luminosities ($>$\,10$^{47}$ erg s$^{-1}$), there will still be sufficient ionising photons that can explain the `excess' \ion{C}{iv} emission observed in the X-ray weak quasars with respect to normal sources of similar X-ray luminosities (see Fig. 14 in Paper II).

\subsection{LUCI/LBT observations}
Besides the effects on the \ion{C}{iv} emission line, we are interested in assessing whether the dearth of X-ray photons could affect other emission lines, such as the [\ion{O}{iii}] $\lambda\lambda 4959,5007$\AA\ doublet, whose production needs at least $\simeq 35$ eV ($\simeq 354$ \AA) photons, and the H$\beta$. Moreover, wavelengths longer than 2200 \AA, not covered by the SDSS spectra in this redshift interval, are key to determine the continuum fluxes at rest-frame 2500 \AA\ and to inspect the UV \ion{Fe}{ii} and \ion{Fe}{iii} emission often included among the characteristic parameters of X-ray weak quasars (e.g., \citealt{Leighly2007b,marzianisulentic2014,luo2015}). Therefore, observations of this spectral interval are important for several reasons. For instance, the analysis of the \ion{Mg}{ii} emission line provides generally more reliable estimates of BH masses than the \ion{C}{iv}-based ones, to test whether any systematic difference between the BH masses and Eddington ratios of the X-ray weak and normal quasars is present. Additionally, we can estimate the \ion{Fe}{ii}/\ion{Mg}{ii} ratio in our sample, which represents a proxy of the gas metallicity at redshift $z\sim 3$.

To investigate these issues, our group has been awarded observing time with the two LBT Utility Cameras in the Infrared \citep[LUCI1 and LUCI2,][]{Ageorges2010} at the 8.4-m Large Binocular Telescope (LBT) located on Mount Graham (Arizona), to carry out near-infrared spectroscopy of the $z\sim3$ quasars in the $zJ$ and $\ks$ bands. 
LUCI observations were performed between November 2018 and April 2021 with the $zJ$ filter coupled with the grism G200 and the $\ks$ filter coupled with the grism G150, covering an observed range of 0.9--1.2 $\mu$m and 1.95--2.40 $\mu$m, respectively.
A slit width of 1$^{\prime\prime}$ was employed, providing a spectral resolution $R=1050$--1200 in $zJ$ and 2075 in $\ks$.
The journal of the observations is shown in Table~\ref{tbl:lucizJ} and in Table~\ref{tbl:luciK}, where we list the seeing measured during the observations and the average signal-to-noise (S/N) in the observed wavelength ranges 1.05--1.10 $\mu$m and 2.04--2.20 $\mu$m of the final flux-calibrated spectra. 

Observations in the $zJ$ band were performed for all the quasars but one (J1507$+$2419), which was too faint to be observed with an exposure time comparable with the rest of the sample. Regarding the $\ks$ observations, a further constraint on the redshift ($z=3.19-3.29$) within the sample is dictated by the requirement that the [\ion{O}{iii}]  emission line falls in a wavelength range ($2.10-2.15$ $\mu$m) with good atmospheric transmission. This condition allowed us to observe only 9 targets, for which the $\ks$ spectroscopic data\footnote{$\ks$ observations were performed with LUCI1 only.}  were acquired from November 2018 to April 2019 with seeing around $0.7^{\prime\prime}-1.0^{\prime\prime}$.

The 2D raw spectra were reduced by the LBT Spectroscopic Reduction Center at INAF -- IASF Milano, with a reduction pipeline optimised for LBT data \citep[]{scodeggio2005,gargiulo2022sipgi} performing the following steps.
For each source, calibration frames are created for both LUCI1 and LUCI2. Imaging flats and darks are used to create a bad pixel map, applied to every single observed frame, along with a correction for cosmic rays. Dark and flat-field corrections are applied independently to each observed frame through a master dark and a master flat, obtained from a set of darks and spectroscopic flats. A master lamp describes the inverse solution of the dispersion to be applied to individual frames to calibrate in wavelength and remove any curvature due to optical distortions. The mean accuracy achieved for the wavelength calibration is 0.26 \AA\ in the $\ks$ band and 0.22 \AA\ (0.25 \AA) for LUCI1 (LUCI2) in the $zJ$ band.
The 2D wavelength-calibrated spectra are then sky-subtracted following the method described in \cite{davies2007}. The flux calibration is then applied to the 2D spectra through the sensitivity function, obtained from the spectrum of a star observed close in time and air-mass to the scientific target. Finally, wavelength- and flux-calibrated, sky-subtracted spectra are stacked together and the 1D spectrum of the source is extracted.
We checked the $zJ$ ($\ks$) flux calibration by convolving each spectrum with the 2MASS $J$ ($\ks$) filter to compute its $J$ ($\ks$) band Vega magnitude, and then corrected to match to observed value reported by SDSS DR16v4 \citep{ahumada202016th}. The uncertainty on the flux calibrated spectrum was estimated as the squared sum of a statistic term given by the mean rms of left and right telescopes and a calibration factor \rev{$\Delta m/m = (m_{\rm LBT}-m_{\rm 2MASS})/m_{\rm 2MASS}$}.

There is no overlapping region between the LBT $zJ$ spectra and the SDSS ones, leaving a small gap in between. Since the two segments of each spectrum were not collected simultaneously, part of the observed flux gap can be due to some intrinsic variation of the emission, but we expect this contribution to be rather small since more luminous quasars tend to be less variable in time \citep[e.g.,][]{uomoto1976image,cristiani1995optical,wilhite2008variability}. Systematics in the flux calibration could also be involved, but in principle their contribution should be small since SDSS spectrophotometric calibration is accurate to 4\% rms for point sources \citep{adelman2008ma}. \rev{LBT recalibration factors in flux with respect to the 2MASS $J$ ($\ks$) filter are in the range $0.6-1.9$ ($0.4-1.7$) with a mean value of $1.02$ ($0.93$).} 


\begin{table*}
\caption{\label{tbl:lucizJ}Log of the LUCI/LBT $zJ$ observations.}
\centering
\begin{tabular}{lccccc}
\hline\hline
Name & Obs. Date\tablefootmark{a} & $t_{\rm exp}$\tablefootmark{b} & S/N\tablefootmark{c} & Seeing\tablefootmark{d} & Instr.\tablefootmark{e} \\
\hline
J0301$-$0035 &  2019 Sep 28 & 1320 & 49 & 1.0 & LUCI1 \\
J0304$-$0008 &  2019 Sep 28 & 1680 & 43 & 1.0 & LUCI1 \\        
J0826+3148 &  2019 Oct 17 & 1920 & 28 & 1.0 & LUCI1 \\       
J0835+2122 &  2019 Oct 17 & 1680 & 35 & 1.1 & LUCI1 \\
J0900+4215 &  2020 Feb 02 & 480  & 22 & 1.2 & LUCI2 \\
J0901+3549 &  2020 Feb 02 & 1200 & 24 & 0.8 & LUCI2 \\
J0905+3057 &  2020 Feb 03 & 1680 & 27 & 1.1 & LUCI2 \\
J0942+0422 &  2020 Nov 12 & 960  & 24 & 1.2 & LUCI1$+$LUCI2 \\
J0945+2305 &  2020 Dec 26 & 6000 & 37 & 1.0 & LUCI1$+$LUCI2 \\
J0947+1421 &  2020 Dec 26 & 840  & 33 & 0.9 & LUCI1$+$LUCI2 \\
J1014+4300 &  2020 Dec 26 & 720  & 46 & 0.8 & LUCI1$+$LUCI2 \\
J1027+3543 &  2020 Dec 26 & 840  & 17 & 0.8 & LUCI1$+$LUCI2 \\
J1111$-$1505 &  2021 Apr 02 & 2700 & 26 & 0.9 & LUCI1$+$LUCI2 \\
J1111+2437 &  2021 Jan 09 & 3360 & 33 & 1.4 & LUCI1$+$LUCI2 \\
J1143+3452 &  2021 Jan 12 & 2160 & 33 & 1.0 & LUCI1$+$LUCI2 \\
J1148+2313 &  2021 Jan 17 & 1440 & 23 & 0.8 & LUCI1$+$LUCI2 \\
J1159+3134 &  2021 Jan 17 & 1800 & 18 & 0.9 & LUCI1$+$LUCI2 \\
J1201+0116 &  2021 Apr 02 & 960  & 35 & 0.9 & LUCI1$+$LUCI2 \\
J1220+4549 &  2021 Jan 31 & 1500 & 35 & 0.8 & LUCI1$+$LUCI2 \\
J1225+4831 &  2021 Feb 01 & 1920 & 18 & 0.7 & LUCI1$+$LUCI2 \\
J1246+2625 &  2020 Jun 23 & 1200 & 25 & 0.7 & LUCI2 \\
J1246+1113 &  2020 Jun 23 & 2400 & 15 & 0.8 & LUCI2 \\
J1407+6454 &  2020 Jun 23 & 1080 & 30 & 0.8 & LUCI2 \\
J1425+5406 &  2020 Jun 26 & 1050 & 29 & 0.9 & LUCI2 \\
J1426+6025 &  2020 Jun 27 & 300  & 15 & 0.8 & LUCI2 \\
J1459+0024 &  2020 Jun 27 & 1650 & 12 & 0.9 & LUCI2 \\
J1532+3700 &  2020 Jun 26 & 900  & 21 & 0.9 & LUCI2 \\
J1712+5755 &  2019 Oct 19 & 960  & 12 & 1.1 & LUCI1 \\
J2234+0000 &  2019 Sep 28 & 1350 & 16 & 0.9 & LUCI1$+$LUCI2 \\
\hline
\end{tabular}
\tablefoot{
\tablefoottext{a}{Observation Date.}
\tablefoottext{b}{Total exposure.}
\tablefoottext{c}{Signal-to-noise ratio in the observed wavelength range 1.05--1.10 $\mu$m.}
\tablefoottext{d}{Average seeing of the observation in arcseconds.}
\tablefoottext{e}{LUCI camera used for the observation.}
}
\end{table*}

\begin{table*}
\caption{\label{tbl:luciK}Log of the LUCI1/LBT $\ks$ observations.}
\centering
\begin{tabular}{lcccc}
\hline\hline
Name & Obs. Date\tablefootmark{a} & $t_{\rm exp}$\tablefootmark{b} & S/N\tablefootmark{c} & Seeing\tablefootmark{d}  \\
\hline
J0303$-$0023  &  2018 Nov 06     &  2400 & 25 & 0.9 \\ 
J0304$-$0008  &  2018 Nov 07     &  2420 & 15 & 1.0 \\
J0945$+$2305  &  2018 Nov 08--09 &  4960 & 18 & 0.7 \\
J0942$+$0422  &  2018 Nov 11     &  1650 & 38 & 1.0 \\
J1220$+$4549  &  2019 Jan 26     &  2875 & 10 & 1.0 \\
J1111$+$2437  &  2019 Jan 26     &  2645 & 21 & 1.0 \\
J1201$+$0116  &  2019 Jan 28     &   800 & 29 & 1.7 \\
J1425$+$5406  &  2019 Feb 28     &  2415 & 42 & 0.8 \\
J1426$+$6025  &  2019 Apr 27     &   600 & 51 & 1.0 \\
\hline
\end{tabular}
\tablefoot{
\tablefoottext{a}{Observation Date.}
\tablefoottext{b}{Total exposure.}
\tablefoottext{c}{Signal-to-noise ratio in the observed wavelength range 2.04--2.20 $\mu$m.}
\tablefoottext{d}{Average seeing of the observation in arcseconds.}
}
\end{table*}

\section{Analysis}
\label{analysis}
\subsection{Fitting procedure}
\label{fit_procedure}
The spectral fits of both the $zJ$ and $\ks$ data were performed through  a custom-made code, based on the IDL MPFIT package \citep{Markwardt2009}, which takes advantage of the Levenberg-Marquardt technique \citep{more1978levenberg} to solve the least-squares problem.

The region roughly between 2400 \AA\ and 3200 \AA\ corresponding to the rest frame of the $zJ$ spectra is mainly characterised by \ion{Fe}{ii} and \ion{Fe}{iii} emission lines (which blend in a pseudo-continuum), the Balmer continuum, and the \ion{Mg}{ii} 2798 \AA\ emission line. In this region there are only two small continuum windows between 2650--2670 \AA\ and 3030--3070 \AA\ \citep{mejia2016active}, but the former was often included in the atmospheric absorption band which affects the range $\sim$1.11--1.16 $\mu$m in the observed frame. This, together with the lack of other bright emission lines and/or other continuum windows, led to the problem of having only one narrow interval to anchor the power-law continuum, thus leading to a degeneracy between the slope of the power law and the strength of iron emission.

\rev{In order to break this degeneracy, we followed a similar approach to the one described in \citet{vietri2018}. Thus, we adopted the same value of the continuum power-law slope as the one derived from the SDSS spectra in Paper II for each source. In the six cases where multiple SDSS observations were available (J0303--0023, J0303--0008, J111--1505, J1159+3134, J1407+6454, J1459+0024), we assumed the average of the individual best-fit spectral indices.} We then accounted for the remaining emission with the required number of \ion{Fe}{ii} templates, as produced by different synthetic photoionization models through the \textssc{CLOUDY} simulation code \citep{ferland2013}, and convolved with different Gaussian profiles with a velocity dispersion of up to 7,000 km s$^{-1}$. Broad (FWHM\,$>$\,1,000 km s$^{-1}$) Gaussian components were added in some cases, to provide a more faithful description of the iron profile. This procedure gave satisfactory results for most of the spectra (see appendix~\ref{LBTzJspectraatlas}). Furthermore, in Appendix \ref{fe_mgii_check} we checked the reliability of fixing the power-law slope in order to match the one of the continuum underlying the \ion{C}{iv} line and the continuum windows at bluer wavelengths, and we compared our way of estimating the strength of \ion{Fe}{ii} with archival data.

\begin{figure}[h!]
\includegraphics[width=\linewidth,clip]{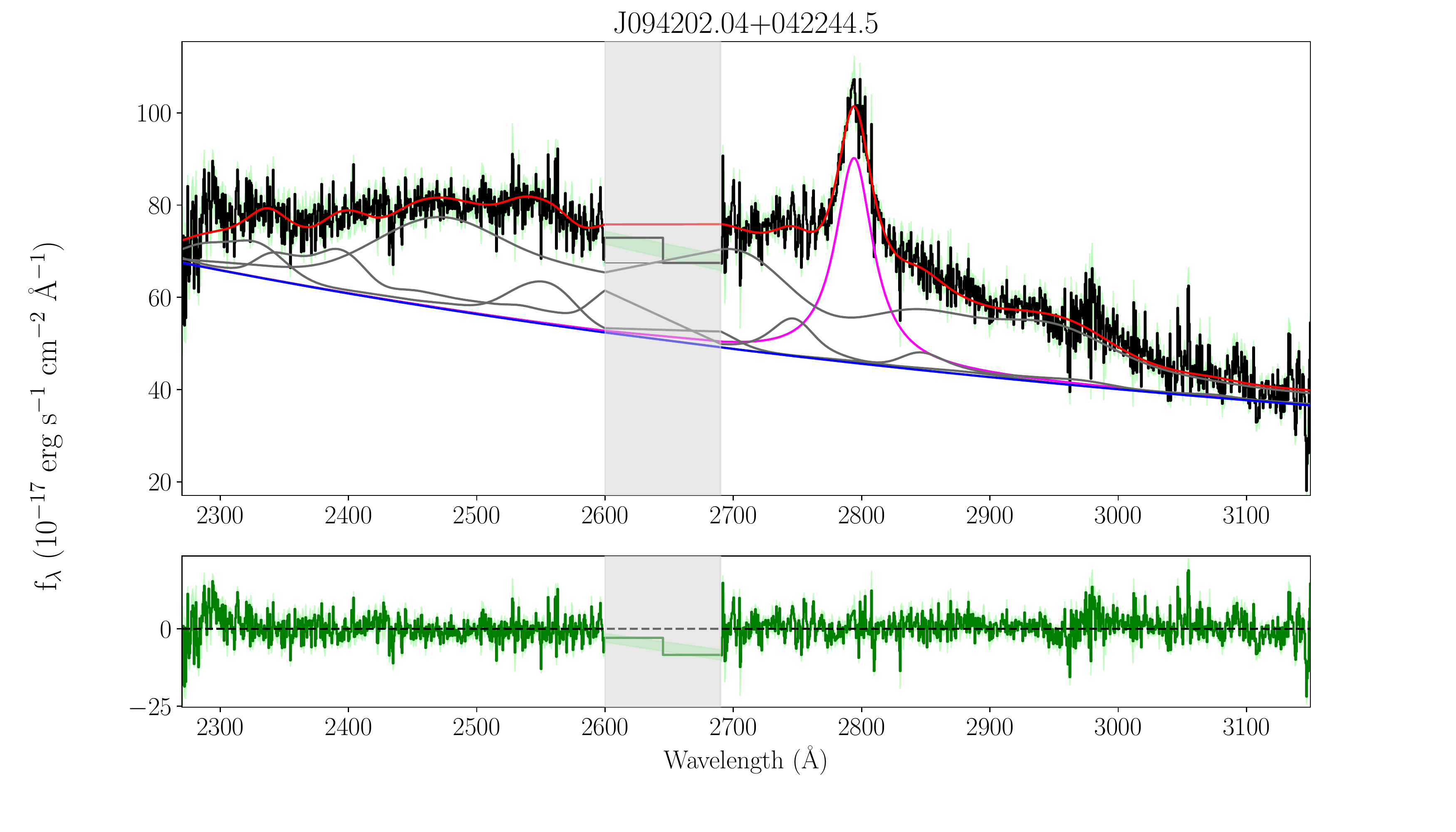}
 \caption{\textit{Top panel}: spectrum (black), global model (red) and model components of a typical LBT $zJ$ fit (J0942+0422). The continuum power law is shown in blue, the \ion{Mg}{ii} line in magenta, the iron pseudo-continuum in dark grey. The shaded light grey band corresponds to the observed-frame atmospheric absorption window at 1.11--1.16 $\mu$m and was not included in the fit. 
 \textit{Bottom panel}: global model residuals (green).}
 \label{fig:ex_fit_lbtj}
\end{figure}
\begin{figure}[h!]
\includegraphics[width=\linewidth,clip]{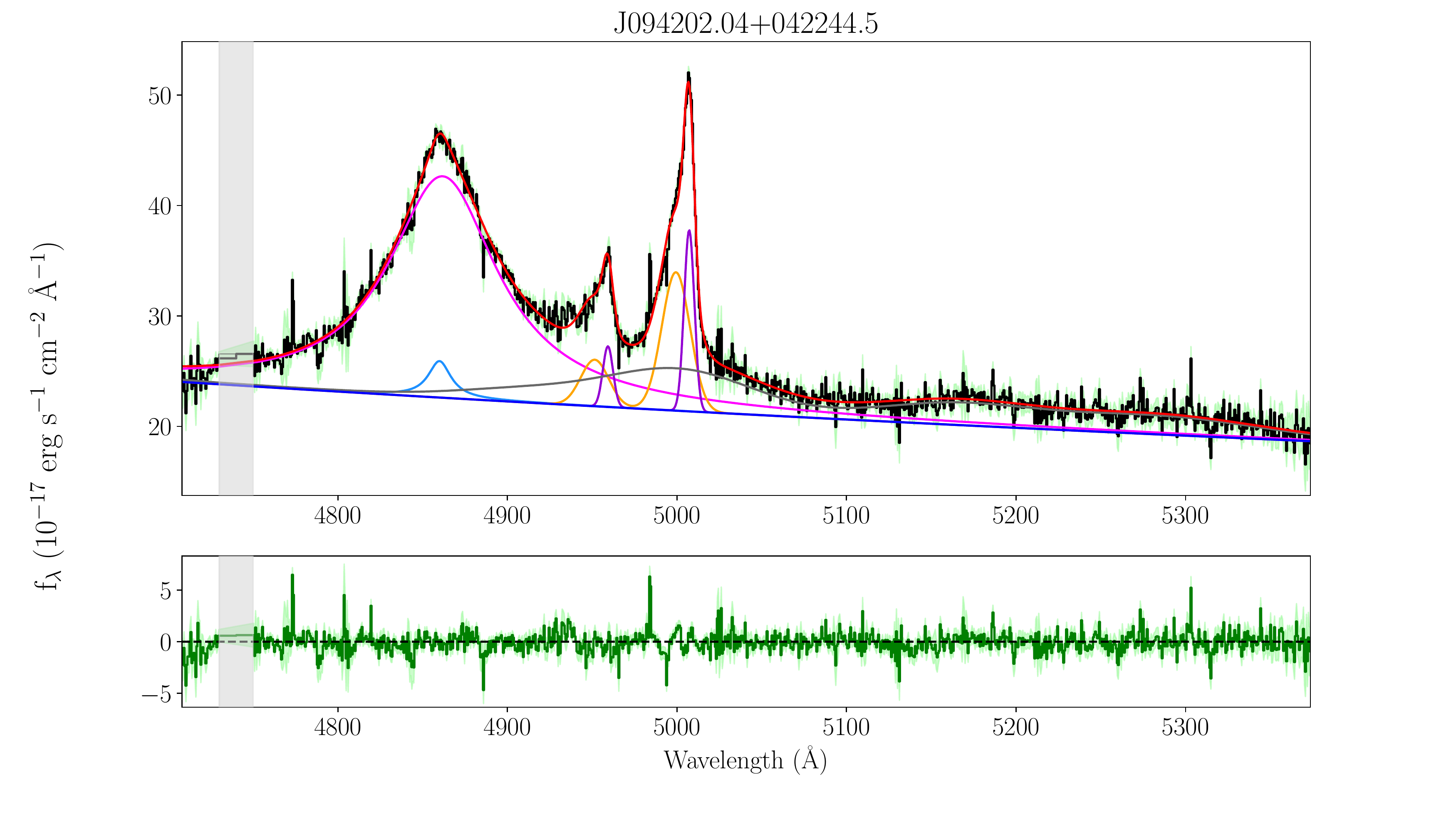}
\caption{\textit{Top panel}: spectrum (black), global model (red) and model components of a typical LBT $\ks$ fit (J0942+0422). The continuum power law is shown in blue, the broad and narrow H$\beta$ in magenta and turquoise, the [\ion{O}{iii}] doublet in orange and violet, the iron pseudo-continuum in dark grey. The shaded light grey band is affected by poor atmospheric transmission and was not included in the fit. 
\textit{Bottom panel}: global model residuals (green).}
\label{fig:ex_fit_lbtk}
\end{figure}

The model adopted to fit the $\ks$ spectra included a multi-Gaussian (one broad, one or two narrow) deconvolution for the emission lines (i.e., H$\beta$ $\lambda4861$ and [\ion{O}{iii}] $\lambda\lambda 4959,5007$), and \ion{Fe}{ii} templates to account for the optical iron emission. The ratio between the core [\ion{O}{iii}] $\lambda$4959 and $\lambda$5007 components was fixed to be equal to 1/3, and a blue component for both [\ion{O}{iii}] lines was also included \rev{to account for possible outflows from the narrow line region (NLR)}.
Examples of the fits performed on typical LBT $zJ$ and $\ks$ band spectra are shown, respectively, in Fig. \ref{fig:ex_fit_lbtj} and Fig. \ref{fig:ex_fit_lbtk}.

\subsection{Composite spectra}
To build the composite LBT spectra for the X-ray weak and X-ray normal quasars, we followed a similar procedure to the one described in \citet{lusso2015}. For the sake of consistency with the assumptions made in Paper II, we excluded from the stack the radio-bright (J0900$+$4215), the two BAL (J0945$+$2305, J1148$+$2313), and the reddest (J1459$+$0024) quasars. As the latter three are X-ray weak, we want to avoid an enhanced flatness of the resulting composite spectrum as a result of their BAL/red nature. This further selection brought the X-ray weak group of LBT $zJ$ ($\ks$) spectra down to 7 (4) objects. In particular, to build the composite spectra:

\begin{enumerate}
 \item We corrected the quasar flux density\footnote{In the following, we will use the word ``flux'' to mean the flux density, i.e., flux per unit wavelength, unless specified otherwise.} $f_{\lambda}$ for Galactic reddening by adopting the $E(B-V)$ estimates from \citet{schlegel98} and the Galactic extinction curve from \citet{fitzpatrick1999correcting} with $R_V=3.1$.

 \item We generated a rest-frame wavelength array with fixed dispersion for the $zJ$ ($\ks$) spectra with $\Delta\lambda$ equal to 2.25 \AA\, (2.39 \AA), roughly corresponding to the resolution at the central wavelength of the observed spectra ($R$=1125 at 
 1.05 $\mu$m and $R$=2075 at 
 2.105 $\mu$m, respectively, for $zJ$ and $\ks$), shifted to the rest frame according to the mean quasar redshift.
 
 \item Each quasar spectrum was then shifted to the rest frame and linearly interpolated over the rest-frame wavelength array with fixed dispersion $\Delta\lambda$, while conserving its flux.

 \item We normalized every spectrum by their integrated flux over the wavelength ranges 2400--3100 \AA\ ($zJ$) and 4800--5300 \AA\ ($\ks$), which are covered by all the spectra. 
 
 \item In each spectral channel we extracted the median value of the normalized fluxes. The uncertainty on the median flux in a spectral channel was estimated as the 95\% semi-interquartile range of the fluxes divided by the square root of the number of spectra in that channel.

 \end{enumerate}

\begin{figure}[h!]
\includegraphics[width=\linewidth,clip]{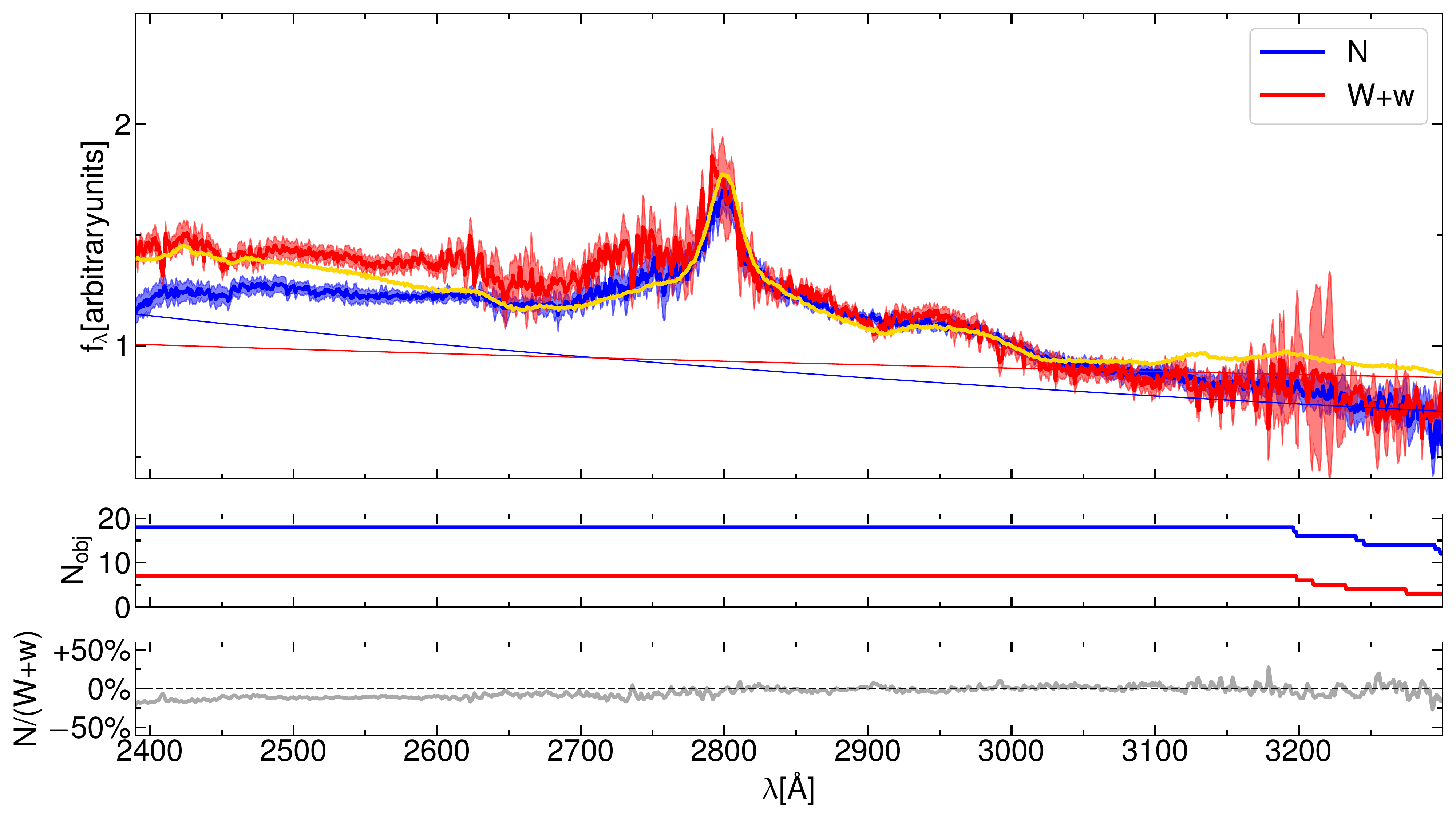}
 \caption{\textit{Top panel}: median LBT $zJ$ spectra for the X-ray normal ($N$, in blue) and the X-ray weak (\textit{W+w}, in red) subsamples. The gold spectrum is the average quasar spectrum from \citet{vandenberk2001}. Fluxes are normalized by their value at 3000 \AA. The continuum power laws are the extrapolation of the ones found at UV wavelengths (i.e., SDSS).
\textit{Middle panel}: number of spectra contributing to each spectral channel, according to the same colour code.
 \textit{Bottom panel}: ratio between the $N$ spectrum and the \textit{W+w} one.}
 \label{fig:median_lbtj}
\end{figure}

\begin{figure}[h!]
\includegraphics[width=\linewidth,clip]{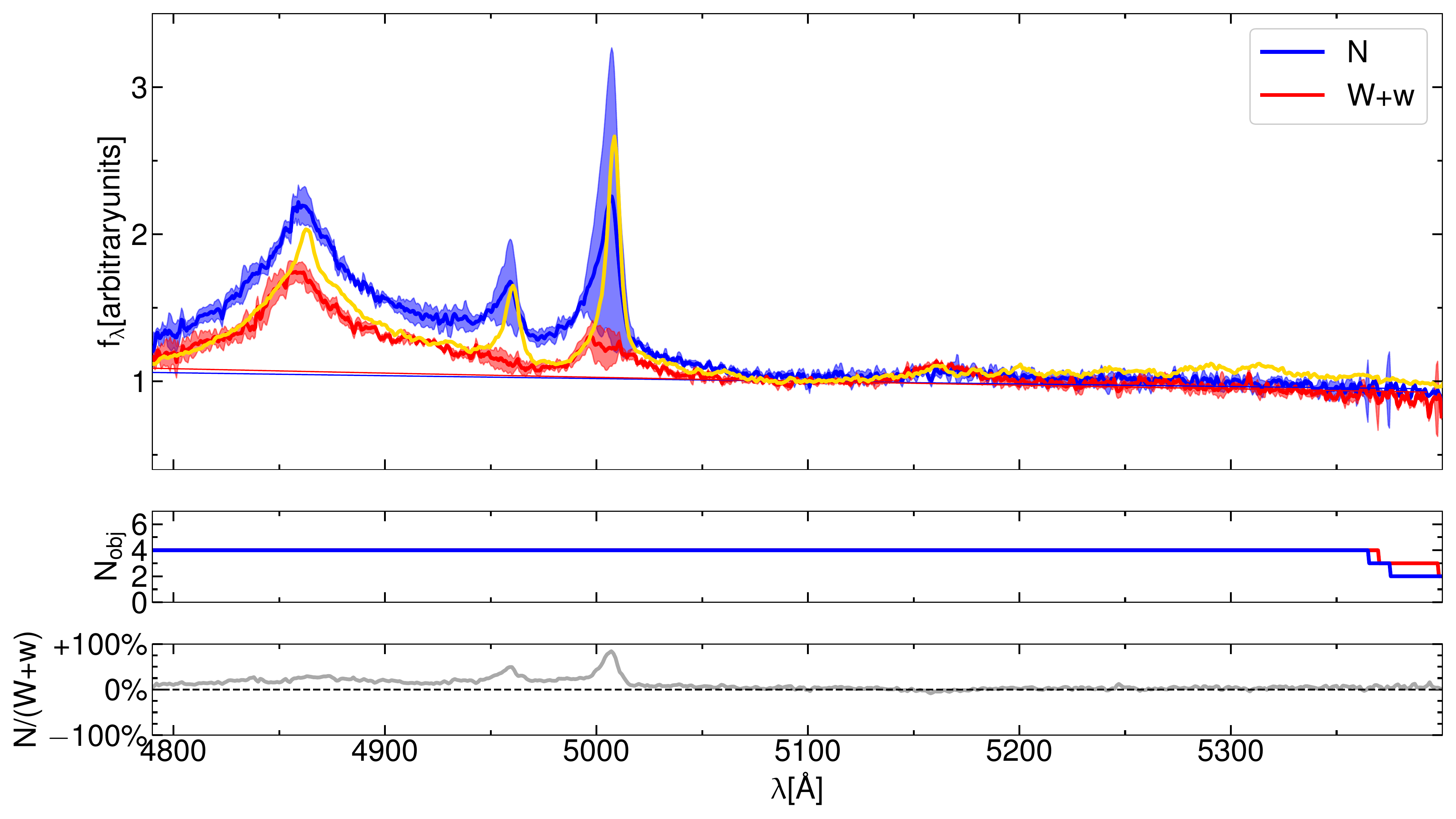}
\caption{\textit{Top panel}: median LBT $\ks$ spectra for the X-ray normal and the X-ray weak subsamples. The color code is the same as in Figure \ref{fig:median_lbtj}. Fluxes are normalized by their  value at 5100 \AA. 
\textit{Middle panel}: number of spectra contributing to each spectral channel, according to the same colour code.
\textit{Bottom panel}: ratio between the $N$ spectrum and the \textit{W+w} one. }
\label{fig:median_lbtk}
\end{figure}

The spectral stacks obtained with the procedure described above are shown in Figures \ref{fig:median_lbtj} and \ref{fig:median_lbtk} for the $zJ$ and the $\ks$ data, respectively. As a reference, we also overplot the average quasar spectrum from \citet{vandenberk2001}, which is built from 2204 SDSS spectra spanning a redshift range $0.044\leq z \leq 4.789$.
If we extrapolate the slope of the continuum found in Paper II to the wavelengths covered by the LBT spectra, the flatter continuum in the X-ray weak composite hints at a larger EW of \ion{Fe}{ii} compounds, as suggested also by the analysis of the individual sources.
We will further discuss this point in Section~\ref{results}. Despite the emission line differences, both composites are in broad agreement with the reference one, thus implying no strong evolution of the general spectral properties of the $z\sim3$ sample with respect to AGN at other redshifts.

\subsection{Black-hole masses and Eddington ratios}
\label{sec:BH}

We computed single-epoch BH masses from the available emission lines with known reliable virial relations for each object. In the SDSS spectra, the \ion{C}{iv}\,$\lambda 1549$ line is available for the whole sample, whereas the \ion{Mg}{ii}\,$\lambda 2798$ line, present in all the LBT $zJ$ spectra, in 18/29 cases is fully or marginally hidden by atmospheric absorption, hindering a reliable determination of its FWHM. H$\beta$ was used for the 9 objects of the LBT $\ks$ subsample. 
It is well-known that BH masses from different lines have a different reliability: H$\beta$-based masses are generally regarded as the benchmark \citep[see for instance][]{denney2012outflows,shen2013mass,db2020}, but in this case they are only available for a minority of sources. \ion{Mg}{ii}-based masses are consistent, within their non-negligible systematic uncertainty ($\sim$0.3 dex; \citealt{shen2011}), with the H$\beta$ ones. 
On the other hand, while \ion{C}{iv}-based masses only provide a very rough estimate of the mass of the black hole powering the AGN, this emission feature is present for every object of the sample, and in 18/29 objects this is the only available tool to estimate the BH mass. The inclusion of \ion{C}{iv}-based BH masses comes with two caveats.
First, the \ion{C}{iv} emission line centroid can shows blueshifts up to
about 10,000  km s$^{-1}$ with respect to the systemic redshift, suggesting the presence of an outflowing phase \citep{2005MNRAS.356.1029B,richards2006spectral}, not compatible with the virial assumption under which BH masses are estimated. This can cause an overestimate of the BH mass by up to an order of magnitude \citep[e.g.,][]{kratzer2015mean}.
Secondly, \ion{C}{iv} calibrations should be considered with caution since they are affected by significant scatter due to different systematics, and allow us to derive BH masses only at the price of large uncertainties, $\gtrsim 0.4$ dex \citep[][]{shen2011,shenliu2012,rakshit2020spectral,wu2022catalog}. To mitigate these issues, BH masses based on the \ion{C}{iv} FWHM were estimated by adopting the corrections described in \citet{coatman2017}, suitable for  objects with a blueshifted \ion{C}{iv} emission. We employed the offset velocities reported in Paper II to perform the correction of the FWHM and subsequently of $M_{\rm BH}$ for objects whose blueshifts are positive, whereas the correction factors were set to unity otherwise (3/29 objects), since the authors themselves cautioned against the use of their correction in case of negative blueshift (i.e., redshift) of the \ion{C}{iv} line centroid.

We estimated single-epoch masses based on the continuum luminosity ($\lambda L_{\lambda}$) evaluated close to the considered emission line and its FWHM through the following expression:
\begin{equation}
  \log\left(\frac{M_{\rm BH}}{M_{\odot}} \right) = A+B \, \log \left(\frac{\lambda L_{\lambda}}{10^{44} \textrm{erg} \, \textrm{s}^{-1}}\right)+2\, \log\left(\frac{\rm FWHM}{\rm km \, s^{-1}}\right),
\end{equation}
where the $A$ and $B$ coefficients have been calibrated by different authors for each line as: 
\begin{equation}
  (A,B) =
    \begin{cases}
      (0.71,0.53) \,\, $\ion{C}{iv}$ \, 1549 \, \AA  & \text{\citet{coatman2017}}\\
      (0.74,0.62) \,\, $\ion{Mg}{ii}$ \, 2798 \, \AA & \text{\citet{shen2011}}\\
      (0.70,0.50) \,\, \rm{H\beta} & \text{\citet{bongiorno2014m}.}
    \end{cases}   
    \label{eq:eq1}
\end{equation}    
The wavelengths  1350 \AA, 3000 \AA, and 5100 \AA\ have been adopted to estimate the continuum luminosity for the \ion{C}{iv}, \ion{Mg}{ii}, and H$\beta$ emission lines, respectively. The data availability from the UV to the visible for 9 objects allowed us to verify the reliability of the calibrations by comparing pairs of BH mass estimates. 
SMBH masses from the \ion{C}{iv} line were compared to the ones already estimated in \citet{wu2022catalog}. To this purpose, we computed the distribution of the differences $\Delta \log(M_{\rm BH}) = \log(M_{\rm BH,CIV})-\log(M_{\rm BH,SDSS})$. The mean value and the standard deviation of this distribution are $\langle \Delta \log(M_{\rm BH}) \rangle = -0.05$ and $\sigma_{\Delta \log(M_{\rm BH})}=0.3$. The minor offset of the distribution could be due to the prescription for the evaluation of the BH mass (in \citealt{wu2022catalog} the authors adopted the calibration from \citealt{vestergaard2006determining}) and/or to the fitting procedure, but in general we do not find strong outliers. 

\begin{figure}[h!]
\includegraphics[width=\linewidth,clip]{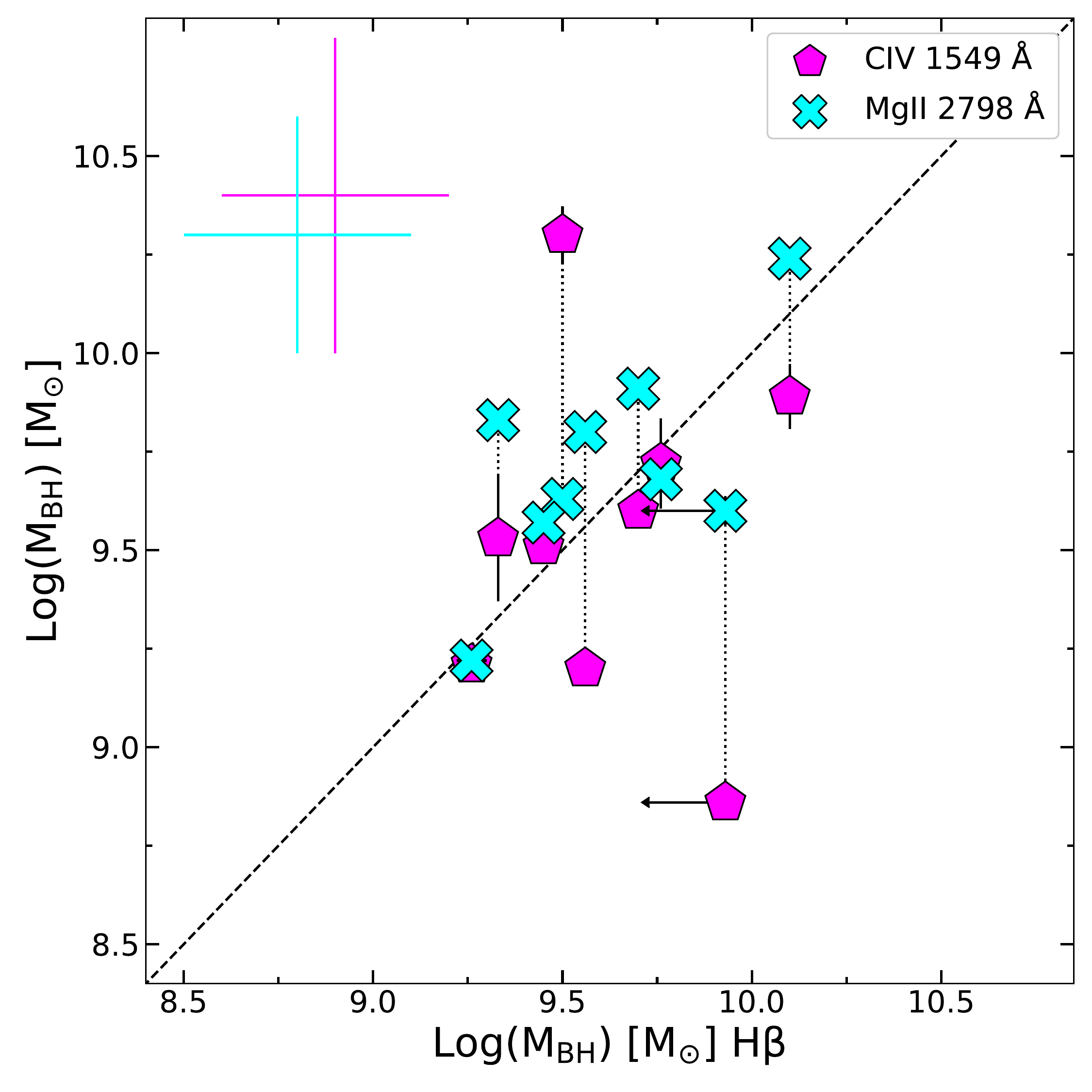}
\caption{Single-epoch $M_{\rm BH}$ comparison. The estimates based on \ion{C}{iv}\,$\lambda1549$ (magenta pentagons) and \ion{Mg}{ii}\,$\lambda2798$\ (cyan crosses) are shown against the broad H$\beta$\,$\lambda4861$ ones. The typical systematic uncertainty of the calibrations is shown in the top left corner with the same colour code, while the dashed black line represents the 1:1 relation. \rev{Dotted lines join BH mass estimates for the same object according to different emission lines}. The H$\beta$ BH mass for J1111$+$2437 is denoted as an upper limit since the line profile is not well defined (see Appendix \ref{individual_notes} for details).}

\label{fig:bh_masses}
\end{figure}

For all the objects whose spectra included the H$\beta$ emission line, also the \ion{Mg}{ii} $\lambda2798$ line and the $3000$ \AA\ luminosity were available, therefore it was possible to provide an additional estimate of the BH masses and compare them with the H$\beta$-based ones.
\ion{C}{iv}- and \ion{Mg}{ii}-based BH masses against H$\beta$-based are shown in Figure \ref{fig:bh_masses}.
The uncertainty is dominated by the systematic term \rev{(0.4 dex for \ion{C}{iv} and 0.3 dex for \ion{Mg}{ii} and H$\beta$-based masses)}, being the statistical uncertainty on the BH masses on average 17\% for the \ion{C}{iv}, 5\% for the H$\beta$, and 2\% for the \ion{Mg}{ii} estimates. The fiducial mass for each object was derived as a weighted mean, using as weight the total uncertainty on each mass estimate, given by the square root of the squared sum of the systematic term and the statistical one. BH masses are listed in Table A\ref{tbl:TA1} together with the best-fit line and continuum parameters.

Eddington ratios ($\lambda_{\rm Edd} = L_{\rm bol}/L_{\rm Edd}$) were calculated by assuming the standard definition of $L_{\rm Edd} = 1.26\times 10^{38}\,(M_{\rm BH}/M_{\odot})$ erg s$^{-1}$. The $L_{\rm bol}$ value was computed for each object as stated in Paper I, by employing the 1350 \AA\ monochromatic luminosity available from SDSS photometry and the bolometric correction of \citet{richards2006spectral}.
Considering the uncertainties on the BH masses, as well as the ones on the bolometric conversion factors, which in the case of the 1350 \AA\ luminosity can be up to 50\% \citep{richards2006spectral}, we can only give crude estimates of the Eddington ratios. We found that, \rev{on average, our sources are close to the Eddington limit, with a median $\lambda_{\rm Edd}$ of 0.9}, which is expected given the very high luminosities observed in the $z\sim3$ sample.

\section{Results}
\label{results}
\subsection{\ion{Mg}{ii} and \ion{Fe}{ii} emission}
Intense \ion{Fe}{ii} emission and high \ion{Fe}{ii}/\ion{Mg}{ii} ratios are typically observed for X-ray weak sources (e.g., PHL 1811, \citealt{Leighly2007b}).
We thus estimated the rest-frame equivalent width for both the \ion{Mg}{ii} line and the \ion{Fe}{ii} emission complex to assess whether possible differences arise between the X-ray weak and X-ray normal quasars in our sample. However, the significance\footnote{Throughout this work, the significance is reported as $m \sigma$ with $m = |<x_N>-<x_{W+w}>|/\sqrt{\sigma_{x_N}^2/n_N +  \sigma_{x_{W+w}}^2/n_{W+w}}$, where $<x_i>$ is the mean value, $\sigma_i$ the standard deviation, and $n_i$ the size of the \textit{i-th} sample.} of any difference between the two samples is limited by the small statistics 
(18 $N$ vs.~10 \textit{W+w} objects -- excluding the radio bright J0900+4215 -- in the LBT $zJ$, and 4 $N$ vs.~5 \textit{W+w} objects in the LBT $\ks$ spectral samples).

We found that the mean value of EW\,\ion{Mg}{ii} with the standard error of the mean for the $N$ group is $\langle$EW\,\ion{Mg}{ii}$\rangle_{N}$=59$\pm$9 \AA, while $\langle$EW\,\ion{Mg}{ii}$\rangle_{W+w}$=66$\pm$13 \AA\ for the \textit{W+w} group,  without any statistically significant difference between the two samples. However, potential differences could be diluted by the blend of \ion{Mg}{ii} with the Fe emission. In this analysis, we considered all the \ion{Mg}{ii} lines for which the full profile was available, including the BAL J0945$+$2305 (the other BAL and the reddest object did not show an analysable \ion{Mg}{ii} profile). However, the results do not change if we exclude the latter source.

The mean EW\,\ion{Fe}{ii} of the $N$ sample is $\langle$EW\,\ion{Fe}{ii} $\rangle_{N}$=292$\pm$40 \AA, which is 
somewhat smaller 
than the value obtained for the \textit{W+w} sample, $\langle$EW\,\ion{Fe}{ii}$\rangle_{W+w}$=469$\pm$70 \AA. The difference is statistically significant at the $2.2 \sigma$ level, and a Kolmogorov-Smirnov (KS) test provides a p-value=0.017, implying that the two distributions are indeed different at the 98.3\% level. In the estimate of \ion{Fe}{ii} mean values we neglected the source J1225$+$4831, whose power-law continuum, as extrapolated from the SDSS spectrum, is significantly steeper than the one required to adequately fit the \ion{Mg}{ii} emission line (see appendix~\ref{individual_notes} for details).
Even attempting a free-slope fit, there is a very faint \ion{Fe}{ii} contribution. 

Both EW\,\ion{Fe}{ii} and EW\,\ion{Mg}{ii} were estimated by normalising the integrated flux of the line by the continuum flux at 3000 \AA, as generally done for such ratios \citep[e.g.,][]{sameshima2020mg}.
The ratio between the equivalent widths of \ion{Fe}{ii} and \ion{Mg}{ii} shows, on average, a higher value for the \textit{W+w} sample, $\langle$\ion{Fe}{ii}/\ion{Mg}{ii}$\rangle_{W+w}$=8.4$\pm$1.4, than for the $N$ one, $\langle$\ion{Fe}{ii}/\ion{Mg}{ii}$\rangle_{N}$=4.4$\pm$0.5.

\begin{figure}[h!]
\includegraphics[width=\linewidth,clip]{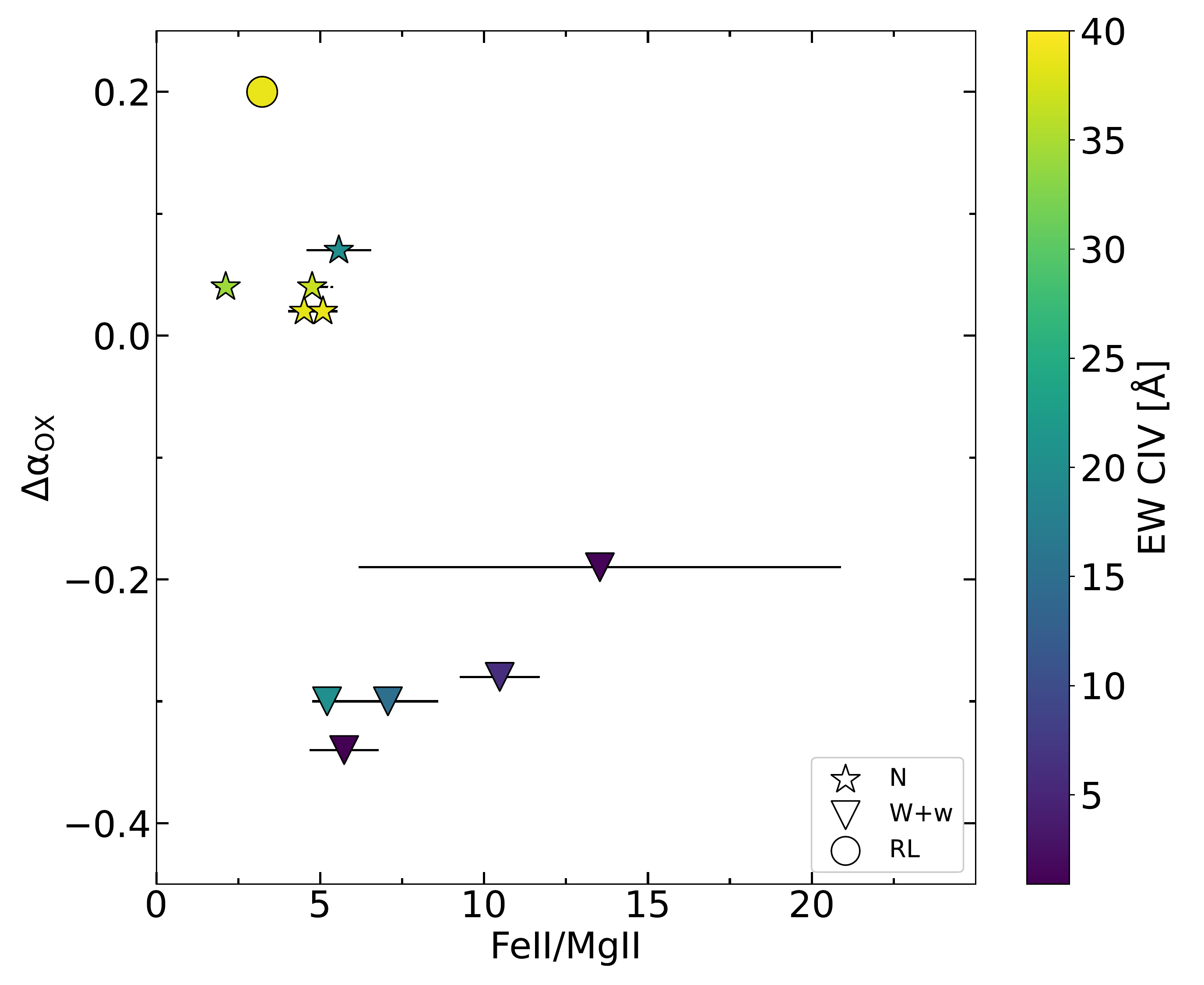}
\caption{$\Delta \alpha_{\rm OX}$\,--\,(\ion{Fe}{ii}/\ion{Mg}{ii}) plane for our $N$ (stars) and $W+w$ (triangles) sources. The radio--loud quasar J0900+4215 is marked as a circle. We observe an increasing trend of the \ion{Fe}{ii}/\ion{Mg}{ii} ratio with increasing X-ray weakness. In colour code, the EW\,\ion{C}{iv} value is reported to compare the newly observed properties of the \ion{Mg}{ii} region with the relative weakness of the \ion{C}{iv} emission.}
\label{fig:daox_fe_mgii}
\end{figure}

Figure \ref{fig:daox_fe_mgii} shows the $\Delta \alpha_{\rm OX}$ (the difference between the observed $\alpha_{\rm OX}$ and that predicted from the $\alpha_{\rm OX}-L_{\rm UV}$ relation for objects within the same redshift interval of our sample),\footnote{For a more detailed discussion about how the values of $\Delta \alpha_{\rm OX}$ are evaluated, we refer to Section 2.3 of Paper II, and references therein.} an index of X-ray weakness,  as a function of the \ion{Fe}{ii}/\ion{Mg}{ii} ratio. On average, the \ion{Fe}{ii}/\ion{Mg}{ii} ratio is higher in X-ray weak quasars with respect to X-ray normal ones. We also added EW\,\ion{C}{iv} in colour code to compare this trend with the modest decrease of \ion{C}{iv} with increasing X-ray weakness observed in Paper II. \rev{We assessed this trend by means of a Spearman's rank test, which yielded a correlation index of $r_S=-0.7$. A possible origin for such a trend is discussed in Sec. \ref{sec:discuss}, in terms of increased \ion{Fe}{ii}\textsubscript{UV} emission in X-ray weak quasars associated with outflow-induced shocks and turbulence.}
The statistical uncertainty on the \ion{Fe}{ii}/\ion{Mg}{ii} ratio was estimated by fitting 100 mock spectra for each source: the flux in every spectral channel was created by adding a random value to the actual flux, extracted from a Gaussian distribution whose amplitude was set by the uncertainty value in that spectral channel. After fitting every mock sample, we computed the distribution of the \ion{Fe}{ii}/\ion{Mg}{ii} values, and set the uncertainty as the standard deviation of the distribution, \rev{after applying a 3$\sigma$ clipping}.

\begin{figure}[h!]
\includegraphics[width=\linewidth,clip]{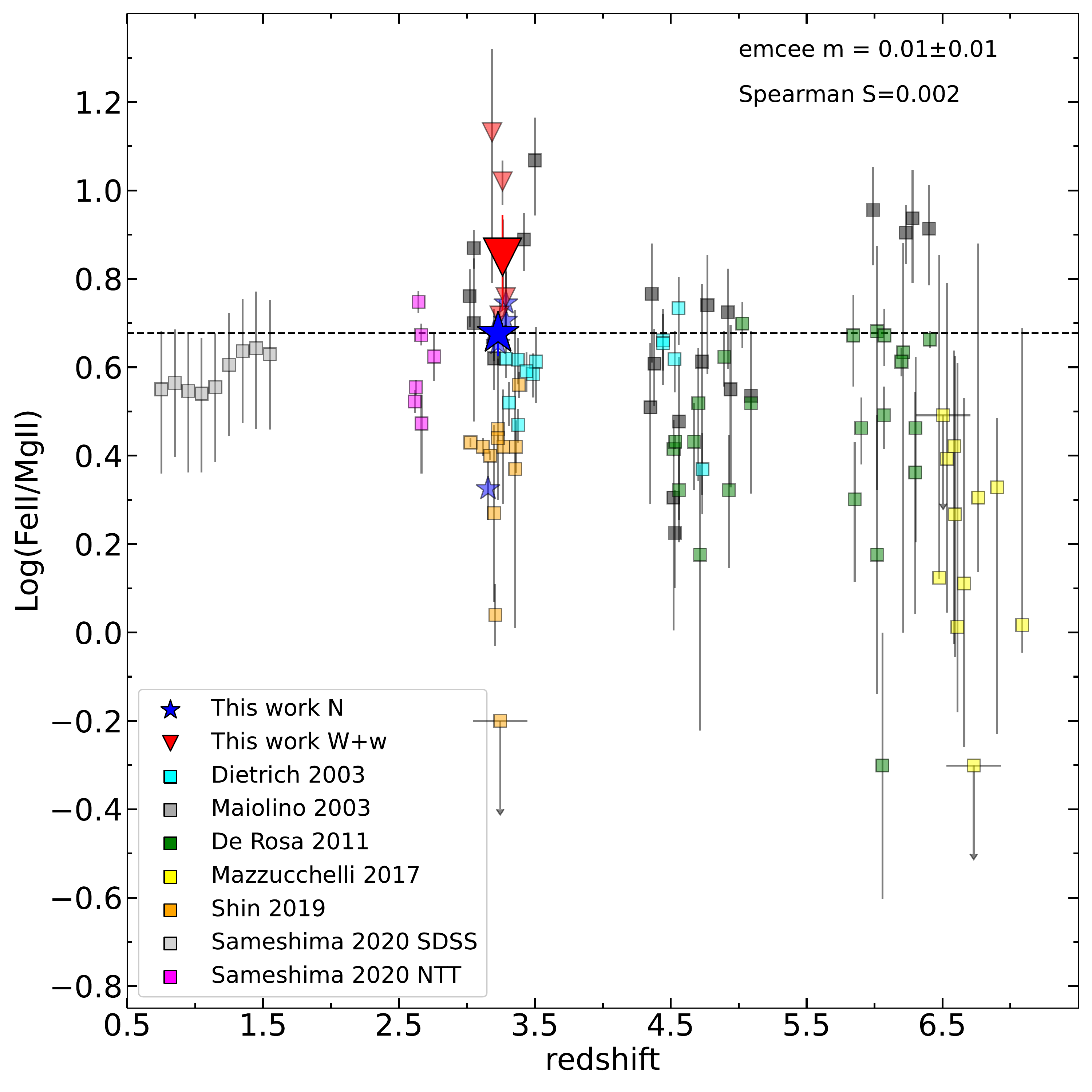}
\caption{Log(\ion{Fe}{ii}/\ion{Mg}{ii})\,--\,$z$ plane. $N$ sources (blue stars) follow the expectations for objects at similar redshift, whereas \textit{W+w} sources (red triangles) are located at the top of the distribution around $z\sim 3$. The black dashed line represents the mean \ion{F}{ii}/\ion{Mg}{ii} ratio of our $N$ quasars extrapolated over the entire redshift range.}

\label{fig:feii_mgii}
\end{figure}

In Figure \ref{fig:feii_mgii} we show the estimates of \ion{Fe}{ii}/\ion{Mg}{ii} for our sample of 10 quasars where the \ion{Mg}{ii} line was observed (excluding J0900$+$4215, which is flagged as radio-bright), in comparison with other literature samples probing different redshift intervals \citep[][]{dietrich2003fe,maiolino2003early,de2011evidence,mazzucchelli2017physical,shin2019fe,sameshima2020mg}. We found that, on average, there is no clear trend for an evolution in the \ion{Fe}{ii}/\ion{Mg}{ii} ratio across cosmic time, which implies already chemically enriched BLR regions at high redshift. To verify quantitatively this trend, we performed a Spearman's rank order probability test by using the X-ray normal quasars only together with the other samples, in order to avoid any possible bias introduced by the boosted \ion{Fe}{ii}/\ion{Mg}{ii} ratios of X-ray weak sources. We then performed a linear fit of all the data sets together with \emcee \citep{emcee}. 
The Spearman's test yielded a correlation coefficient of $-$0.26, revealing a mild trend of decreasing \ion{Fe}{ii}/\ion{Mg}{ii} ratio with redshift. The fit of the \ion{Fe}{ii}/\ion{Mg}{ii}$-z$ relation then provides evidence for a flat slope, confirming a non-significant evolution of this ratio across cosmic time. Although the highest redshift sample from \citet{mazzucchelli2017physical} exhibits systematically lower values than the ones at lower redshift, the same authors note that the uncertainties are so large that the consistency with a non-evolving \ion{Fe}{ii}/\ion{Mg}{ii} ratio cannot be ruled out. Indeed, if we exclude this sample from the regression analysis, we find an even lower correlation coefficient (0.002), and again a slope virtually consistent with zero ($m=0.01 \pm 0.01$). However, it is possible that, at very high redshift ($z \gtrsim 6$), we might be observing a genuine depletion of iron, symptomatic of the presence of young stellar populations in the galaxies hosting these quasars. 

Our results further confirm the interpretation that samples at high redshifts, biased towards high luminosities (i.e., $L_{\rm bol}>10^{46}$ erg s$^{-1}$), presumably host SMBHs in already chemically-mature galaxies \citep[e.g.,][]{kawakatu2003protoquasars,juarez2009metallicity}. Indeed, the values of their \ion{Fe}{ii}/\ion{Mg}{ii} ratios are consistent with the ones at lower redshifts. Possible systematic effects and a consistency check on our ability to reproduce the iron emission in our objects are discussed in Appendix \ref{fe_mgii_check}.

A strong correlation between the \ion{Fe}{ii}/\ion{Mg}{ii} ratio and the Eddington ratio has been observed, but not with the BH mass \citep[][see also \citealt{dong2011controls}]{sameshima2017chemical}. 
Since the $N$ and the \textit{W+w} samples are fairly homogeneous in terms of BH masses and Eddington ratios (see Sec. \ref{fig:bh_masses}), it is unlikely that one of these parameters is the fundamental driver of the observed difference. 
Theoretical works using \textsc{cloudy} simulations showed that several physical properties can impact the \ion{Fe}{ii}/\ion{Mg}{ii} ratio, such as gas density and microturbulence \citep{verner2003revisited,baldwin2004origin,sameshima2017chemical,temple2020fe}. We will perform a detailed photoionization modelling of the SDSS and LBT data in a dedicated publication.  

\subsection{H$\beta$ properties}
\label{sec:hbeta}

The 9 rest-frame optical spectra enabled us to investigate the properties of the H$\beta$--[\ion{O}{iii}] complex.
We found that, on average, X-ray weak sources display a \rev{weaker} H$\beta$ emission than X-ray normal ones. The mean value of the EW for the \textit{W+w} group is $\langle$EW\,H$\beta\rangle_{W+w}$=\rev{73$\pm$8 \AA, } whereas it is $\langle$EW\,H$\beta\rangle_{N}$=\rev{108$\pm$11} \AA\, for the \textit{W+w} sample, a difference statistically significant at the $2.6 \sigma$ level. 

In order to assess whether the Fe emission of X-ray weak sources is enhanced also in the optical band, we checked the \ion{Fe}{ii}/H$\beta$ ratio, which is a standard optical indicator of the metallicity, generally evaluated as the intensity ratio between the integrated flux of \ion{Fe}{ii} between 4434 \AA\ and 4684 \AA\ and the H$\beta$ one \citep[e.g.,][]{bg1992,marzianisulentic2014}. 
The relative intensities of the optical Fe emission are statistically consistent, as we found that the average equivalent width of the optical \ion{Fe}{ii} of the $N$ sample is $\langle$EW\,\ion{Fe}{ii}$_{\rm opt}\rangle_{N}$=16$\pm$6 \AA, whereas the \textit{W+w} sample yielded $\langle$EW\,\ion{Fe}{ii}$_{\rm opt}\rangle_{W+w}$=\rev{24$\pm$9 \AA.} 
The ratio is $\langle$\ion{Fe}{ii}$_{\rm opt}$/H$\beta \rangle_{N}$=0.14$\pm$0.06 for the $N$ sources, while the \textit{W+w} ones give $\langle$\ion{Fe}{ii}$_{\rm opt}$/H$\beta \rangle_{W+w}$=0.36$\pm$0.13, \rev{therefore the mild difference between the \ion{Fe}{ii}/H$\beta$ ratios is just mimicking the difference between the H$\beta$ profiles of the two samples}.
Yet, we caution that the 4434--4684 \AA\ interval over which the optical \ion{Fe}{ii} is generally sampled in the literature is not included in our spectra, and we thus had to rely on a full extrapolation for this estimator.
For this reason, we also checked whether any difference could be found in the observed region using the equivalent width of the \ion{Fe}{ii}$_{\rm opt}$ emission between 4900 \AA\ and 5300 \AA, which we directly observed and included in our fits. Also in this case, \rev{the difference remains marginal as the $N$ and \textit{W+w} groups respectively yield $\langle$\ion{Fe}{ii}$_{\rm opt}$/H$\beta \rangle_{N}$=0.31$\pm$0.07 and $\langle$\ion{Fe}{ii}$_{\rm opt}$/H$\beta \rangle_{W+w}$=0.72$\pm$0.32.}

The ratio between UV and optical \ion{Fe}{ii} emission is larger for X-ray weak sources, being \rev{$\langle$\ion{Fe}{ii}$_{\rm UV}$/\ion{Fe}{ii}$_{\rm opt}\rangle_{N}$=13$\pm$4 and $\langle$\ion{Fe}{ii}$_{\rm UV}$/\ion{Fe}{ii}$_{\rm opt}\rangle_{W+w}$=26$\pm$9}, in line with the expectations of higher ratios of UV to optical \ion{Fe}{ii} emission for increasingly weaker SEDs in the extreme UV (EUV), as shown in \citet{Leighly2007b}. We note, however, that our results are not directly comparable with the ones in the third panel of Fig. 24 in \cite{Leighly2007b}, since the \ion{Fe}{ii} emission is evaluated on a much shorter interval (4900--5300 \AA) than the one therein (4000--6000 \AA).

\subsection{[\ion{O}{iii}] properties}
\label{sec:oiii}
In the context of the unified model, the [\ion{O}{iii}] is produced in the NLR, on galactic scales, and it is believed to be an isotropic indicator of the AGN strength in both type I and type II AGN (e.g., \citealt{mulchaey1994,bassani1999three,netzer2009}, and references therein). The [\ion{O}{iii}] luminosity ($L$\textsubscript{[\ion{O}{iii}]}) is a secondary indicator of the nuclear luminosity, depending on the fraction of continuum radiation within the opening angle of the torus reaching the gas in the NLR, but is also influenced by local properties such as the NLR clumpiness, its covering factor, and the amount of dust extinction \citep[e.g.,][]{ueda2015iii}. 

The EW\,[\ion{O}{iii}] value can be considered as a proxy of the inclination of our line of sight to the AGN accretion disc (e.g., \citealt{risaliti2011iii, shenho2014}). \rev{ \citet{bisogni2017inclination} investigated in detail the distribution of EW\,[\ion{O}{iii}] in the SDSS DR7, showing that 
EW\,[\ion{O}{iii}] $>$ 30 \AA\ generally corresponds to high inclination angles, while lower values reflect the intrinsic EW\,[\ion{O}{iii}] distribution. Only one of our objects (J0303$-$0008) displays an EW\,[\ion{O}{iii}] in excess of 30 \AA, thus possibly being observed at relatively large inclination (yet still likely within $\theta\lesssim60\degr$, see Sec. \ref{sec:discuss}). The mean (median) value of EW\,[\ion{O}{iii}] in our sample is 14.4$\pm$7.6 (3.7) \AA, or 6.6$\pm$2.3 \AA\ after excluding the strong [\ion{O}{iii}] emitter, consistent with the median value of the EW\,[\ion{O}{iii}] distribution from the current SDSS release \citep{wu2022catalog}, which is 14.1 \AA, and well below the 30 \AA\ threshold where inclination effects should become relevant.}

\begin{figure}[h!]
\includegraphics[width=\linewidth,clip]{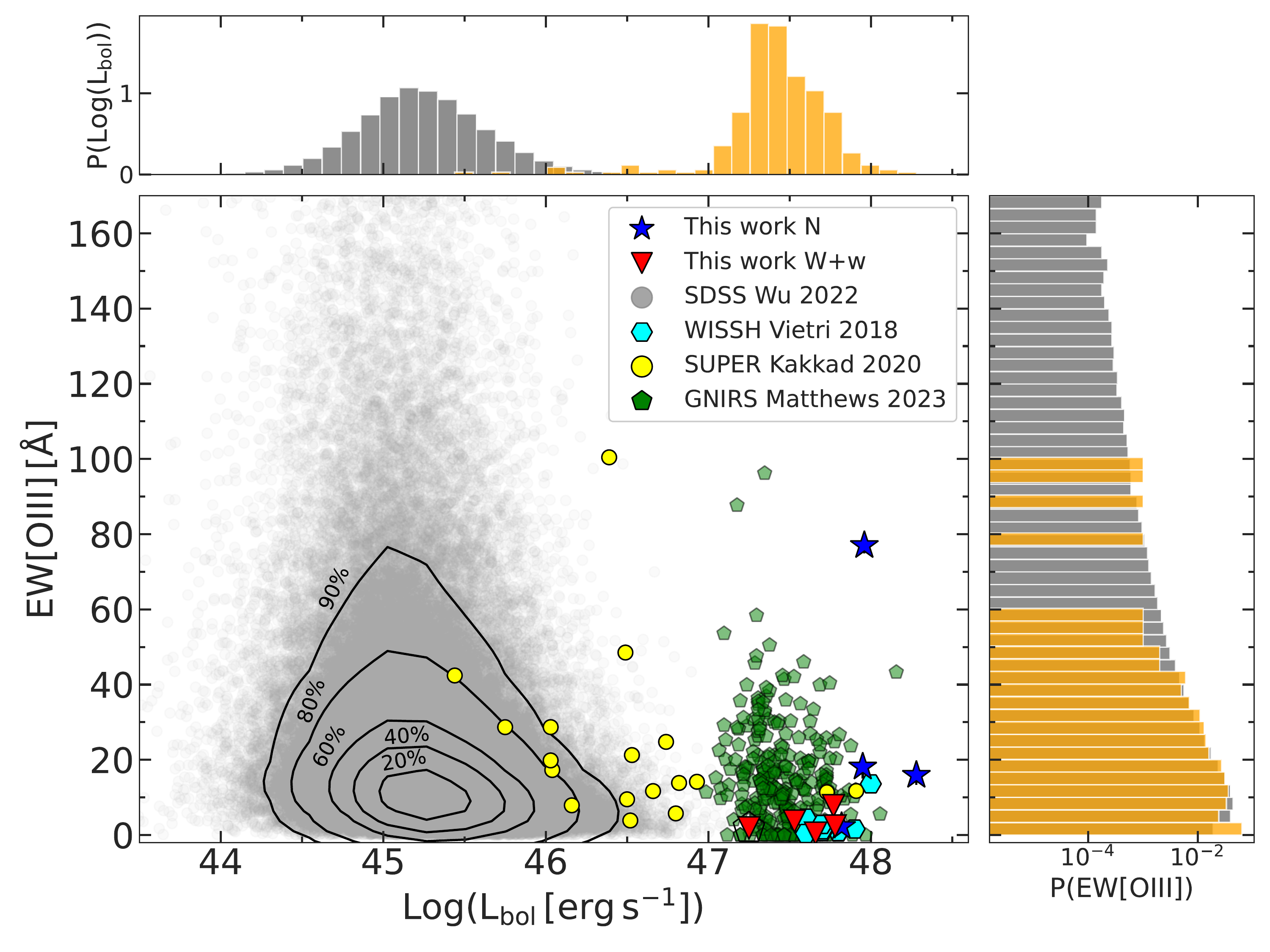}
\caption{\rev{Distribution of EW\,[\ion{O}{iii}] of our and other reference samples, as a function of bolometric luminosity. Side panels show the probability density of the quantities represented in the plot. The BAL J0945+2305 is marked with a hollow pentagon. All but one of our objects show low ($<$\,30 \AA\,) EW\,[\ion{O}{III}], suggesting a non-edge-on line of sight. Generally, luminous high-redshift quasars tend to cluster around low EW\,[\ion{O}{iii}] values.}}
\label{fig:distr_oiii}
\end{figure}

\rev{With the aim of putting our sample into a broader context, we compared the results concerning the [\ion{O}{iii}] emission with other samples of luminous high-redshift quasars in the literature.
The sample analyzed in \citet{vietri2018}, part of the WISSH survey, exhibits extremely weak [\ion{O}{iii}] profiles, with a median value of 1.5 \AA. The 19 AGN at $z\sim2$ from the SUPER sample described in \citet{kakkad2020super} yield, instead, a median value of 14.2 \AA, although these sources span a wide interval in terms of bolometric luminosity (10$^{45.4}$--10$^{47.9}$ erg s$^{-1}$). When matching the SUPER sample in luminosity with our own quasars (i.e., considering only the objects with $L_{\rm bol}$ exceeding 10$^{46.9}$ erg s$^{-1}$), the median value of EW\,[\ion{O}{iii}] decreases slightly to 11.6 \AA.\footnote{We also note that for the SUPER sample EW [\ion{O}{iii}] is not directly reported by the authors, so it was estimated here as $L$\textsubscript{[\ion{O}{iii}]}$/L_{5100\,\AA}$. This normalization could produce a slightly overestimated EW\,[\ion{O}{iii}] in case of a steeply decreasing optical continuum. However, by assuming an average continuum slope $\alpha_{\lambda}=-0.42$ from \citet{vandenberk2001}, normalizing at 5100 \AA\ rather than at 5007 \AA\ has a negligible effect on EW\,[\ion{O}{iii}].}
Lastly, the median EW\,[\ion{O}{iii}] for the GNIRS--DQS quasars \citep{matthews2023gemini}, equals 12.7 \AA, noting that for this sample $\sim 15 \% $ of the sources have an unreliable EW measurements because of the weak [\ion{O}{iii}] profile.
Although the cumulative EW\,[\ion{O}{iii}] distribution of these luminous high-redshift samples clusters around the same peak of the global SDSS quasar distribution \citep{wu2022catalog}, objects with high equivalent width become relatively rarer, as shown in the right side panel in Fig. \ref{fig:distr_oiii}. For instance, the fraction of objects with EW\,[\ion{O}{iii}]\,$>$\,50 \AA\, in the SDSS catalogue is $\sim$10\%, while the joint incidence in all the mentioned high-redshift samples is about 2\%, although some systematic effects could marginally modify this estimate given the non-uniform analysis (e.g., continuum fitting windows, different Fe templates) of the different samples.}

\rev{The weaker [\ion{O}{iii}] profiles in very luminous quasars with respect to the global SDSS population is
likely related to the luminosity evolution of this parameter. Indeed, it has been shown that the EW of the [\ion{O}{iii}] core component anti-correlates with $L_{5100\,\AA}$, which can be regarded as a proxy of $L_{\rm bol}$ \citep{shenho2014}. Two effects might come into play. Assuming that the intrinsic intrinsic $L$\textsubscript{[\ion{O}{iii}]} does not evolve significantly with $L_{\rm bol}$ (or $z$), the observed anti-correlation would be mainly driven by inclination, whereby high-$L_{\rm bol}$, high-$z$ sources are preferentially observed with a ``face-on'' line of sight.
Otherwise, if $L$\textsubscript{[\ion{O}{iii}]} is increasing more slowly than $L_{\rm bol}$ (e.g., \citealt{shen2016rest}), we would also witness a trend of decreasing EW\,[\ion{O}{iii}], as for the standard Baldwin effect (\citealt{baldwin1977}; see also Sec. 4.1 of \citealt{ueda2015iii} for other possible effects).}

\rev{As for other emission lines (\ion{C}{iv}, \ion{H}{$\beta$}), we generally observe less prominent line profiles in \textit{W+w} objects also for the [\ion{O}{iii}]\,$\lambda \lambda$4959,5007 doublet {(see also \citealt{green1998})}. The mean values for the two subsamples are $\langle$EW\,[\ion{O}{iii}]$\rangle_{N}$=16.6$\pm$5.6 \AA\ and $\langle$EW\,[\ion{O}{iii}]$\rangle_{W+w}$=4.5$\pm$1.5 \AA, giving a $\sim$2.1$\sigma$ tension. In this computation we conservatively excluded from the $N$ sample the only high-[\ion{O}{iii}] emitter, J0303$-$0008, for which inclination effects might be non negligible. The EW\,[\ion{O}{iii}] values of all the other sources suggest that these are likely seen at relatively low inclination instead, hence the tentative difference between the strength of the [\ion{O}{iii}] emission in $N$ and \textit{W+w} quasars could be due to intrinsic effects. At this stage, it is therefore more informative to consider $L$\textsubscript{[\ion{O}{iii}]}, rather than EW\,[\ion{O}{iii}]. As we do not expect any systematic difference between the geometrical (size, covering factor) or physical (metallicity, density) properties of the NLR in X-ray weak and X-ray normal quasars, we argue that the main driver of $L$\textsubscript{[\ion{O}{iii}]} is the line emissivity (e.g., \citealt{baskin2005controls}), which ultimately depends on the shape of the EUV/soft-ray SED.}

\begin{figure}[h!]
\includegraphics[width=\linewidth,clip]{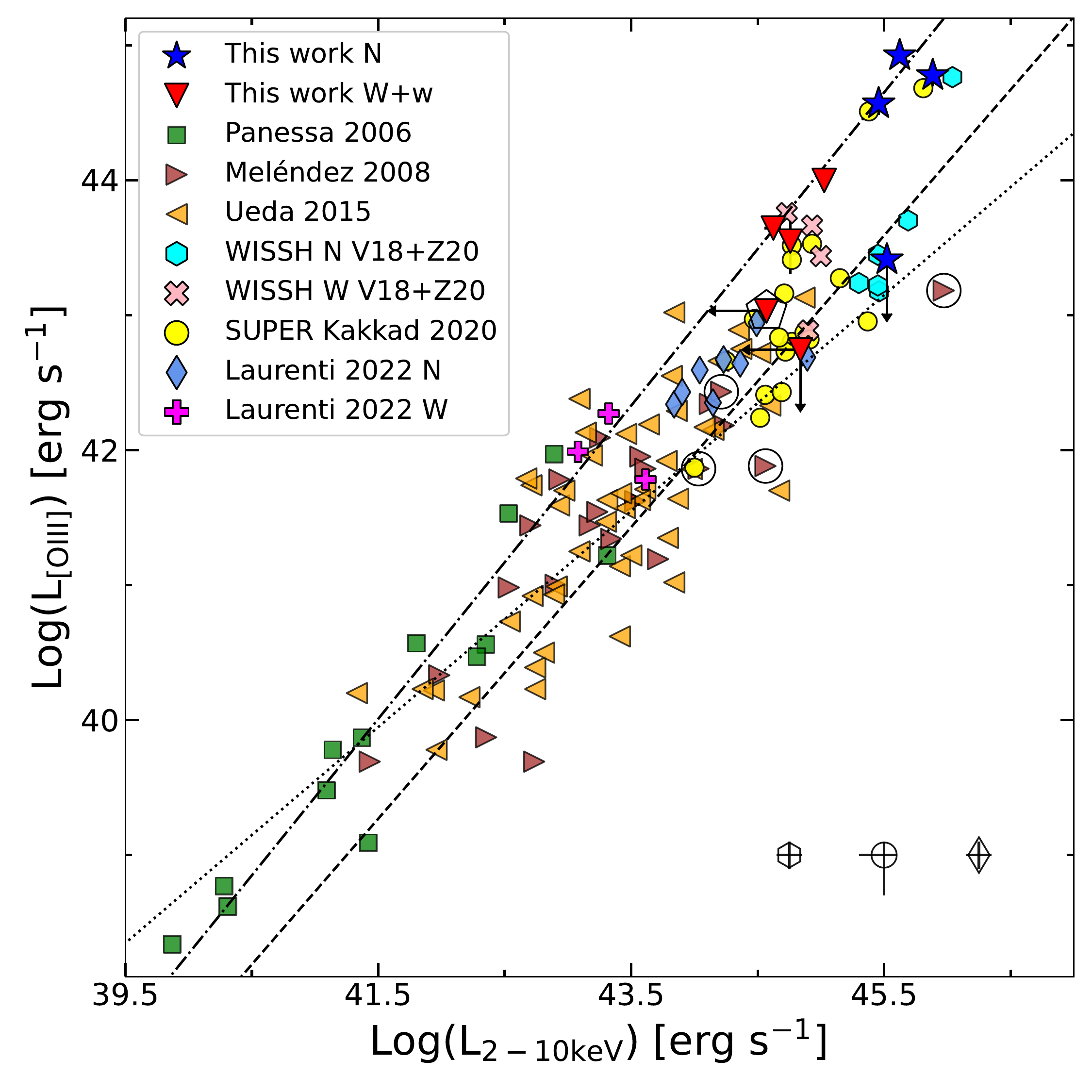}
\caption{Log($L$\textsubscript{[\ion{O}{iii}]})--log($L$\textsubscript{2--10 keV}) plane. The dot-dashed, dotted and dashed lines represent the best fits for the \citet{panessa2006x}, \citet{melendez2008new}, and \citet{ueda2015iii} samples, respectively. \rev{Other luminous high-redshift samples with currently available [\ion{O}{iii}] and X-ray data from the WISSH (\citealt{vietri2018, zappacosta2020}) and SUPER \citep{kakkad2020super} surveys, as well as the less luminous sample from \citet{laurenti2021x} including some X-ray weak objects, are also shown}.
Our objects where the [\ion{O}{iii}] is barely detectable (i.e., J0303$-$0023 and J1425+5406) are represented as upper limits. Circled data are defined as radio-loud by the authors. BAL J0945+2305 is marked by a hollow pentagon. \rev{The typical uncertainties for the WISSH, SUPER, and \citet{laurenti2021x} samples are respectively shown at the bottom of the plot.}}
\label{fig:loiii_lhx}
\end{figure}

\rev{If the range of [\ion{O}{iii}] intensities that we observe is caused by intrinsic differences in the unobservable portion of the SED, a correlation with the X-ray emission should be expected, as the level of the latter determines the SED steepness. Several studies have investigated the relation between [\ion{O}{iii}] and X-ray emission, finding a high degree of correlation  (e.g., \citealt{mulchaey1994, panessa2006x, melendez2008new, ueda2015iii}). Indeed, being both proxies of the intrinsic power of the central engine, $L_{\rm X}$ and $L$\textsubscript{[\ion{O}{iii}]} are naturally expected to correlate in the case of type I objects, where the line of sight does not cross the dusty torus (the correlation holds also in type II objects, when considering absorption--corrected $L_{\rm X}$). In Figure \ref{fig:loiii_lhx} we show the relation between the hard (2--10 keV) X-ray and the [\ion{O}{iii}] luminosity of our quasars together with other reference samples \citep{panessa2006x,melendez2008new,ueda2015iii}. We also include the objects with currently available [\ion{O}{iii}] and X-ray data from \citet{laurenti2021x} and, in the high-luminosity tail, also those from the WISSH and SUPER samples. For the WISSH and the \citet{laurenti2021x} samples we also divided the sources in X-ray normal and X-ray weak according to their $\Delta \alpha_{\rm OX}$ value, adopting a conservative threshold of $\Delta \alpha_{\rm OX} \leq -0.3$ to define X-ray weakness. Objects whose [\ion{O}{iii}] emission is barely detectable are labelled as upper limits. Our sample seems to follow the trend of less luminous objects, even though, because of its very faint [\ion{O}{iii}], J0303$-$0023 is slightly below the other $N$ sources. The $L_{\rm X}$ values of J0945+2305 and J1425+5406 are labelled as upper limits, being marginally detected as stated in Paper I. Remarkably, despite their weak [\ion{O}{iii}] profiles, X-ray weak objects do not drop out from the main trend of the $L$\textsubscript{[\ion{O}{iii}]}-- $L_{\rm X}$ relation.}

We also tested, at least qualitatively, the possibility that the lower $L$\textsubscript{[\ion{O}{iii}]} in X-ray weak objects were due to dust extinction in the NLR. We did not have the possibility to use the Balmer decrement, the most straightforward way to assess local extinction, because the H$\alpha$ line is not covered by our spectra. To check if $N$ and \textit{W+w} quasars had intrinsically similar [\ion{O}{iii}] emission lines, but the latter subsample suffered from a larger dust extinction, we considered $N$ quasars with average [\ion{O}{iii}] emission (i.e., we excluded J0303$-$0023, whose emission is barely detectable, and J0304$-$0008, the brightest [\ion{O}{iii}] emitter) and allowed for increasing dust extinction. The detailed procedure is explained in Appendix \ref{oiii_ext_check}. Even in the case of significant reddening, $E(B-V)=0.2$, [\ion{O}{iii}] is still clearly detectable. A high degree of extinction is also disfavoured by the locus occupied in the $\rm{\Gamma_{1450-3000\,\AA}-\Gamma_{0.3-1\,\mu m}}$ plane (see Fig. 2 of Paper II) by the whole sample, which was selected in order to include blue unobscured quasars.

\rev{We finally note that the [\ion{O}{iii}] profile does not generally reveal signatures of strong outflows, with the only exception of J0942$+$0422, where a fairly broad ($\sim$1200 km s$^{-1}$) component is detected, blueshifted with respect to the core component by $\sim$470 km s$^{-1}$ (Fig. \ref{fig:ex_fit_lbtk}). It is possible, however, that in some cases, where the [\ion{O}{iii}] profile is weak and blended with \ion{Fe}{ii}\textsubscript{opt}, any blueshifted [\ion{O}{iii}] component, resulted undetectable even if present.}

\subsection{Relation to accretion parameters}
\label{sec: accretion_parameters}

The quasars at high redshift constituting this sample have been chosen so as to display a high degree of homogeneity in the UV, being very luminous with a blue spectrum, according to the criteria described in Paper I. We then found that on the X-ray side the sources are far less homogeneous, since the sample also includes a significant fraction of X-ray weak objects. We then focused on the spectroscopic optical/UV properties to find any evidence for differences accompanying the X-ray weakness in the \textit{W+w} sample.

\begin{figure}[h!]
\includegraphics[width=\linewidth,clip]{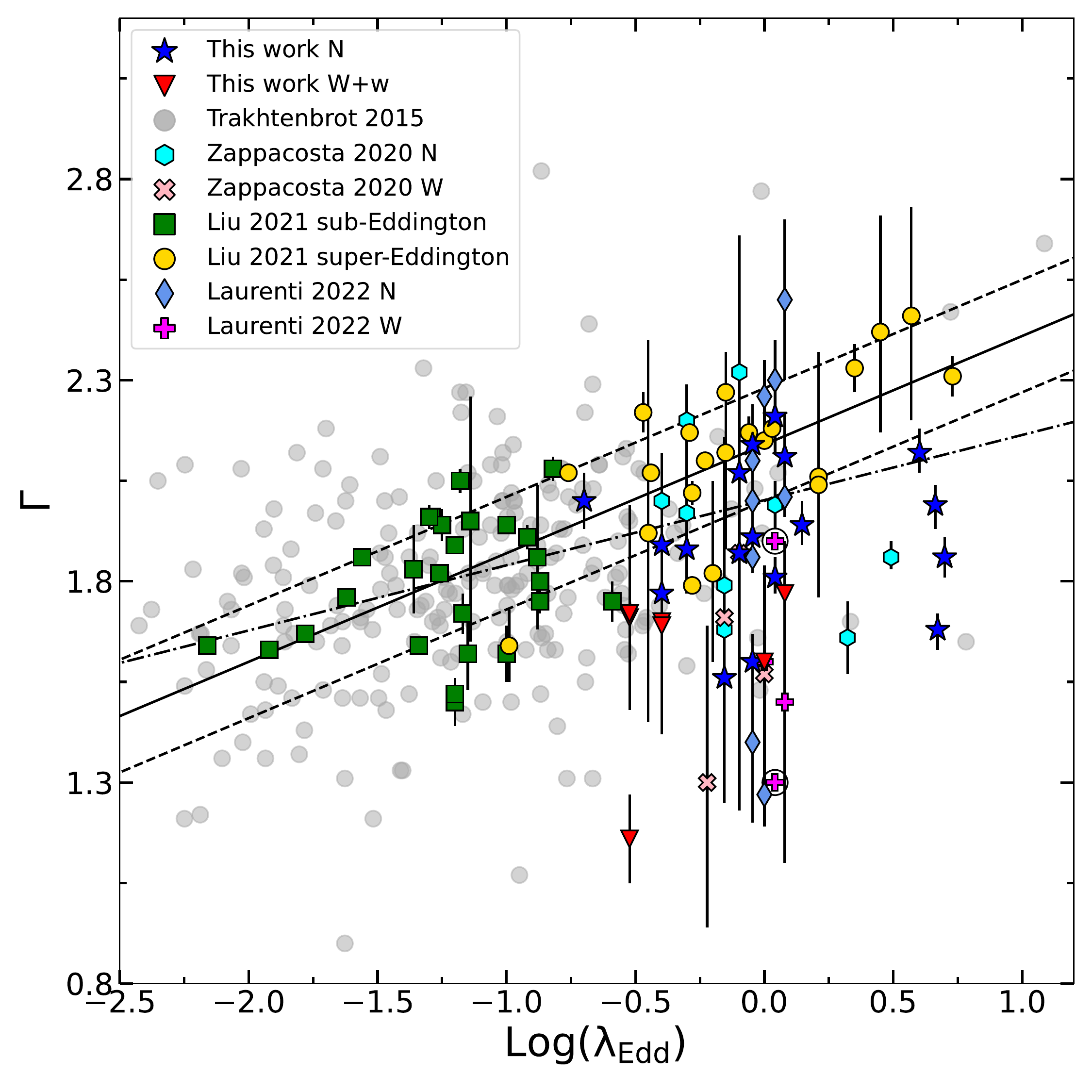}
\caption{$\Gamma$--log($\rm{\lambda_{Edd}}$) plane. Highly accreting samples are labeled as indicated in the legend. Circled sources have a fixed $\Gamma$ in their X-ray analysis. The black continuous line represents the best fit from \citet{liu2021observational}, which produced a slope $\beta=0.27\pm 0.04$, while the dashed line represents the 1$\sigma$ dispersion. The dot dashed line is our best fit including all the samples but the \citet{trakhtenbrot2017bat} one, for which we had no information about possible X-ray weak sources, which results in a flatter slope $\beta=0.16\pm 0.03$.}
\label{fig:gamma_lambda}
\end{figure}

The newly estimated (and generally more reliable for 11/29 objects) BH masses as well as the Eddington ratios allowed us to investigate further what the key accretion parameters of the sample are, and test the possibility that the observed differences between the two X-ray groups could be associated to their location in this parameter space. 
In Figure \ref{fig:gamma_lambda} we present the relation between the Eddington ratio and the X-ray photon index for our sample, together with other recent samples in the literature formed by both super- and sub-Eddington accretors. Although with a huge scatter, a steeper $\Gamma$ is observed as $\lambda_{\rm Edd}$ increases.
The Spearman's rank test for our $z\sim3$ sample, together with other non-weak sources (shown in Fig. \ref{fig:gamma_lambda}), produced a correlation index $r_S=0.47$, lower than the ones reported by \citet{liu2021observational}, i.e., $r_S=0.72$ with a p-value$=1.27\times10^{-8}$ for their full sample, and $r_S=0.60$ with a p-value$=0.004$ for the super-Eddington subsample, which represents the tightest correlation found so far for highly accreting sources.

We performed a linear regression using the module \emcee allowing for intrinsic dispersion, removing known X-ray weak objects and sources whose $\Gamma$ was fixed in the X-ray spectral analysis. The former were removed as their coronal emission is possibly experiencing a ``non-standard'' phase, the latter because the poor quality of the data did not allow a simultaneous estimate of both $\Gamma$ and $N_{\rm H}$. The resulting slope is $\beta=0.16\pm0.03$, somewhat flatter than other previous findings ($\beta = 0.31\pm 0.01$, \citealt{shemmer2008hard}; $\beta = 0.31\pm 0.06$, \citealt{risaliti2009sloan};  $\beta = 0.57\pm 0.08$, \citealt{jin2012combined}; $\beta = 0.32\pm 0.05$, \citealt{brightman2013statistical}), but fully consistent with \citet{trakhtenbrot2017bat}, who report $\beta = 0.17 \pm 0.04$ for hard X-ray selected AGN in the {\it Swift}/BAT spectroscopic survey (BASS).

For 11/29 objects the UV/optical analysis yielded more robust estimates of the BH masses based on H$\beta$ and/or \ion{Mg}{ii}. We did not find any statistically significant difference in the average $M_{\rm BH}$ between $N$ and \textit{W+w} sources, being the mean BH mass for the $N$ group $\langle \log(M_{\rm BH} / M_{\odot}) \rangle_{N}$ = 9.7$\pm$0.1, and $\langle \log(M_{\rm BH} / M_{\odot}) \rangle_{W+w}$ = 9.8$\pm$ 0.1 for the \textit{W+w} one.
As this sample was selected to have high and uniform (with a standard deviation of $\sim$0.1 dex) bolometric luminosity, and the BH masses are nearly identical, we expect that also the Eddington ratios should not differ significantly. Indeed, the mean (median) values are $\langle \lambda_{\rm Edd}\rangle_{N}$ = 1.6 $\pm$ 0.4 (0.9) and $\langle \lambda_{\rm Edd}\rangle_{W+w}$ = 0.5 $\pm$ 0.1 (0.4). The slight difference between the mean (median) values is likely due to the marginally higher bolometric luminosity of the $N$ sample, $\langle \log(L_{\rm bol}/{\rm erg \, s}^{-1} \rangle_{N}$ = 47.84 $\pm$ 0.04, with respect to the \textit{W+w} one, $\langle \log(L_{\rm bol} / {\rm erg \, s}^{-1} \rangle_{W+w}$ = 47.58 $\pm$ 0.06.

\begin{figure}[h!]
\includegraphics[width=\linewidth,clip]{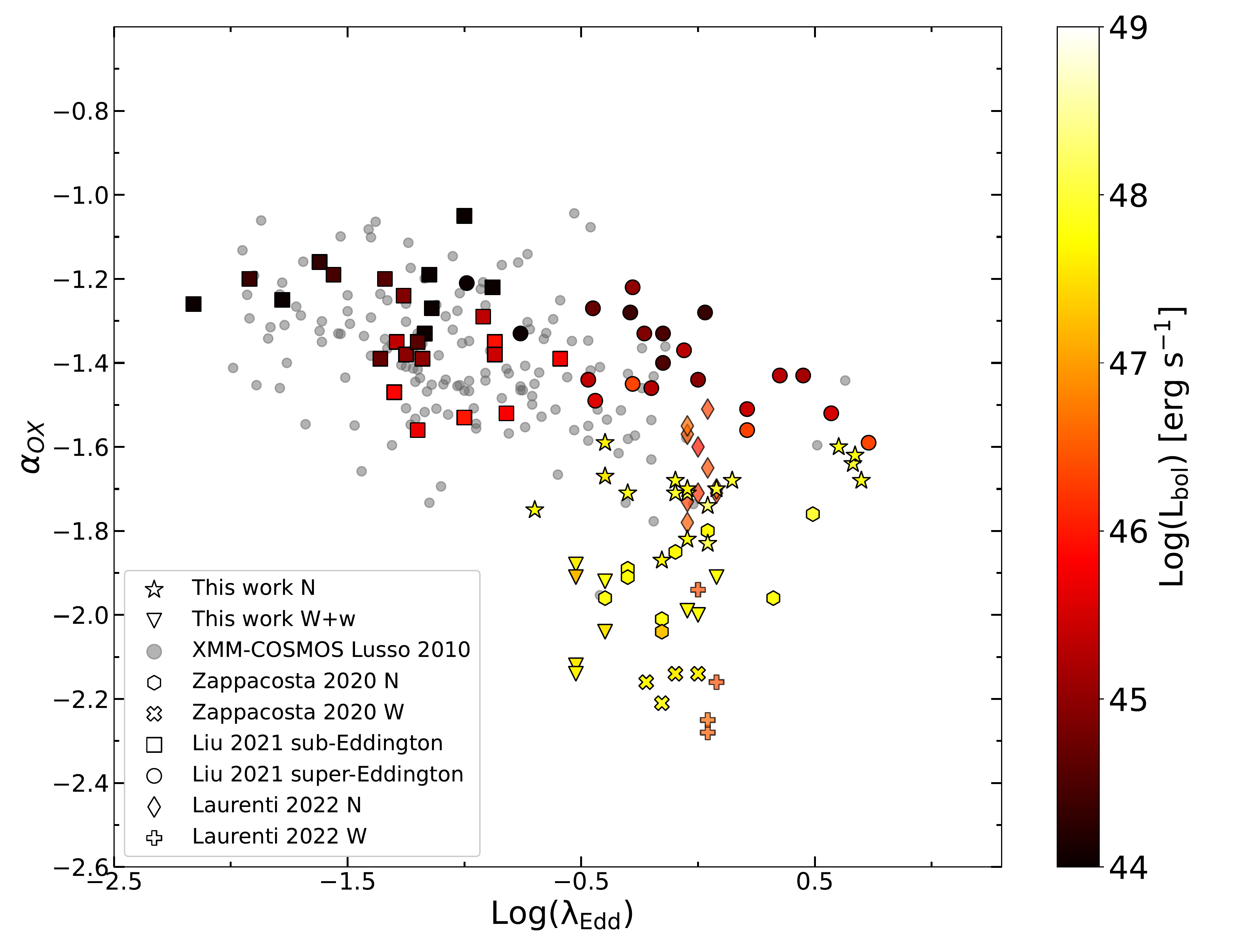}
\caption{$\alpha_{\rm OX}$ against log($\lambda_{\rm Edd}$). Different symbols refer to  different samples as listed in the legend. The bolometric luminosity is color coded and, for a fixed Eddington ratio value, spans 3--4 orders of magnitude. }
\label{fig:lambda_aox}
\end{figure}

The higher fraction of X-ray weak objects reported in high-$\lambda_{\rm Edd}$ quasar samples (Paper I; \citealt{zappacosta2020,laurenti2021x}) compared to the general AGN population \citep{pu2020fraction} hints at some modification of the interplay between the corona and the disc emission in the high-accretion regime. 
The strength of the coronal emission with respect to the disc one can be roughly estimated through the indicator $\alpha_{\rm OX}$. In Figure \ref{fig:lambda_aox} we show $\alpha_{\rm OX}$ as a function of the Eddington ratio for different AGN samples. The resulting anti-correlation (\citealt{lusso2010}; see also Section 4.2 in \citealt{liu2021observational}, and references therein) displays a rather large spread, since sources with similar $\lambda_{\rm Edd}$ can span several orders of magnitude in terms of bolometric luminosity (which ultimately governs the SED steepness).

\begin{figure}[h!]
\includegraphics[width=\linewidth,clip]{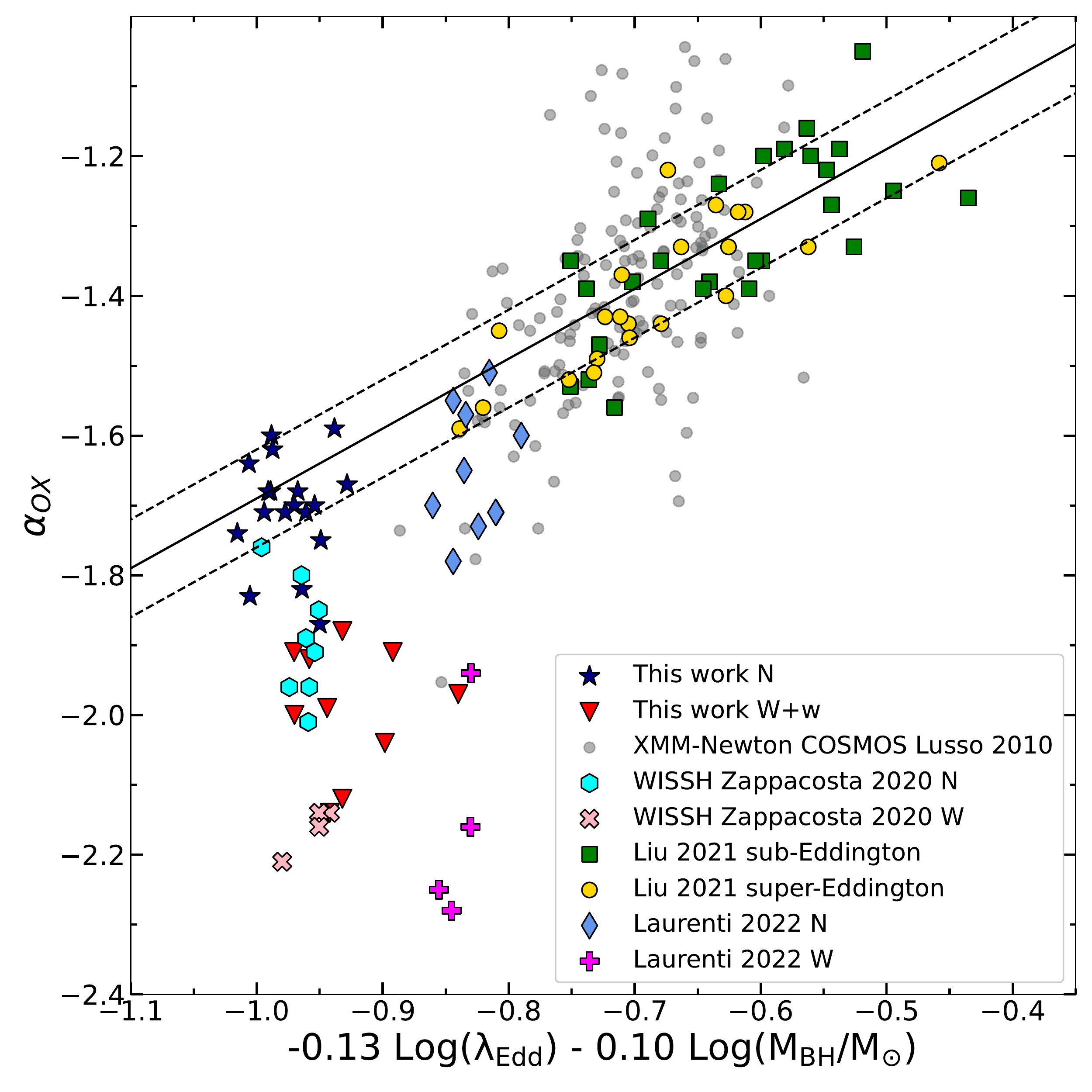}
\caption{Edge-on view on the $\alpha_{\rm OX}-\lambda_{\rm Edd}-M_{\rm BH}$ space. The various samples are listed in the legend. The solid line is the best regression as found by \citet{liu2021observational}, while the dashed lines represent the 1$\sigma$ dispersion. Quasars with high BH masses and high accretion rates extend the range of the relation. The bulk of our X-ray normal sources follows the extrapolation of the relation calibrated on less luminous sources at lower redshifts.}
\label{fig:aox_lambda_mbh}
\end{figure}

Nonetheless, the relation should naturally tighten if we include the BH mass into the parameter space: for a given $\lambda_{\rm Edd}$, the higher $M_{\rm BH}$, the higher $L_{\rm bol}$. \citet{liu2021observational} proposed that the scatter of the relation is due to a ``non-edge-on'' view on a more fundamental plane in the $\alpha_{\rm OX}-\lambda_{\rm Edd}-M_{\rm BH}$ space. In Figure \ref{fig:aox_lambda_mbh} we superimpose to the relation found by \citet{liu2021observational} all the type I AGN samples shown in Figure \ref{fig:lambda_aox}, which include the extreme BH hole mass tail of the quasar population at high redshift. Remarkably, the bulk of our $N$ quasars sits nicely on the extrapolation of the best fit regression line, whereas the \textit{W+w} group drops far below, exhibiting lower $\alpha_{\rm OX}$ than the other sources in the same $\lambda_{\rm Edd}$--$M_{\rm BH}$ domain.
The $\alpha_{\rm OX}$ parameter is still very steep also for other groups of very luminous quasars, such as those from \citet{zappacosta2020} and \citet{laurenti2021x}, but also the bulk of X-ray normal sources of these samples is slightly below (by $\sim -0.2$) the prediction. Very luminous sources could behave differently from the expectations, suggesting that, if a universal relation existed, it could be steeper, or that the high and low ends of the BH mass--Eddington ratio distribution cannot be jointly described by a linear relation with $\alpha_{\rm OX}$.
The sources of \citet{liu2021observational}, whose fit we extrapolated, were selected with criteria akin to ours, being radio-quiet, non-BAL, and with good quality (S/N\,$>$\,6) X-ray data. However, no criteria about their optical/UV emission (e.g., luminosity, photometric indices) were applied. Moreover, in 7/48 objects with multiple X-ray data available, the authors chose the high-flux ones, and this choice can somewhat affect the comparison.
Moreover, the sources analysed in \citet{zappacosta2020} were chosen to study the impact of outflows driven by hyper-luminous quasars. It is thus possible that the mismatch between the selection criteria can also reduce the general agreement of the results.
Overall, a better sampling of the very-luminous, highly accreting side of the distribution will be key to assess whether the accretion mechanism is the same on very different scales of its governing parameters.

\section{Discussion }
\label{sec:discuss}

As a general consideration, two different flavours of X-ray weakness exist, as we can find both intrinsic and apparent X-ray weak sources. The former are likely associated with the physics of the corona, which is not efficiently producing the ``normal'' X-ray emission, and we witness a wide range of weakness factors and photon indices. PHL 1811 is the most extreme case of the first kind, given its very steep photon index $2.0<\Gamma<2.6$ (but see \citealt{wang2022nustar}), the variability, and the $\sim$2 orders of magnitude weakness. The latter objects are instead related to some kind of absorption (slim disc, failed/clumpy wind, warm absorber, etc.) along the line of sight. Our objects do not obviously fit in either of these two classes, being in some way intermediate between them, showing relatively flatter ($\Gamma \lesssim 1.7$) spectra in the absence of clear absorption and moderate weakness factors. This is why it is important to understand their prevalence among the AGN population and their underlying physics.

The analysis of the X-ray data in Paper I revealed an anomalously high fraction ($\approx$25\%) of X-ray weak ($\ga$\,3--10 times fainter than expected) objects. Although in some cases (the few marginal detections, plus J1201+01 and J1459+0024) the quality of the data is not good enough to completely rule out some level of absorption, in general a column density exceeding $N_H(z)>3\times 10^{22}$ cm$^{-2}$ can be ruled out. This is a consequence of the lack of any cutoff around 1 keV associated to X-ray absorption.

In Paper II, we found a correlation between X-ray and \ion{C}{iv} luminosity, which holds for X-ray normal objects over the whole range probed by an extended control sample ($10^{43.5}-10^{45.5}$ erg s$^{-1}$ in $L_{2\,\rm keV}$ and $10^{43}-10^{46}$ erg s$^{-1}$ in $L$\textsubscript{\ion{C}{iv}}), whereas X-ray weak sources lie below that main sequence. In absolute terms, X-ray weak objects present more luminous \ion{C}{iv} lines than expected for quasars with similar X-ray luminosity. Still, X-ray weakness might lead to a less efficient population of the excited level ($\sim$8 eV above the ground state) which is collisionally excited because of the X-ray heating of the gas. 
We argue that such a relation (holding for the $N$ sources) is probably a by-product of the more general $\alpha_{\rm OX}-L_{\rm UV}$ relation.

\textit{W+w} quasars do not show strikingly different blueshifts of the \ion{C}{iv} line with respect to $N$ ones, but shallower (i.e., broader and less prominent) line profiles on average. This could be due to (at least) two reasons: (i) fast, mostly equatorial outflows are present among X-ray weak quasars, but our lowly inclined line of sight offsets their observational footprints; (ii) the difference in \ion{C}{iv} emission between $N$ and \textit{W+w} objects is caused by some mechanism affecting the region where the photons producing the \ion{C}{iv} come from.
By analysing $\sim$190,000 spectra from the seventeenth SDSS data release, \citet{temple2023} find that large ($\geq$\,1,000 km s$^{-1}$) median \ion{C}{iv} blueshifts are observed only in quasars with high SMBH masses ($M_{\rm BH}\geq10^9\,M_\odot$) and high Eddington ratios ($\lambda_{\rm Edd}\geq0.2$). The $z\sim3$ sample analysed here fulfils both requirements of high BH masses and high Eddington ratios, so all the sources could be, in principle, in the outflow phase.
The fact that we observe \ion{C}{iv} blueshifts around or slightly less than 1,000 km s$^{-1}$, on average, can be explained as a projection effect, whereby mostly equatorial winds are observed \rev{under low inclination angles}. In this framework, the fraction of X-ray weak objects could be related to the wind duty cycle (e.g., \citealt{fiore2023dyn}), \rev{which is otherwise poorly known. We can set a lower limit to the wind persistence based on the concomitant [\ion{O}{iii}] weakness. Assuming the relation for the single-zone [\ion{O}{iii}]-emitting region described in \citet{baskin2005controls}, that is $R_{\rm NLR} = 40\,(\lambda L_{\lambda}/10^{44})^{0.45}$ pc, where $\lambda L_{\lambda} \sim 10^{47}$ erg s$^{-1}$ for our objects, we get $R_{\rm NLR}\approx$\,1 kpc. Hence, it would take $t \sim R_{\rm NLR}/c \sim $3000 yr for the NLR to respond to changes in the nuclear activity. Considering that the longer light-travel time within the NLR would dilute any more rapid nuclear flickering, the wind lifetime should be at least $\approx$10$^4$ yr.}

\rev{In addition, we notice a qualitative similarity between the X-ray normal and weak UV stacks in the top panel of Fig. 7 in Paper II, and the \ion{C}{iv} core and wind-dominated stacks depicted in Fig. 5 of \citet{temple2021high}, whereby the $N$ (\textit{W+w}) stack is more akin to the \ion{C}{iv} core-dominated (wind-dominated) one. Indeed, the \textit{W+w} and wind-dominated stacks share the broader and shallower line profile and the flatter continuum, although the latter spectrum exhibits more extreme properties in terms of lower line equivalent width and higher offset velocity than ours.}

In principle, the lack of seed photons causing the lower intensity of the \ion{C}{iv} emission could be due to some kind of \textit{shielding} triggered by the high accretion regime: a puffed up disc (e.g., \citealt{luo2015}) could prevent ionising photons from reaching the BLR as well as intercept part of the X-ray emission, causing absorption along the line of sight. According to this model, our line of sight should cross the absorber, thus favouring a more \textit{edge-on} geometry (still preserving the type I nature of these AGN). Considering the thickness of the disc bulge and its density, we should thus detect absorption in the X-rays well in excess of $N_{\rm H}\gtrsim 10^{23}$cm$^{-2}$, but our sample does not fulfill this condition. 

\rev{In Section \ref{sec:hbeta} we reported higher values of EW\,H$\beta$ for the X-ray normal sources with respect to the weak ones. We checked other samples containing X-ray weak sources with available optical data, finding a similar behaviour. The average values for $N$ and $W$ objects in the WISSH sample described in \cite{vietri2018} are $\langle$EW\,H$\beta\rangle_{N}$=62$\pm$7 \AA\ and $\langle$EW\,H$\beta\rangle_{W}$=47$\pm$7 \AA, different at the $\sim$1.6$\sigma$ level. A statistically more significant difference is found for the objects analysed in \citet{laurenti2021x}, for which the SDSS data from the \cite{wu2022catalog} catalogue yield the values $\langle$EW\,H$\beta\rangle_{N}$=51$\pm$6 \AA\ and $\langle$EW\,H$\beta\rangle_{W}$=25$\pm$4 \AA, giving a $\sim$3.7$\sigma$ tension.}\footnote{We note that in both of these samples the average EW\,H$\beta$ is smaller than for our sources, for which the line falls close to the edge of the $\ks$ spectra and the absolute value of its equivalent width strongly depends on the exact placement of the continuum and on the adopted iron template. However, the relative difference between $N$ and \textit{W+w} objects goes in the same direction in all the three samples.}

We thoroughly discussed the [\ion{O}{iii}] properties and the  relation between the [\ion{O}{iii}] and X-ray emission in our sample in Section \ref{sec:oiii}. \rev{The low ($<$\,30 \AA) EW suggests that inclination does not play a major role in our findings, as it is generally regarded to be significant for EW\,[\ion{O}{iii}]\,$>$\,30 \AA. Highly inclined ($\theta>60\degr$) lines of sight are also disfavoured by a luminosity argument. Assuming, for instance, a 75$^{\circ}$ inclination, which might still not intercept the dusty torus at these luminosities (e.g., \citealt{sazonov2015}), the geometric correction $1/\cos\theta$ would yield a factor of $\simeq 4$, shifting the average $\log(L_{\rm bol}/{\rm erg\,s}^{-1})$ of our $z\sim3$ quasars upwards by 0.6 dex to $\sim$48.4, when only one source in the entire \citet{wu2022catalog} catalogue has a larger luminosity.}

Weaker [\ion{O}{iii}] profiles, as discussed in Section \ref{sec:oiii}, are generally found with increasing $L_{\rm bol}$, which is likely due to an intrinsic decrease of the $L$\textsubscript{[\ion{O}{iii}]} with respect to the continuum and, to a lesser extent, to an inclination effect for which more luminous object are preferentially observed at smaller polar angles. 
Finally, the possibility of a reduced [\ion{O}{iii}] emission because of extinction in the NLR of X-ray weak sources is disfavoured (see Appendix \ref{oiii_ext_check}), and would anyway call for an unknown causal connection between very different scales (i.e., $\sim$\,10$^{-3}$ and 10$^{3}$ pc).

We also observed a trend of higher EW\,\ion{Fe}{ii}\textsubscript{UV} and \ion{Fe}{ii}/\ion{Mg}{ii} ratio for X-ray weak sources, in line with the expectations for prototypical X-ray weak quasars (for instance refer to the analyses of PHL 1811 by \citealt{Leighly2007b,Leighly2007a}). 
\rev{These, and some other features characterizing X-ray weak objects, are consistent with the trends expected in the 4D Eigenvector 1 (4DE1; \citealt{sulentic2000}). The transition between the so-called populations B and A (mainly driven by accretion rate and orientation, see for example \citealt{shenho2014}) happens with the increase of the optical Fe emission, the offset velocity of the \ion{C}{iv} emission line as well as a decrease of the equivalent of the width of the \ion{C}{iv} and EW\,[\ion{O}{iii}] emission lines (see respectively supplementary Figure E4 and Figure 1 in main text in \citealt{shenho2014}). We point out that the exact position of our objects along the 4DE1 main sequence is unclear. Indeed, we cannot evaluate the canonical estimator $R_{\rm Fe}$ ($R_{\rm Fe}$=EW\,\ion{Fe}{ii}\textsubscript{opt}/EW\,H$\beta$), as the 4434--4684\AA\, optical Fe complex falls just bluewards our $\ks$ spectra. Even assuming the 4900--5300 \AA\, interval as a proxy of the canonical one, the two samples are not segregated in the FWHM H$\beta$--$R_{\rm Fe}$ plane. Considering their uncertain position along the $R_{\rm Fe}$ axis and the homogeneity of our sources in terms of accretion rate and orientation, which are believed to be the main drivers along the sequence, it is not straightforward to understand the differences between the $N$ and \textit{W+w} sources within the 4DE1 formalism.}

This notwithstanding, we need to piece together all the available observational evidence in order to find a mechanism capable of explaining the following results on the \textit{W+w} sources:
\begin{itemize}
    \item X-ray weakness in absence of clear X-ray absorption (which can still not be ruled out in some objects);
    \item the generally weaker and shallower line profile of \ion{C}{iv} (and optical lines such as \rev{H$\beta$ and [\ion{O}{iii}]}), although still more luminous than expected for normal quasars with similar X-ray luminosity; 
    \item the tentative trend of higher \ion{Fe}{ii}/\ion{Mg}{ii}. 
\end{itemize}

Also the generally larger incidence of X-ray weak objects in highly accreting samples deserves some further considerations. Several studies have recently pointed out that the fraction of X-ray weak quasars could be enhanced when the objects are accreting at near-Eddington rates. \citet{zappacosta2020} found that about 40\% of their sources (belonging to the WISSH sample) have $\Delta \alpha_{\rm OX}<-0.3$ and also display \ion{C}{iv} shifts of over 5,000 km s$^{-1}$, but the column density of the absorber is generally (in 3/4 obects) below $10^{23}$ cm$^{-2}$. The same trend at lower luminosity ($L_{\rm bol}=10^{46.0}-10^{46.6}$ erg s$^{-1}$) and redshift ($z\simeq0.4$--0.7) is observed in the sample analysed by \citet{laurenti2021x}, where $\sim$30\% of their objects exhibits X-ray weakness without requiring absorption. Conversely, the high-$\lambda_{\rm Edd}$ sample of \citet{liu2021observational} does not contain any X-ray weak object as the lowest value of $\Delta \alpha_{\rm OX}$ is $-$0.14. This sample is made of relatively less luminous ($L_{\rm bol}<10^{46.3}$ erg s$^{-1}$), low-redshift ($z\leq0.25$) quasars. It is worth noting that the authors selected the sample discarding, among the other criteria, X-ray and UV/optical absorbed sources, and choosing the X-ray high-flux state in case of multiple observations. The latter selection criterion could partially explain the lack of X-ray weak objects. 

\begin{figure*}[t!]
\centering\includegraphics[width=0.6\linewidth,clip]{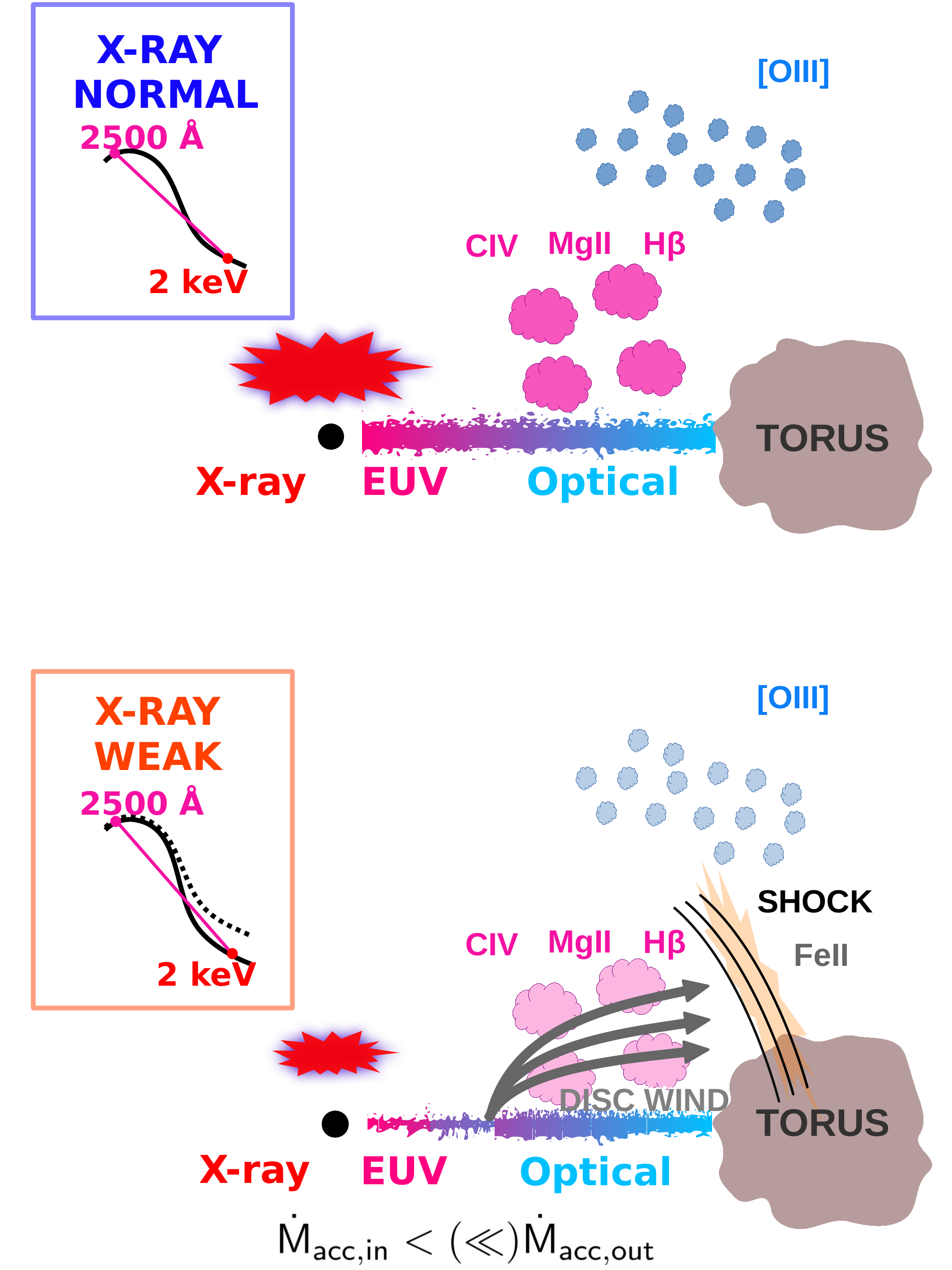}
\caption{A sketch describing the observational features of X-ray normal (top) and weak (bottom) quasars. In the weak case a powerful wind affects the X-ray coronal emission. The outflowing phase depletes the UV-radiating inner disc, and the Comptonization process falls short of seed radiation. The resulting X-ray flux is thus reduced without need for absorption. Shocks at the interface between the disc wind and the BLR gas can enhance the \ion{Fe}{ii}/\ion{Mg}{ii} 
ratio in X-ray weak sources. \rev{The observer is assumed to be on top, but a rather wide range of viewing angles remains compatible with the type--I nature of these sources. Since our line of sight is not expected to be significantly inclined with respect to the disc axis, as suggested by the very high bolometric luminosity of the sample and confirmed by the} low EW\,[\ion{O}{iii}], outflow footprints like absorption dips or strong blue wings in \ion{C}{iv} could be hidden. \rev{On larger scales, the [\ion{O}{iii}] luminosity mimics the X-ray behaviour, being lower in the X-ray weak sample, but fully consistent with the high-luminosity extrapolation of the $L$\textsubscript{[\ion{O}{iii}]}--$L_{\rm X}$ relation. The weaker line emission in X-ray weak sources, here depicted by more tenuous colours, is associated to the difference in the SED shape, as systematic differences between the BLR/NLR in X-ray weak and normal sources are not expected.}
Such a mechanism is more likely to happen in highly accreting sources, where conditions are conducive to the launch of powerful winds, hence the enhanced X-ray weak fraction in high $\lambda_{\rm Edd}$ objects.}
\label{fig:sketch}
\end{figure*}

It is also possible that variability is enhanced in highly accreting objects (e.g., \citealt{ni2020}), and some of the observed weakness could be due to negative fluctuations. However, all the objects with multiple observations (see Sec. 5.1 and Tab. 3 in  Paper I) do not show transitions between $W$ and $N$ states.
Another example of a persistent X-ray weak condition is shown in \citet{laurenti2021x}: J0300$-$08 was targeted in 2011 by the {\it Neil Gehrels Swift Observatory} \citep{gehrels2004swift}, then observed by \xmm in 2018 and again by {\it Swift} in 2021. All these observations ({\it Swift} only provided 3$\sigma$ upper limits) reported an X-ray flux below the expectations. 
Generally speaking, it is possible that a high accretion rate can enhance the possibility of finding a quasar in an X-ray weak state, providing clues about the possible underlying mechanism (e.g., a shielding wind, a photon-trapping disc, a puffed-up slim disc).

Putting together all the pieces of information we have, our best phenomenological guess is represented by the sketch in Figure \ref{fig:sketch}. Specifically, a powerful outflowing phase depletes the inner region of the accretion disc, causing a dearth of the seed UV radiation feeding the coronal X-ray emission, while part of the UV radiation is powering the wind. The change in the local accretion rate can have little impact on the near- to far-UV SED, but a more dramatic effect on the extreme UV (inaccessible to observations) and X-ray emission. If the corona is starved and intrinsically weak, X-ray weakness can be explained without invoking any shielding, which, otherwise, would require moderate to high inclination. The \textit{lowly inclined} perspective is in line with the observed high luminosity, the mild blueshift of the \ion{C}{iv} line (due to the mostly equatorial propagation of radiatively-driven winds; e.g., \citealt{proga2003numerical}), and the small values of EW\,[\ion{O}{iii}] (which, however, also depends on the SED shape).

Despite the reduced X-ray heating, the \ion{C}{iv} emission is weaker but not suppressed since an ample reservoir of ionising photons is still present at large luminosities. Moreover, a crucial contribution to the strength of the \ion{C}{iv} line also comes from the X-ray heating of the gas, which, by means of electron--ion collisions, populates the \ion{C}{iv} excited state. \rev{\cite{krolik1988effects} determined the energy of the continuum photons mainly contributing to the line emission for the most important emission lines, based on different continuum assumptions between $\sim$0.01--2 keV. Both \ion{C}{iv} and H$\beta$ rely on the Lyman continuum photons between 13.6--24.5 eV, and also on the continuum intensity at 300--400 eV and 300--800 eV, respectively, which is likely suppressed in the SED of X-ray weak objects, producing the generally lower equivalent widths observed in such sources.}

On the other hand, the higher incidence of \ion{Fe}{ii}/\ion{Mg}{ii} is rather puzzling. A high value of \ion{Fe}{ii}/\ion{Mg}{ii} \,$\sim$\,15.5 (\citealt{Leighly2007b,Leighly2007a}) is observed in the prototypical intrinsically X-ray weak quasar PHL 1811 but, in that case, it is accompanied by a soft ($\Gamma = 2.3 \pm 0.1$) X-ray emission, which we do not report here for our \textit{W+w} sources. Weakness in our sample is accompanied by flatter photon indices than the average $\Gamma = 1.9$ \citep{piconcelli2005xmm,bianchi09} found in the unobscured quasar population. Following \citet[][see their Figure 24 and the relative SED in Figure 11 in \citealt{Casebeer2006}]{Leighly2007a}, increasingly EUV-deficient SEDs would produce higher values of both \ion{Fe}{ii}/\ion{Mg}{ii} and \ion{Fe}{ii}\textsubscript{UV}/\ion{Fe}{ii}\textsubscript{opt}. 
For our sample, we observe an average \ion{Fe}{ii}/\ion{Mg}{ii} value of 4.4 for $N$ and 8.4 for \textit{W+w}, which is consistent with a softer EUV SED for X-ray weak with respect to X-ray normal objects. Also the  average values for \ion{Fe}{ii}\textsubscript{UV}/\ion{Fe}{ii}\textsubscript{opt} are qualitatively in line with the predictions in \citet{Leighly2007a}, with $13 \pm 4$ for the $N$ and $26 \pm 9$ for the \textit{W+w} subsample (see Sec. \ref{sec:hbeta}). 

There have been claims (e.g., \citealt{sameshima2011implications}, \citealt{temple2020fe}) that the \ion{Fe}{ii} emission could be enhanced by the presence of shocks and microturbulence, as already noted in \citet{baldwin2004origin}, which, in our scenario, could be possibly linked to the outflowing phase. Shocks from the disc wind could thus facilitate the formation of the \ion{Fe}{ii} pseudo continuum in parallel to the main ionization process due to the coronal emission. This could link, at least phenomenologically, X-ray weakness to the higher \ion{Fe}{ii}\textsubscript{UV}/\ion{Mg}{ii} that we observe. Further analyses are needed to thoroughly investigate this suggestion: collecting simultaneous and gapless spectra from the rest-frame optical to the UV (a task currently possible for the $z\sim3$ sample with a ground-based spectrograph such as VLT X-SHOOTER) would definitely improve the reliability of the analysis and our understanding of the impact on the line properties of the different regions of the SED in X-ray weak and normal objects.


\section{Conclusions}
\label{conclusions}

In this third paper of the series, we presented the optical and middle UV analysis of the $z\sim3$ quasars whose X-ray and far UV spectral properties were described, respectively, in \citet{nardini2019} and in \citet{lusso2021most}. The main goal pursued in this work was to investigate additional spectral features that can distinguish X-ray weak (\textit{W+w}) from X-ray normal ($N$) quasars. To this purpose, we were awarded dedicated observations at the Large Binocular Telescope in both the rest-frame UV and optical bands, respectively observed in the $zJ$ and $\ks$ bands between 2018 and 2021.

Our main findings are listed below:

\begin{itemize}
    \item We confirm, by combining \ion{C}{iv}, \ion{Mg}{ii}, and H$\beta$ virial mass estimates, that the black holes hosted in these high-redshift quasars belong to the very high mass tail of the BH mass distribution. Their masses had already grown up to $\sim 10^9-10^{10}\ M_{\odot}$ when the Universe was about 2 Gyr old. Given the bolometric luminosity estimated from the 1350 \AA\ monochromatic flux, we derived the Eddington ratio for each object, confirming that the whole sample is made of highly accreting quasars ($\lambda_{\rm Edd}\sim0.1$--5, with a median of 0.9).
    \item By fixing the slope of the continuum in order to match the one of the bluer SDSS side of the spectrum, we find that \textit{W+w} objects generally exhibit more prominent \ion{Fe}{ii}\textsubscript{UV} emission and, in turn, higher \ion{Fe}{ii}/\ion{Mg}{ii} ratios with respect to $N$ ones. Wind-related microturbulence and shocks in outflows could enhance the \ion{Fe}{ii} production in X-ray weak sources.
    \item Comparing our estimates of the \ion{Fe}{ii}/\ion{Mg}{ii} ratios of the $N$ objects with other literature samples, we found that they are in line with the expectations for quasars at similar redshifts. We also confirm that there is no evidence for an evolving ratio across cosmic time up to $z\sim6.5$, pointing at a prior chemical enrichment of the BLRs.
    \rev{\item EW\,[\ion{O}{iii}] emission is generally low ($<20$ \AA) in all but one of our objects. This suggests that inclination effects do not play a major role, as sources observed under high viewing angles usually display EW\,[\ion{O}{iii}] $\gtrsim$ 30 \AA. The lower EW of several emission lines (\ion{C}{iv}, \ion{H}{$\beta$}, [\ion{O}{iii}]) in X-ray weak quasars could be related to the decrease of EUV photons responsible for the line production. Indeed, the $L$\textsubscript{[\ion{O}{iii}]} of the both X-ray weak and normal quasars are consistent with the high-luminosity extrapolation of the $L$\textsubscript{[\ion{O}{iii}]}--$L_{\rm{2-10\,keV}}$ relation from other samples in literature.}
    \item The presence of a mostly equatorial disc wind could explain all the observational features that we have reported so far. Part of the UV radiation would not be reprocessed in the X-ray corona causing intrinsic (i.e., not due to absorption) X-ray weakness. In the case of modest inclination of the line of sight to the disc, consistent with the EW\,[\ion{O}{iii}] values that we find and with the huge bolometric luminosities, we could be missing the typical footprints of an outflow such as a prominent blue wing or absorption dips in the \ion{C}{iv} profile. A higher \ion{Fe}{ii}\textsubscript{UV} emission in X-ray weak quasars could be attributed to microturbulence and shocked regions at the interface between the outflow and the BLR medium.
    
\end{itemize}

In this third paper of the series, we further demonstrate that X-ray weak and X-ray normal quasars also show different trends in their emission-line properties. Nearly simultaneous observations at rest-frame optical/UV wavelengths and in the X-rays are key to constrain the broadband ionising continuum in both populations. In the future, the results of our analysis need to be confirmed on statistically larger quasar samples, possibly extending at both lower and higher redshifts to better study the evolution of the X-ray weakness fraction and of the related emission-line properties.

\begin{acknowledgements}
We gratefully acknowledge the anonymous referee for the thoughtful comments which resulted in a significantly improved paper. 
We acknowledge financial contribution from the agreement ASI-INAF n.2017-14-H.O. EL acknowledges the support of grant ID: 45780 Fondazione Cassa di Risparmio Firenze. FS is financially supported by the National Operative Program (Programma Operativo Nazionale--PON) of the Italian Ministry of University and Research ``Research and Innovation 2014--2020'', Project Proposals CIR01\_00010. A sincere acknowledgement goes to Ms. Noemi Colasurdo, who first performed the data reduction of the LBT $\ks$ spectra in her Master Thesis and developed the baseline analysis that we used as a benchmark. We also thank Prof. Benjamin Trakhtenbrot for kindly sharing the BASS data. 
\end{acknowledgements}

\bibliographystyle{aa} 
\bibliography{bibl}

\begin{appendix} 
\section{LBT spectra atlas}

\label{LBTzJspectraatlas}

\rev{Figure A\ref{fig:appendix1} shows all the LBT $zJ$ spectra along with their uncertainty and the best-fit model components. The red dashed line is the best-fit model. The blue line is the continuum power law, where the blue square marks the 2500 \AA\, luminosity density. The grey line represents the \ion{Fe}{ii}\textsubscript{UV} pseudo-continuum, while the magenta emission line, if present, denotes the \ion{Mg}{ii} emission feature.
Grey shaded areas have been excluded from the fit. Figure A\ref{fig:appendix2} shows all the LBT $\ks$ spectra. The color code is the same as in Figure A\ref{fig:appendix1} for the best-fit model, continuum (in this case the blue square represents the 5100 \AA\ flux), and iron (\ion{Fe}{ii}\textsubscript{opt}). The green emission lines represent the narrow and broad \ion{H}{$\beta$} components, while the [\ion{O}{iii}] is shown in orange.} Luminosities were calculated assuming a standard concordance cosmology with $H_0=70$ km s$^{-1}$ Mpc$^{-1}$, $\Omega_0=0.3$, $\Omega_{\Lambda}=0.7$.

\section{Optical and UV properties of the $z=3.0$--3.3 sample}
\label{app:appendix_b}
Table A\ref{tbl:TA1} contains the quantities evaluated from the LBT $zJ$ and $\ks$ spectra relevant for the topics discussed throughout the paper.

\section{Possible systematics on \ion{Fe}{ii} emission, and a consistency check}
\label{fe_mgii_check}

It is important to explore, at least qualitatively, the possible systematic differences on the estimate of the \ion{Fe}{ii}\textsubscript{UV} emission with other samples in literature. First, we must take into account that most results are obtained using an iron emission template \citep{vestergaard2001empirical,tsuzuki2006fe} or some modified version of it, whereas ours are computed as the emission of several synthetic BLRs with different ionization parameters. Different fitting approaches may also affect the results. Fixing the spectral slope provides a decent baseline to the iron emission in the majority of the cases, but the modelling of the overall emission is not always optimal and, in one case (J1111+2437), the slope is likely too flat to adequately reproduce the data. 
A detailed study of how these systematics can affect the results can also be found in Section 5.3 of \citet{shin2019fe}. 

Ideally, we would need a larger spectral range to fit the continuum, pivoting on two (or more) continuum windows and covering the interval 2200--3090 \AA, in order to have both a more robust determination of the continuum and a direct estimate of the whole \ion{Fe}{ii}\textsubscript{ii} emission without relying on extrapolations at shorter wavelengths. Whilst the first of these two problems is 
present in our sample, the second should be a minor issue, as the extrapolation on the blue side is generally limited to an interval $\sim$100-\AA\ wide, with the bulk of the iron emission covered by the observations.
Although the issues mentioned above could affect a fair comparison with the other literature samples, the difference between the average values for the $N$ and the \textit{W+w} groups should not be influenced by these possible systematics, as both the $N$ and the \textit{W+w} average would eventually be affected in the same way.

\begin{figure}[h!]
\includegraphics[width=\linewidth,clip]{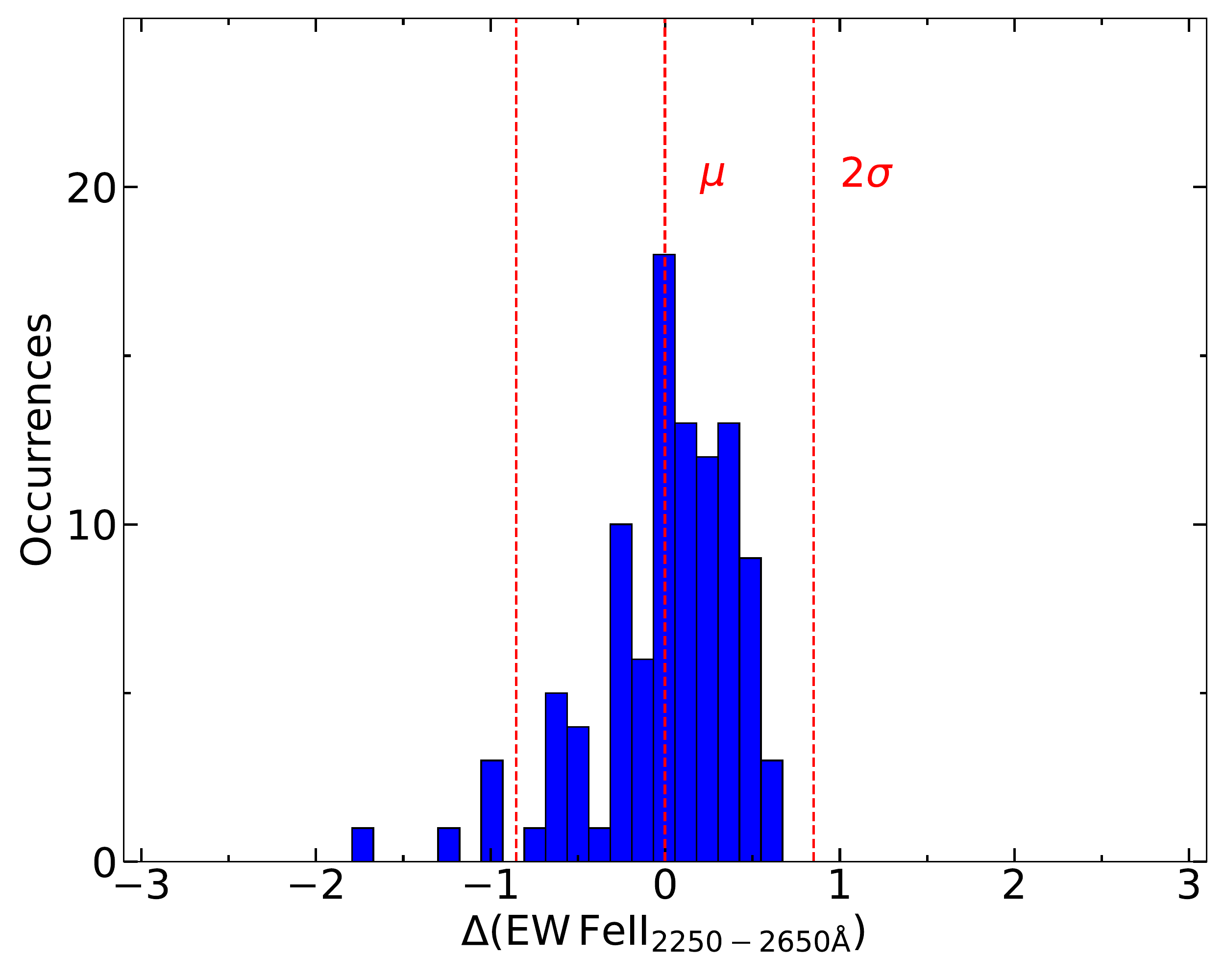}
\caption{Distribution of the residuals between our estimates of  EW\,\ion{Fe}{ii} and the catalogue ones for an SDSS control sample.}
\label{fig:res_distr}
\end{figure}

We performed a consistency check of our capability to reproduce the iron emission. We selected a sample of 100 quasars present in the SDSS DR17 catalogue whose spectra had already been analysed in \citet{wu2022catalog}, and we performed again the fit, focusing on a reliable reproduction of the iron pseudo-continuum. To this purpose, we applied some filters in order to build an \textit{ad hoc} sample: since EW\,\ion{Fe}{ii} in the \cite{wu2022catalog} catalogue is evaluated between 2250--2650 \AA\ we required this wavelength interval to be fully present in all the spectra, together with the \ion{Mg}{ii} and the \ion{C}{iv} emission lines and the nearby continuum windows for a robust estimate of the continuum.
We also required that both the emission lines satisfied the quality cuts suggested by the authors (see their Table 2 and relative discussion in Section 4), that is $F_{\rm line}/\sigma_{F_{\rm line}}>2$, $38<\log\,(L_{\rm line}/{\rm erg\,s}^{-1})<48$ and $N_{\rm pix,line\,complex}>0.5\times N_{\rm pix,line\,complex \, max}$, which sets the minimum fraction of available pixels for the fit. Moreover, we filtered out sources affected by BALs by imposing the BAL\_PROB field equal to 0, since these features could hamper a proper determination of the continuum.
Finally, we also required a detected iron emission by requiring FEII\_UV\_EW$>$0. We sorted out the sources according to their SNR\_MEDIAN\_ALL parameter, and selected 100 of the best quality spectra, avoiding objects affected by bad pixels and spurious features. Successively, we performed a one-by-one analysis, fitting each of them, adopting the same double Gaussian (1 broad, 1 narrow) deconvolution adopted in \citet{wu2022catalog}, but using two sets of our iron templates. After a careful visual inspection of the results, we estimated the EW\,\ion{Fe}{ii} for each object, and compared it against the catalogue ones. The distribution of the relative differences between our estimates and the catalogue ones $\Delta$\ion{Fe}{ii}=(EW\,\ion{Fe}{ii}\textsubscript{ours}$-$EW\,\ion{Fe}{ii}\textsubscript{catalogue})/EW\,\ion{Fe}{ii}\textsubscript{catalogue} is illustrated in Figure \ref{fig:res_distr}. The offset of the distribution is minimal, $\langle \Delta$(\ion{Fe}{ii})$\rangle$=0.002, and most of the sample (95\%) is consistent within a factor $\sim$2 with the catalogue estimates. Even in case of strong iron emission both the continuum and the \ion{Fe}{ii} profile are well reproduced.
This check led to some considerations. First, despite some scatter, in general there is not a strong systematic offset between our estimates and the reference ones. This points to the fact that using the region including the \ion{C}{iv} line and its neighbouring continuum windows to estimate the continuum power law produces reliable results also in the \ion{Mg}{ii} region. Second, we could think of the dispersion of the residuals as a crude estimate of the uncertainty that we get when comparing our \ion{Fe}{ii}/\ion{Mg}{ii} estimates against those evaluated adopting other templates. Third, the good reproduction of the iron emission, even in the case of high EW\,\ion{Fe}{ii}, confirms that our analysis is capable of accurately sampling even the high tail of the EW\,\ion{Fe}{ii} distribution.

\section{A check on [\ion{O}{iii}] extinction}
\label{oiii_ext_check}

\begin{figure}[h!]
\includegraphics[width=\linewidth,clip]{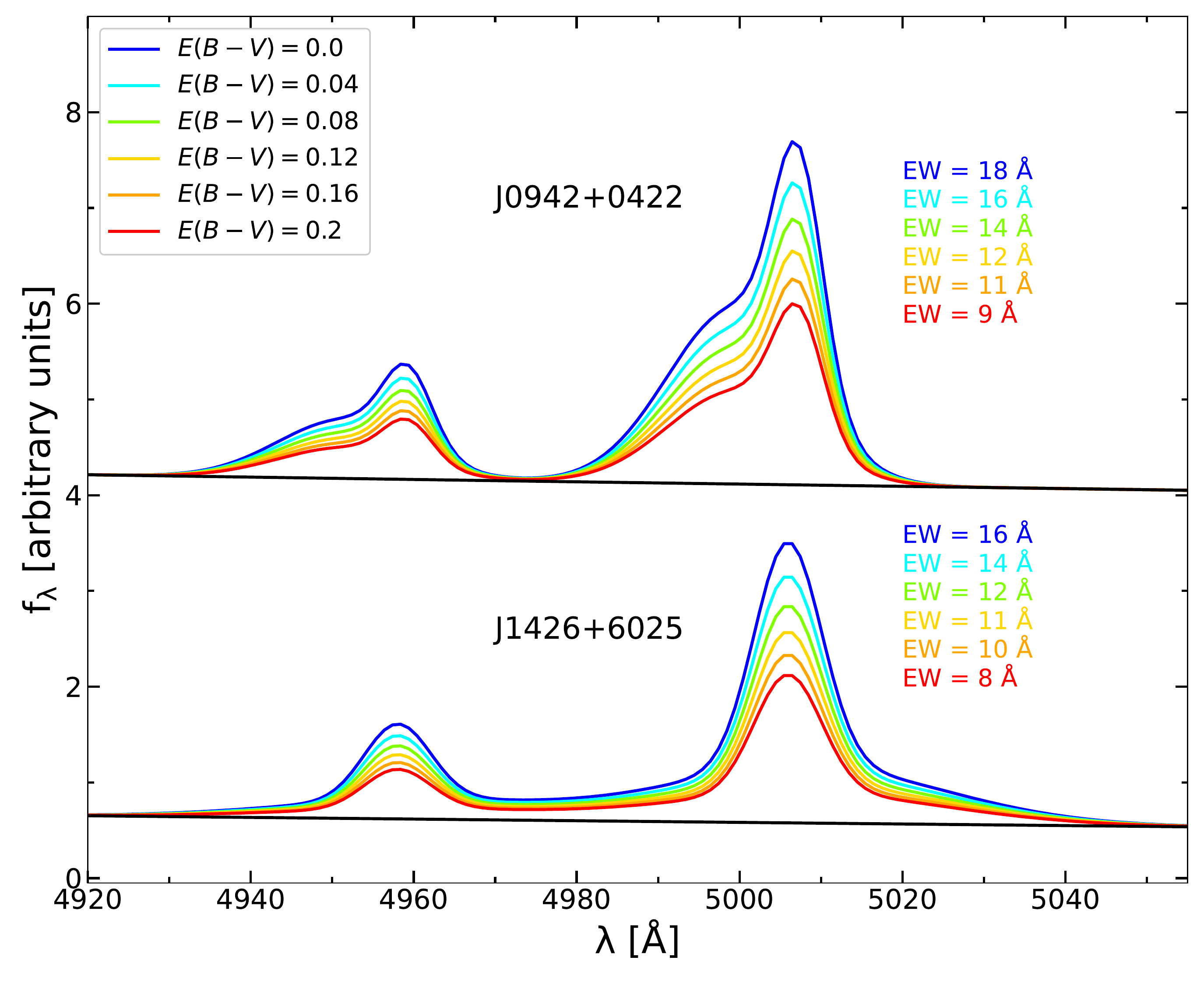}
 \caption{The [\ion{O}{iii}] region of J0942+0422 and J1426+6025 reddened according to the different $E(B-V)$ values, as colour-coded in the legend. All the spectra are normalized by the $f_{5100 \AA}$ of the continuum. The spectra of J0942+0422 were shifted for visualization purposes.}
\label{fig:oiii_ext}
\end{figure}

We checked for the presence of dust extinction affecting the [\ion{O}{iii}] emission in X-ray weak quasars by applying increasing reddening the [\ion{O}{iii}] emission of X-ray normal quasars. Indeed, if the dearth of [\ion{O}{iii}] emission in \textit{W+w} quasars were not intrinsic, but instead due to extinction, the average line profile of $N$ quasars, once reddened, would be similar to the \textit{W+w} one. To rule out this possibility, we performed the following exercise: adopting the best fit model for the [\ion{O}{iii}] emission, we reddened the line profile with increasing values of $E(B-V)$ and evaluated at each step the EW of the line, checking whether this was compatible with the average EW\,[\ion{O}{iii}] for X-ray weak sources. To this purpose, we only considered the $N$ quasars with ``average'' emission, excluding J0303$-$0023 whose [\ion{O}{iii}] is barely detectable and J0304$-$0008 whose [\ion{O}{iii}] is the brightest. We assumed the extinction curve from \citet{gordon2016panchromatic} available in the ``$dust\_extinction$'' package of the Astropy software (\citet{robitaille2013astropy}), adopting an absolute to selective extinction ratio $R_V = A_V/E(B-V)$ = 3.1 \rev{and setting the parameter $f_{\rm A}$ equal to 0, that is, choosing a SMC-like extinction curve}. We then corrected the rest-frame 4900--5200 \AA\ emission for increasing values of extinction $E(B-V)$ from 0 (unobscured) to 0.2 (reddened), which is twice the value adopted for the selection of blue continuum emission for a standard quasar SED \citep{richards2006spectral}, with steps of $\Delta E(B-V)=0.02$. The results of such procedure are shown in Fig. \ref{fig:oiii_ext}. Even in the case of significant reddening the [\ion{O}{iii}] profile is not suppressed to the level observed in the \textit{W+w} subsample, and the EW\,[\ion{O}{iii}] of the reddened spectra are still well above the median value for the X-ray weak sample (2.4 \AA). We thus believe that the [\ion{O}{iii}] emission of X-ray weak quasars is unlikely to be affected by extinction.

\section{Notes on individual objects}
\label{individual_notes}

\end{appendix}
\subsection*{J0303$-$0023}
\rev{This is the only X-ray normal object with faint [\ion{O}{iii}] emission. There are different scenarios which could justify the such a weak line profile without invoking a significant difference in the ionizing SED (a list of possible explanations is examined in Sec. 4.2 of \citealt{agostino2023new}, and references therein). For instance, a low covering fraction of the NLR could be able to obliterate the [\ion{O}{iii}] emission without altering the observed X-ray properties. Another possible reason for a weak [\ion{O}{iii}] profile in presence of normal X-ray emission is the case of NLR gas depletion, due to galactic activity (e.g., major mergers, gas stripping). In principle, for an individual object, it would be also important to consider the difference in size and characteristic timescales between the X-ray corona and the much farther region producing the [\ion{O}{iii}] emission. Although unlikely, since none of our sources deviate from the $L$\textsubscript{X}--$L$\textsubscript{[\ion{O}{iii}]} relation, we cannot completely exclude that the AGN in J0303$-$0023 has experienced a sudden re-brightening to which the NLR has not yet been able to respond.}


\subsection*{J0942+0422}
\rev{This source presents a broad (FWHM=1173$\pm$27 km s$^{-1}$), blueshifted ($v_{\rm off}$=469$\pm$17 km s$^{-1}$) component in the [\ion{O}{iii}] profile. The width and offset velocity are compatible with a galactic scale outflow. However, being this source an X-ray normal quasar, such feature does not relate with any exotic X-ray behaviour.}

\subsection*{J1111+2437}
The H$\beta$ found in the optical spectrum of this object does not show a clear line profile and was thus fitted adopting a broken power law profile convolved with a Gaussian kernel. The resulting shape is very skewed with a red side which extends well below the [\ion{O}{iii}] doublet. In this case the derived FWHM of the H$\beta$ profile is considered as an upper limit rather than a proper estimate.

\subsection*{J1148+2313}
A visual inspection of the spectrum suggests that the slope of the continuum power law may be inverted with respect to the one fitting the SDSS data, and this is confirmed if we try to fit the spectrum without fixing the slope. This leads to a higher value of EW\,[\ion{Fe}{ii}] than the one that we would find by allowing the slope to vary. However, being the \ion{Mg}{ii} emission of this object affected by atmospheric extinction, it is not included in the sample used to evaluate the \ion{Fe}{ii}/\ion{Mg}{ii} ratio, and does not alter the subsequent results.

\subsection*{J1225+4831}
The slope derived on the blue SDSS side of the spectrum of this object is not able to reproduce the shape of the LBT $zJ$ counterpart. Even adopting a free slope 
for fitting this spectrum, the result is modest as we are not able to include any iron emission, and we thus exclude this objects from the considerations in which such feature is implied.

\subsection*{J1426+6025}
The optical spectrum of this object shows a bump redwards the H$\beta$ line, partially embedding the [\ion{O}{iii}]\,$\lambda$4959\AA\ emission. We interpreted such emission as produced by optical Fe, as it seems unlikely that H$\beta$ could display such a broad redshifted component. At the same time, we cannot completely rule out the presence of intense blueshifted [\ion{O}{iii}] wings, as the resulting profile is degenerate for the combination of such wings and the Fe emission.

\onecolumn

\begin{figure}[h!]

    \includegraphics[width=.33\textwidth]{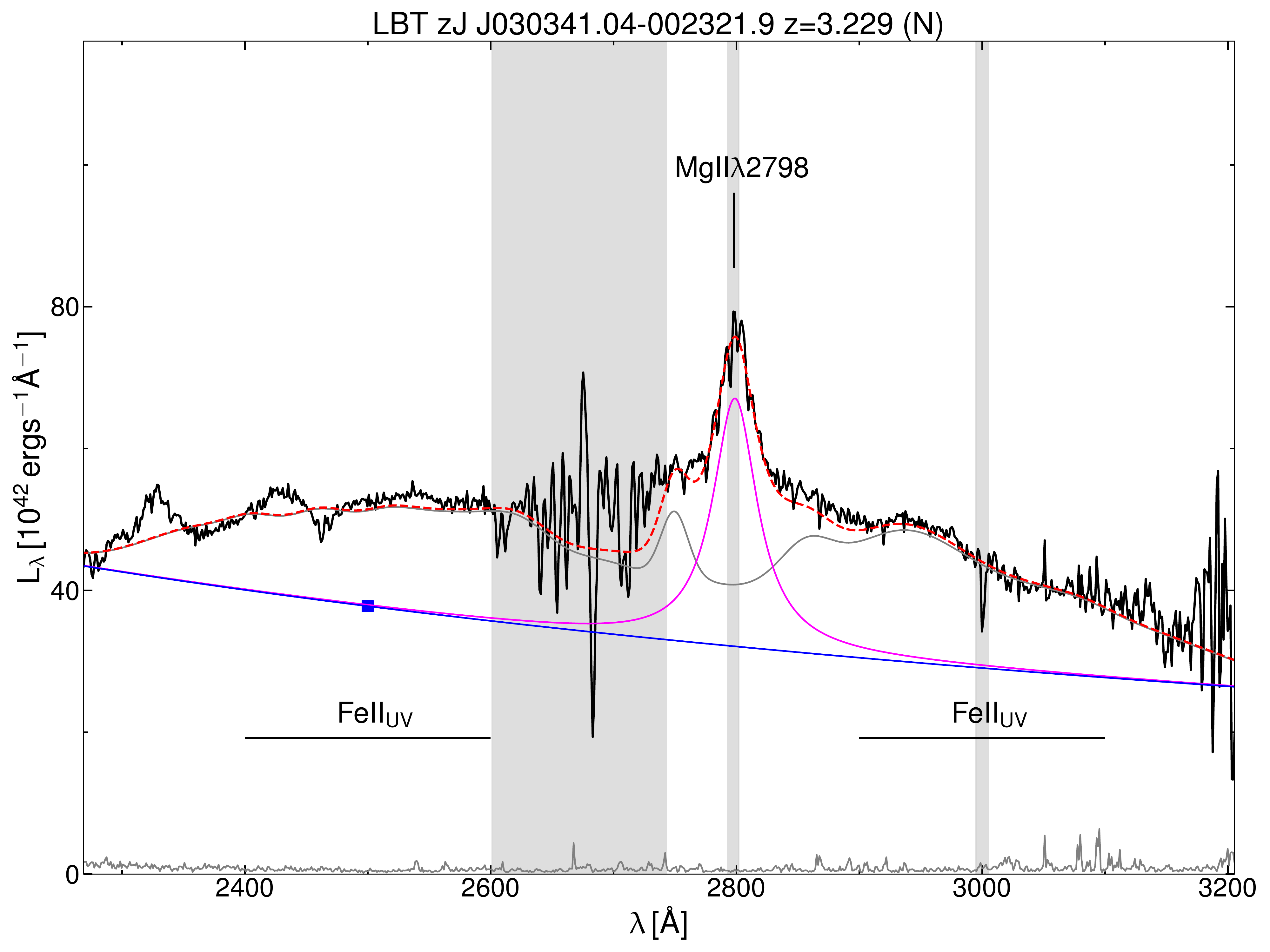}\hfill
    \includegraphics[width=.33\textwidth]{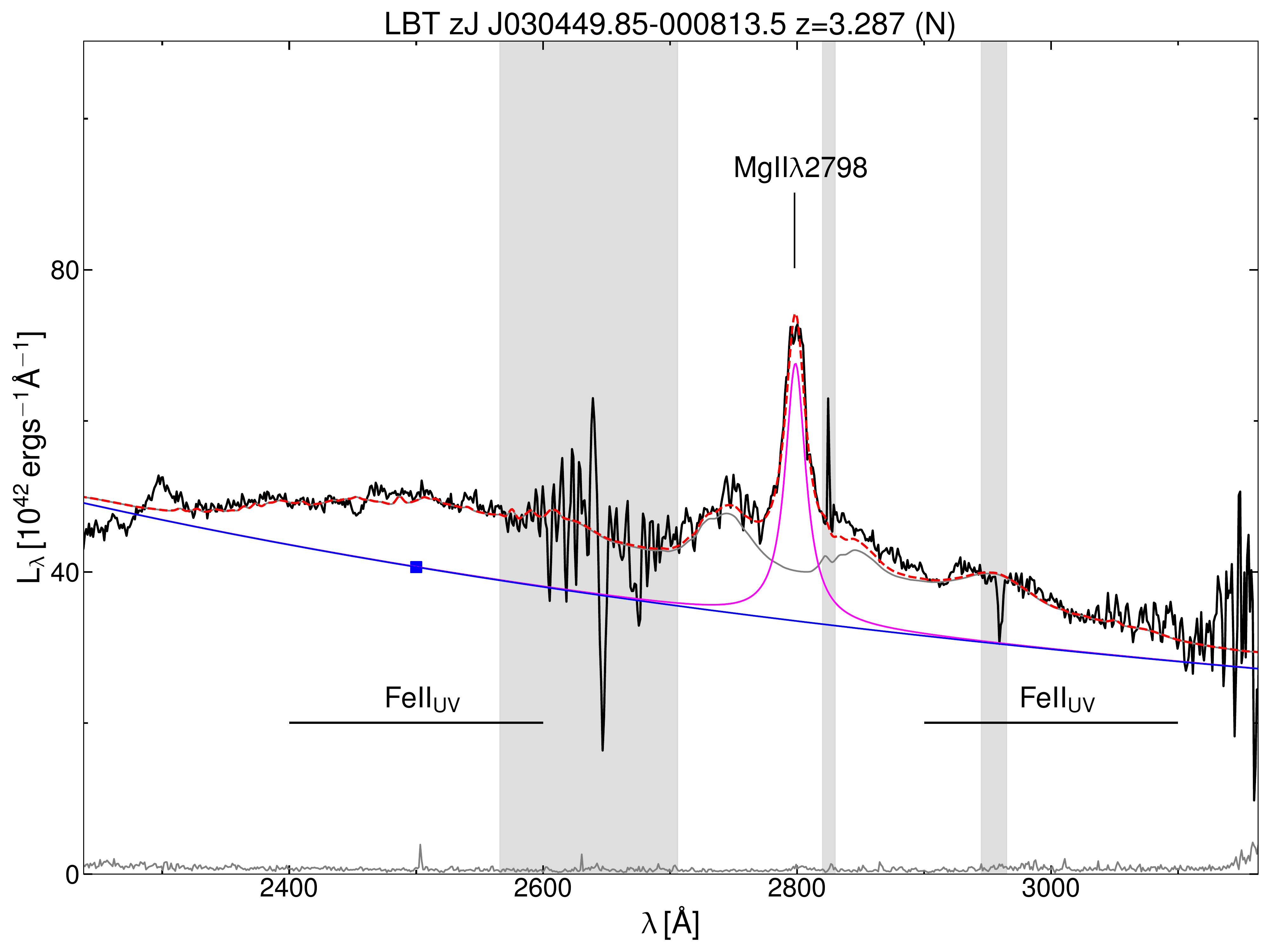}\hfill
    \includegraphics[width=.33\textwidth]{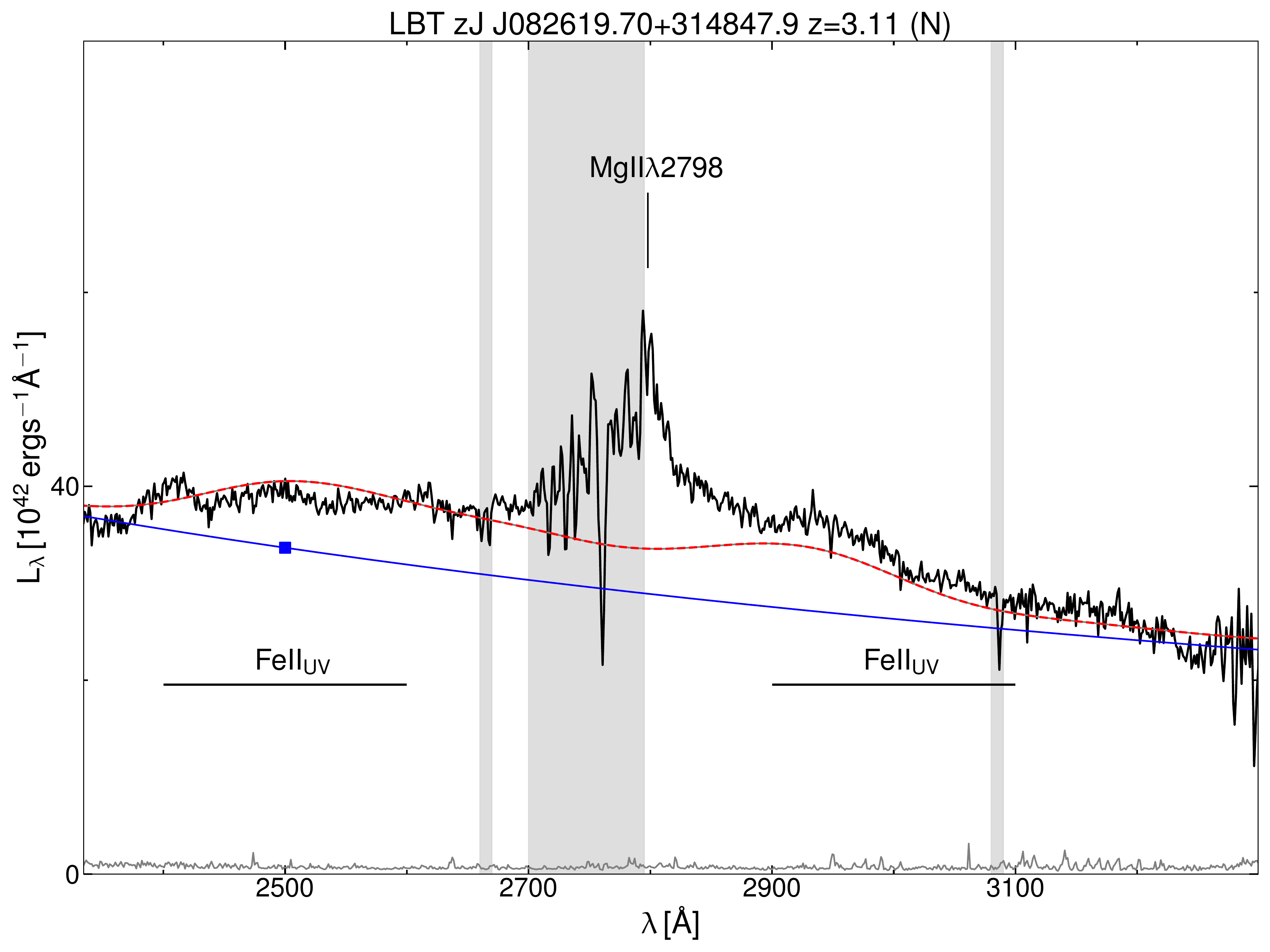}

    \includegraphics[width=.33\textwidth]{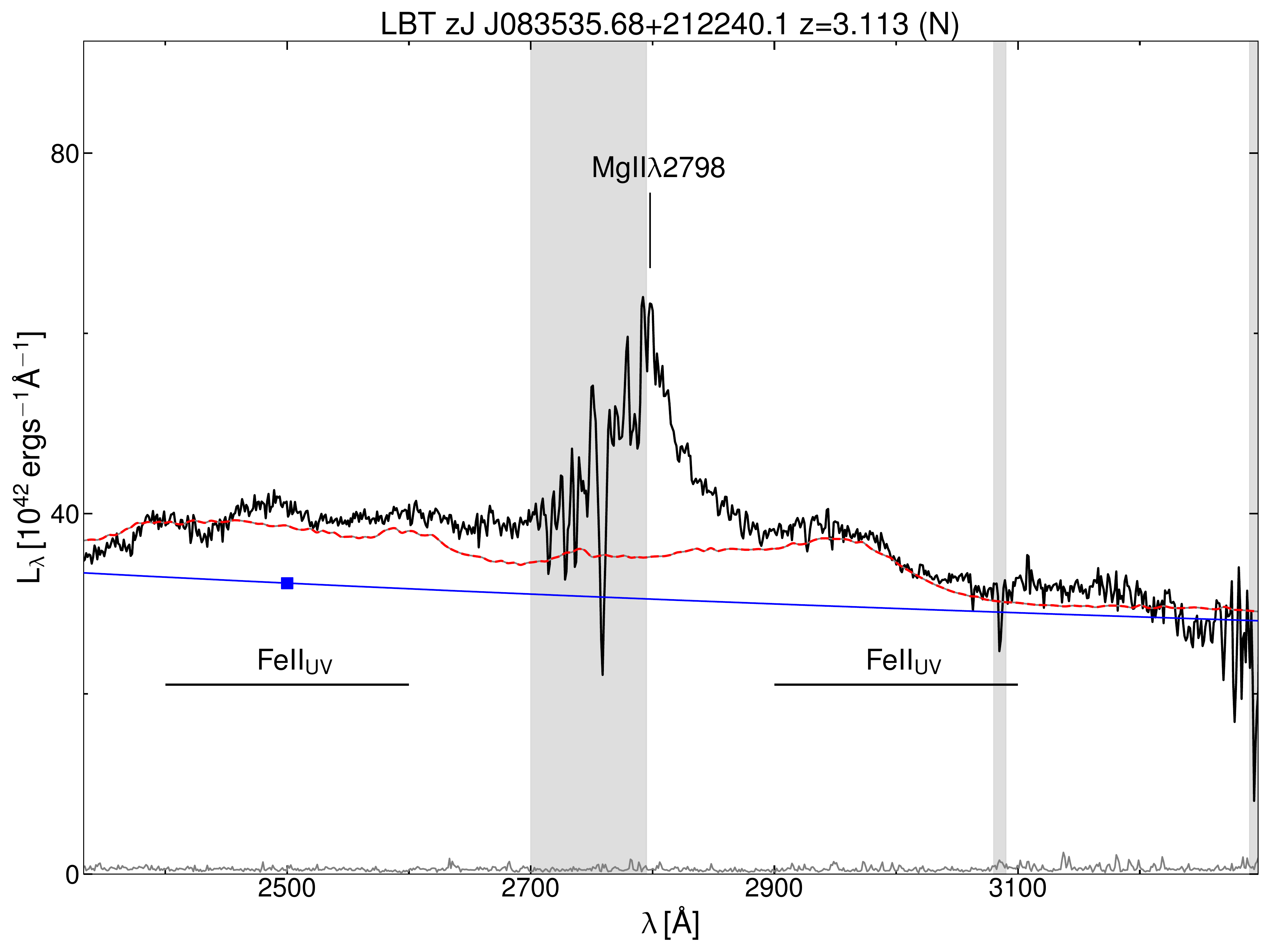}\hfill
    \includegraphics[width=.33\textwidth]{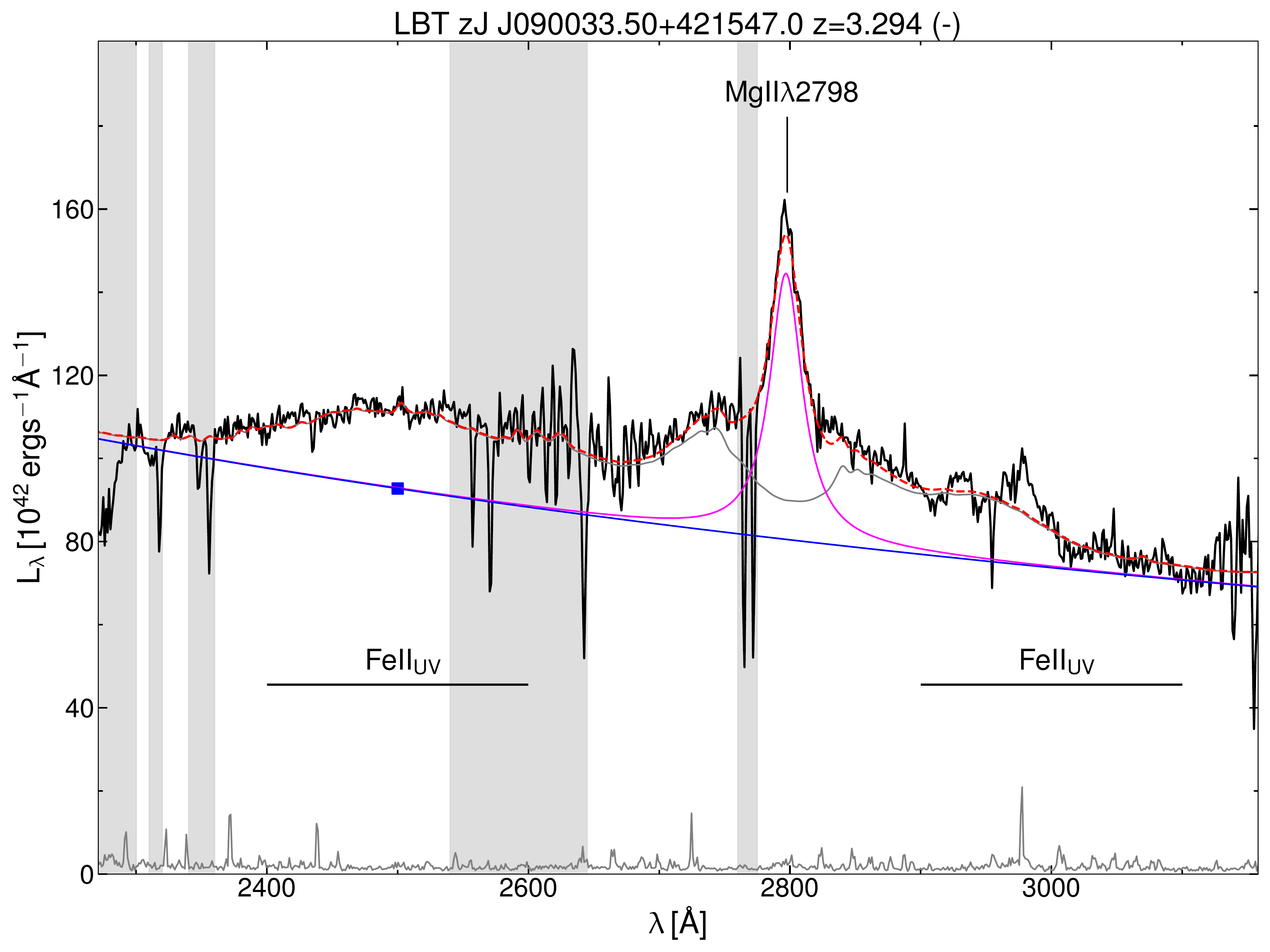}\hfill
    \includegraphics[width=.33\textwidth]{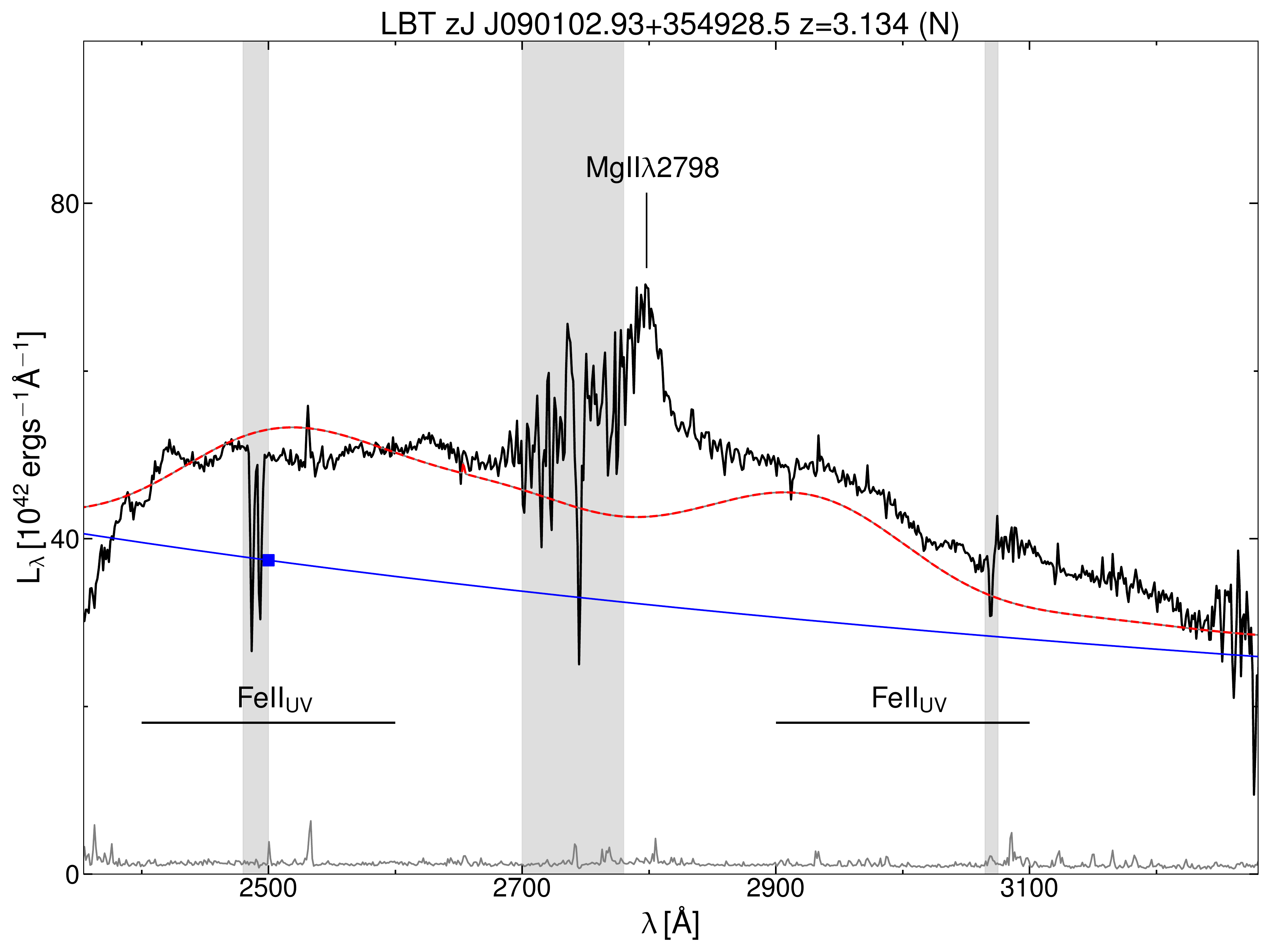}

    \includegraphics[width=.33\textwidth]{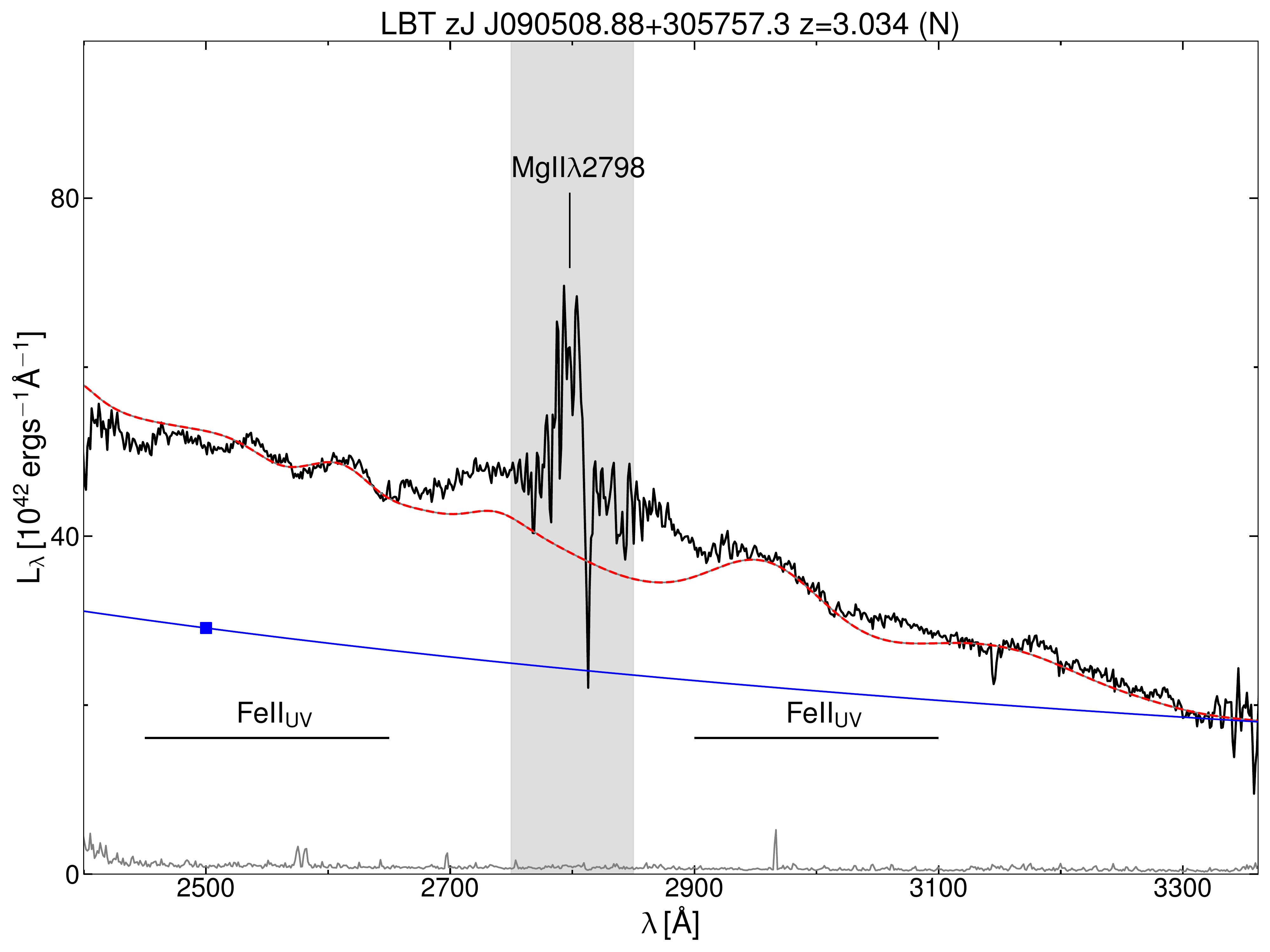}\hfill
    \includegraphics[width=.33\textwidth]{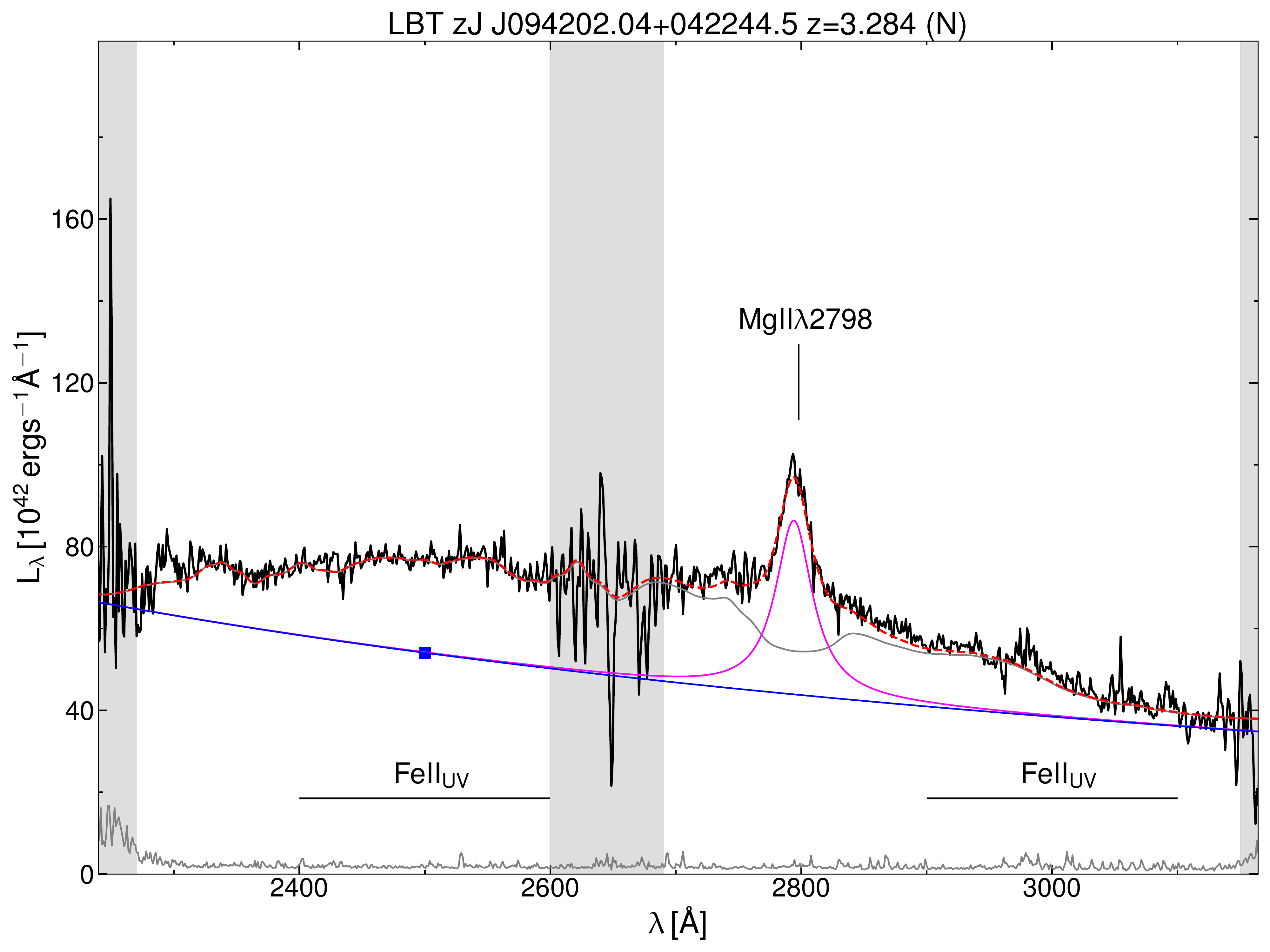}\hfill
    \includegraphics[width=.33\textwidth]{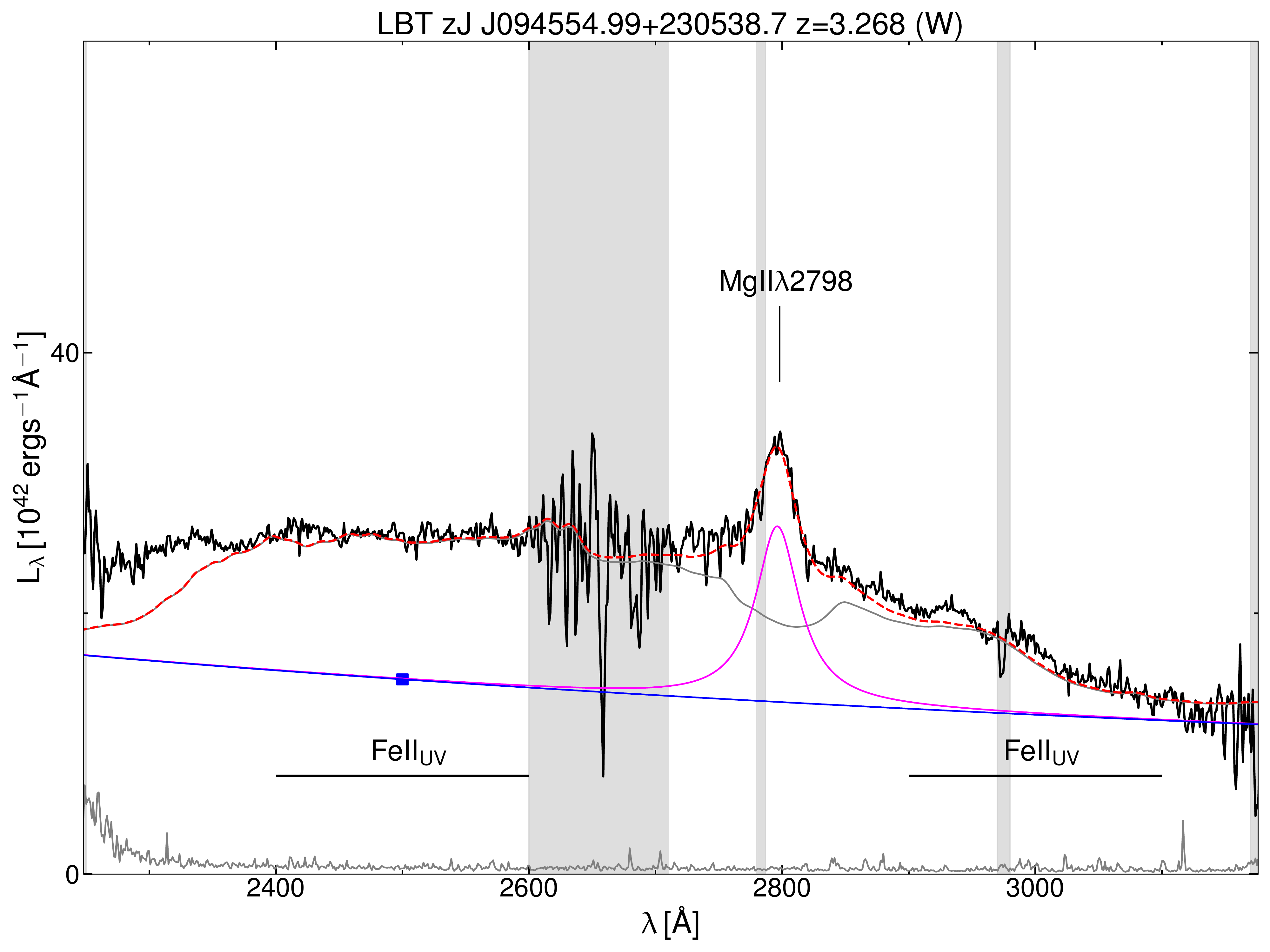}

    \includegraphics[width=.33\textwidth]{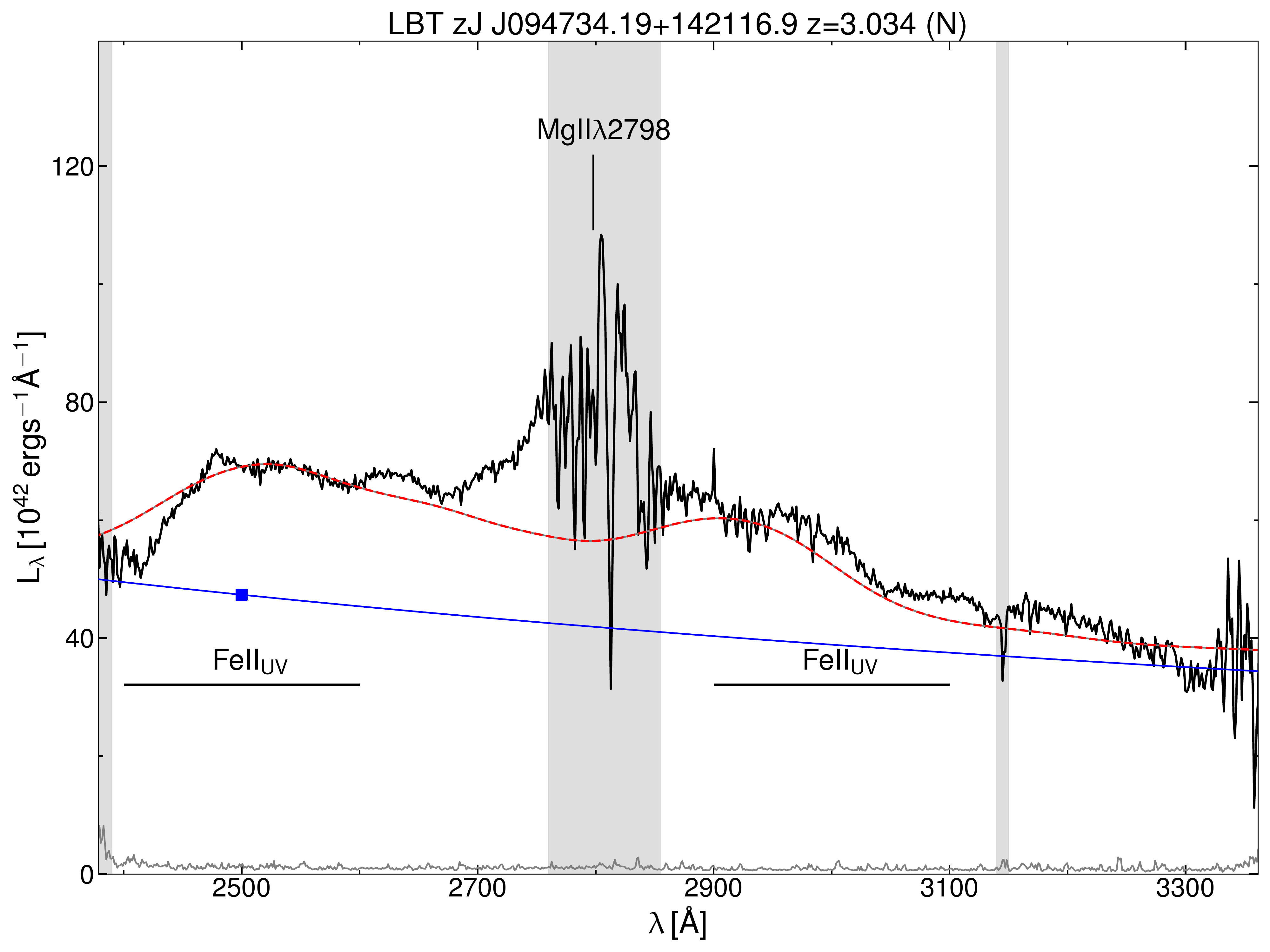}\hfill
    \includegraphics[width=.33\textwidth]{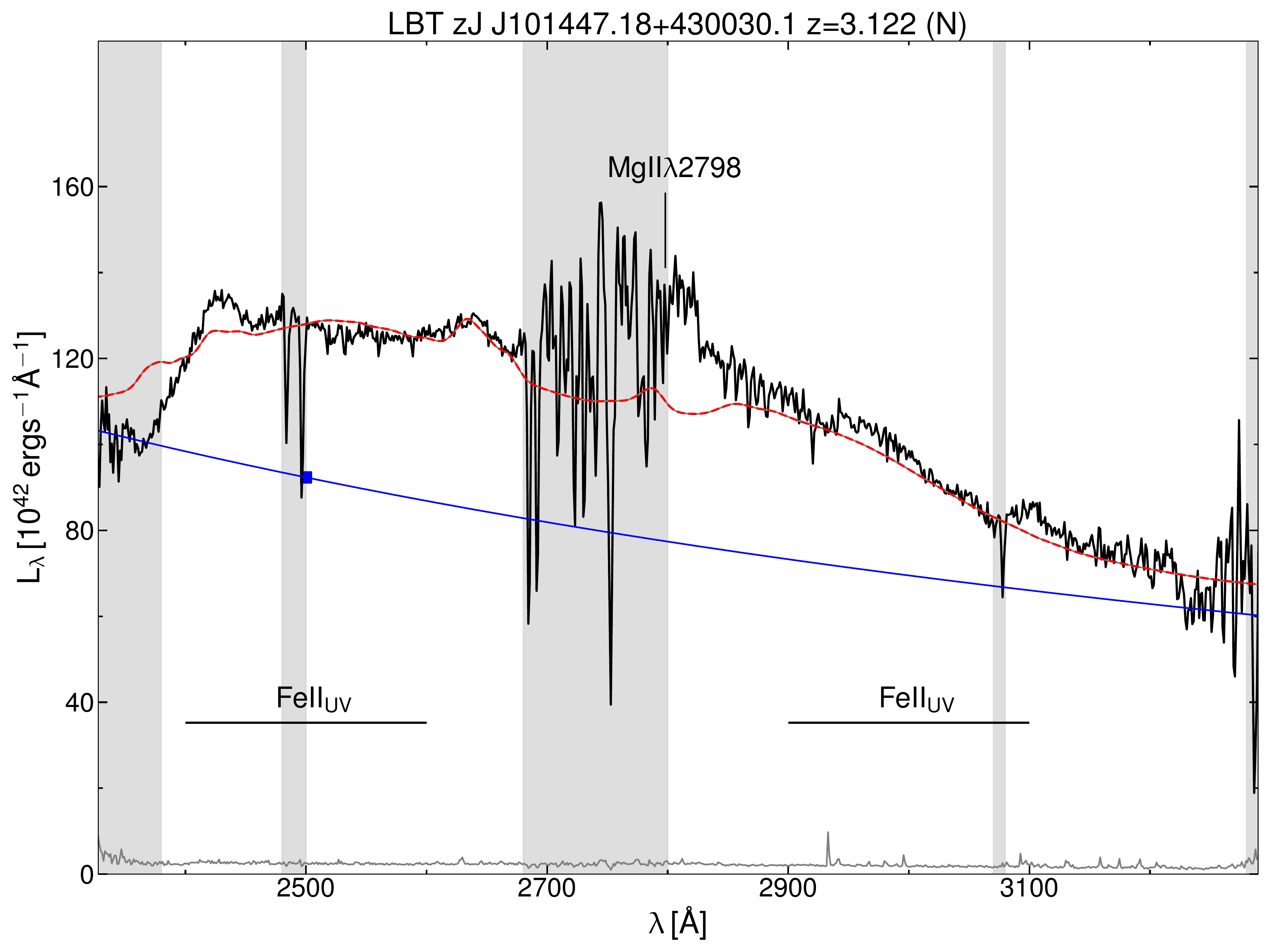}\hfill
    \includegraphics[width=.33\textwidth]{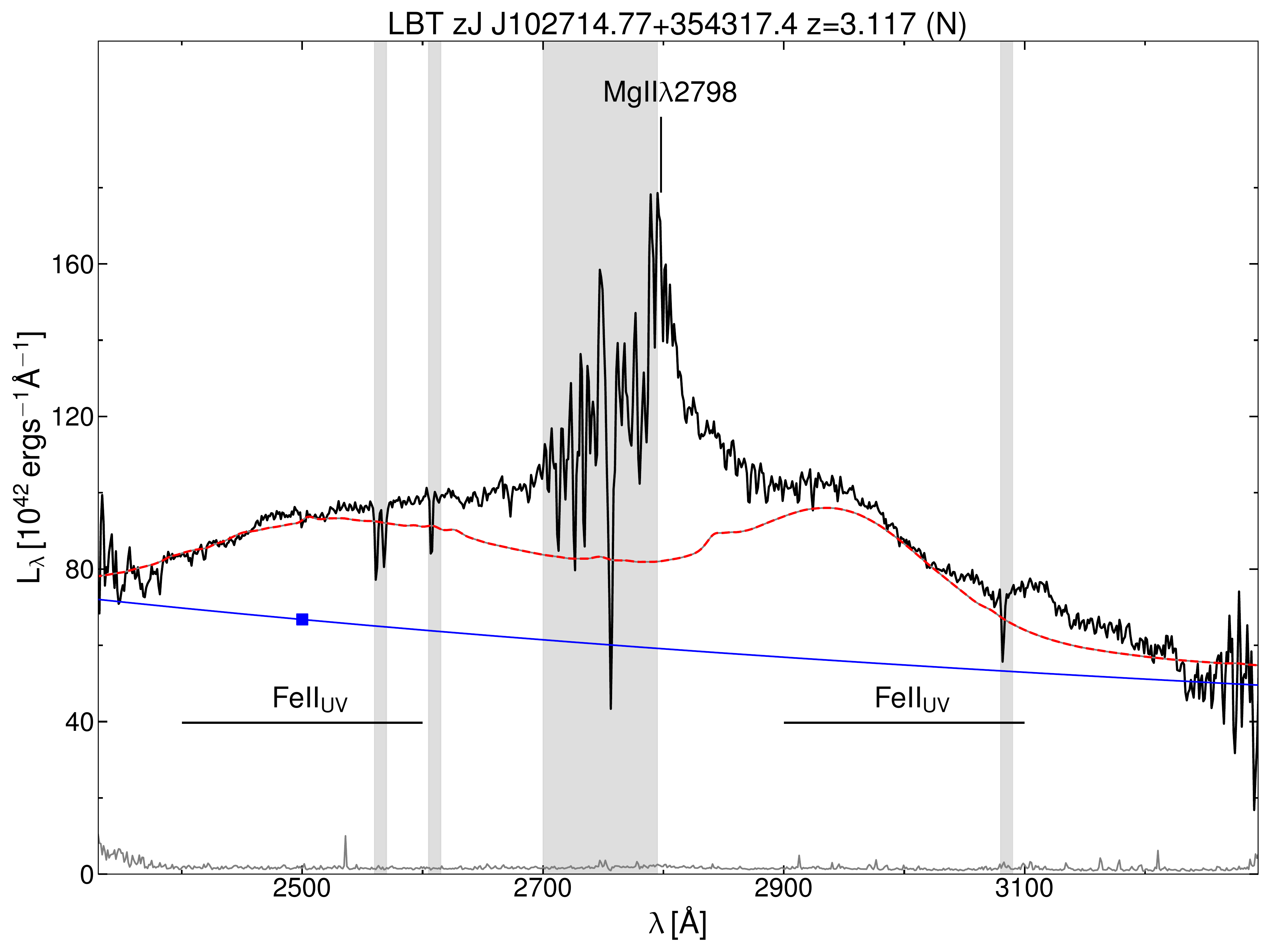}

    \includegraphics[width=.33\textwidth]{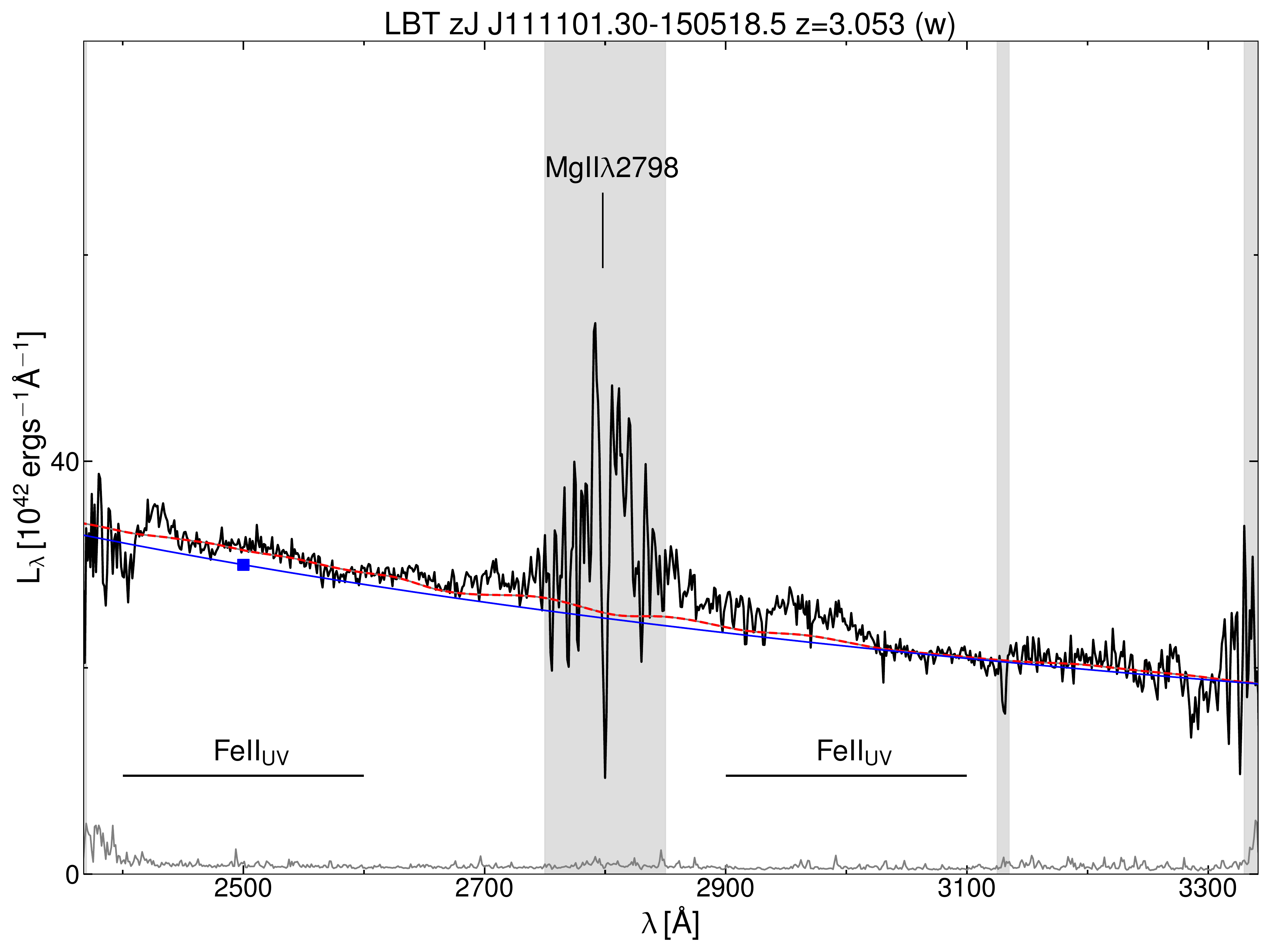}\hfill
    \includegraphics[width=.33\textwidth]{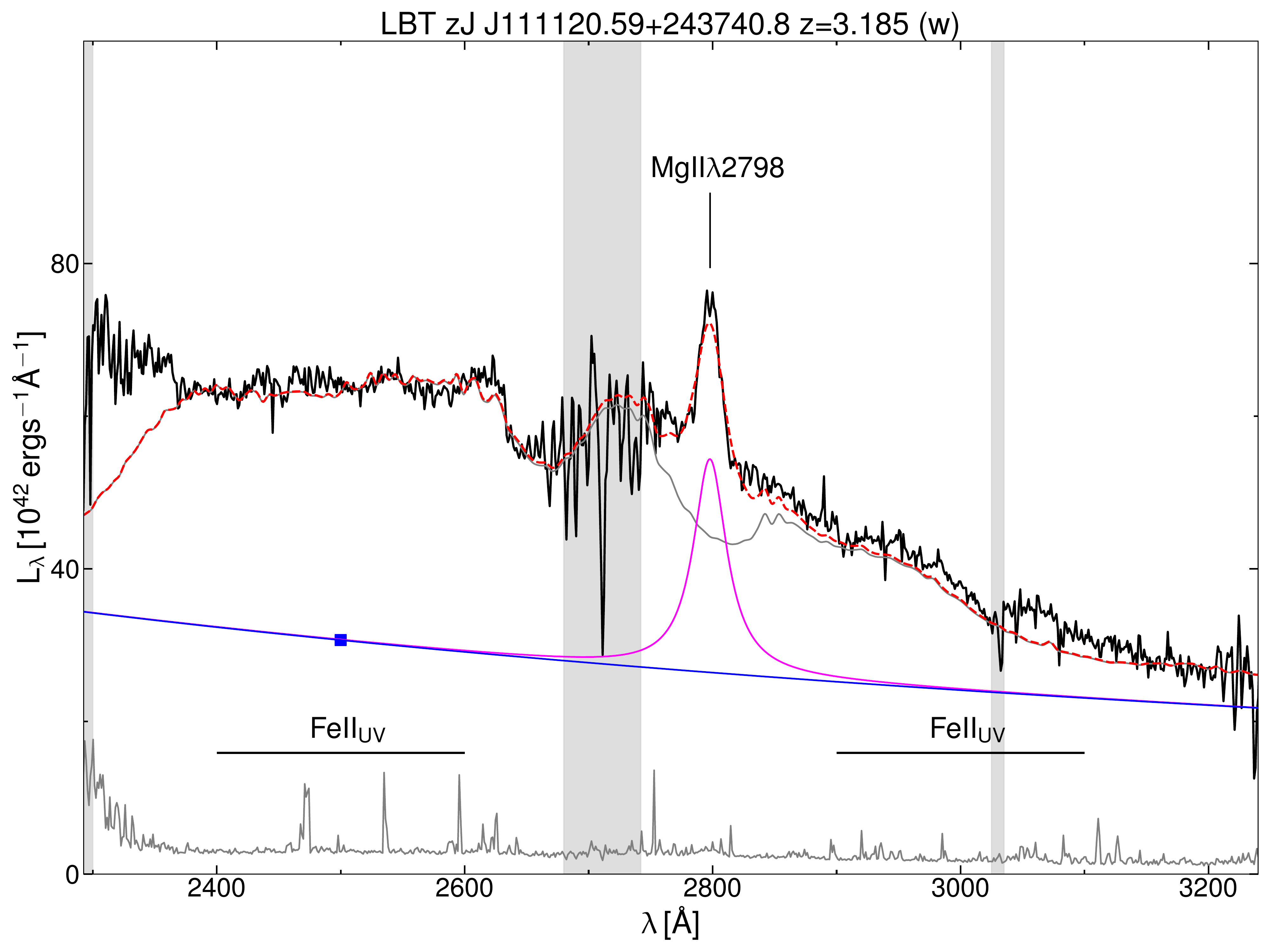}\hfill
    \includegraphics[width=.33\textwidth]{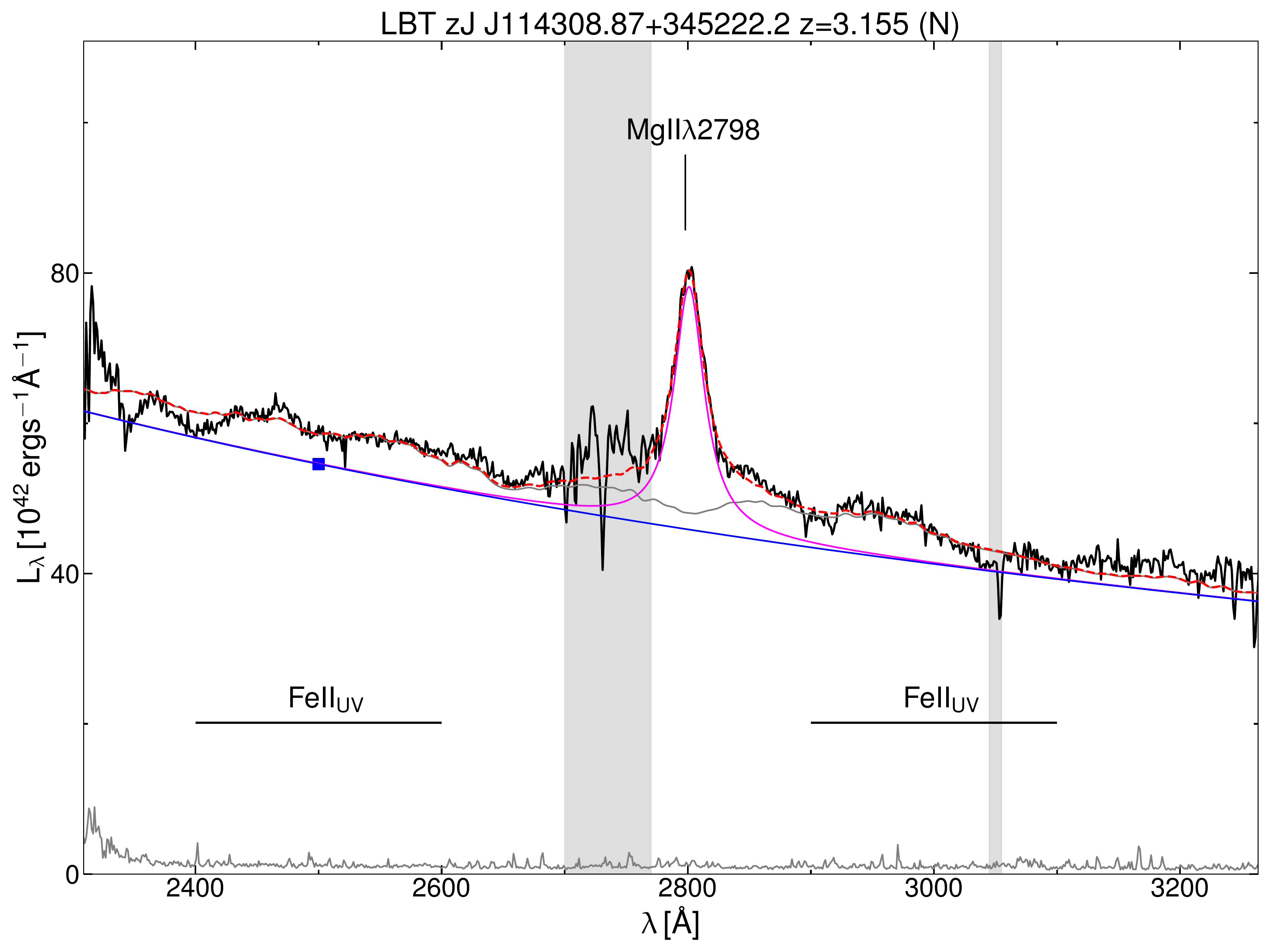}
    
\captionsetup{labelformat=empty}
\caption{Fig. A1: LBT $zJ$ spectra in black. The colour code is described in Appendix \ref{app:appendix_b}.}
\label{fig:appendix1}    

\end{figure}

\newpage
\clearpage
\begin{figure}
    \includegraphics[width=.33\textwidth]{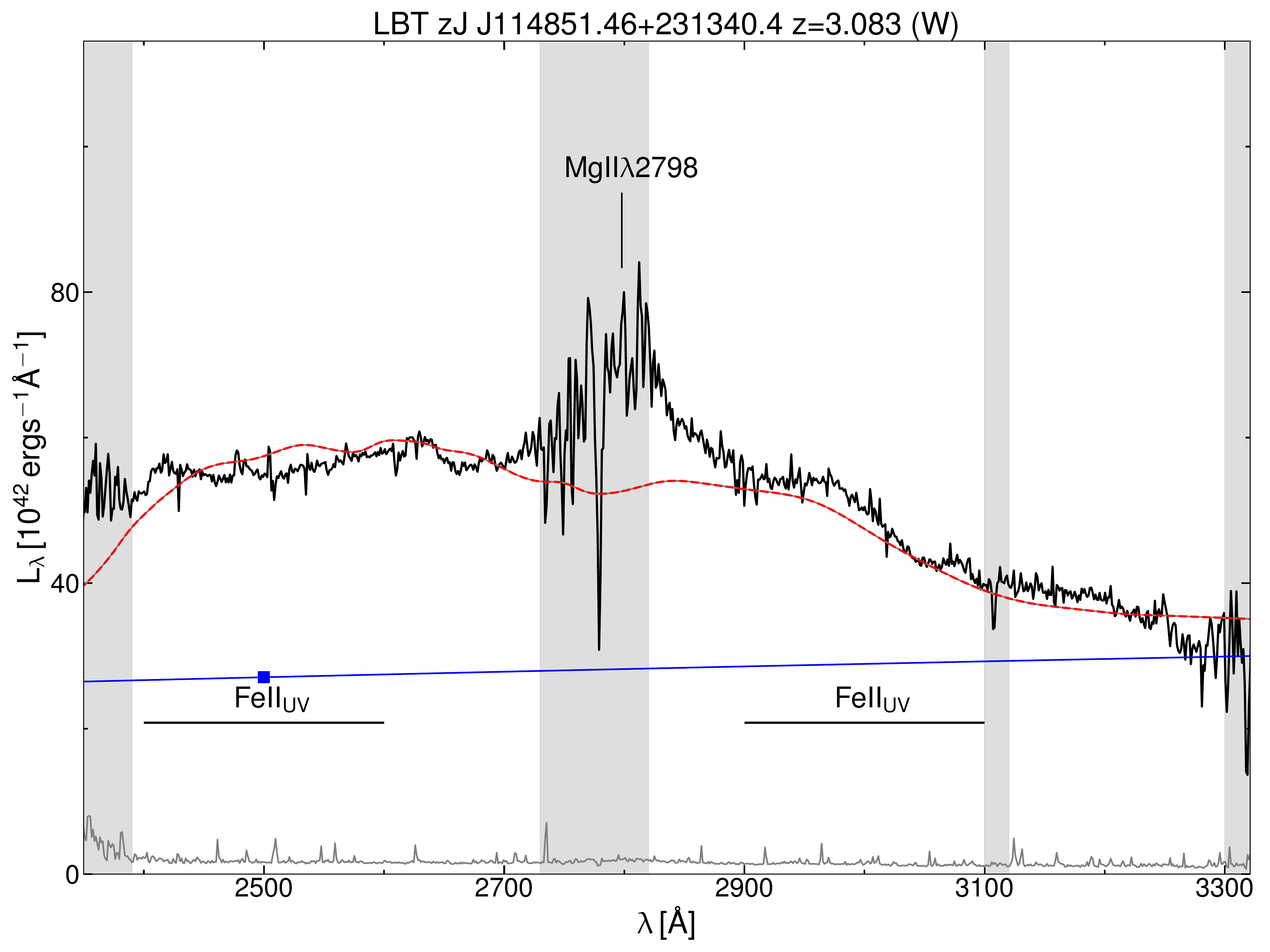}\hfill
    \includegraphics[width=.33\textwidth]{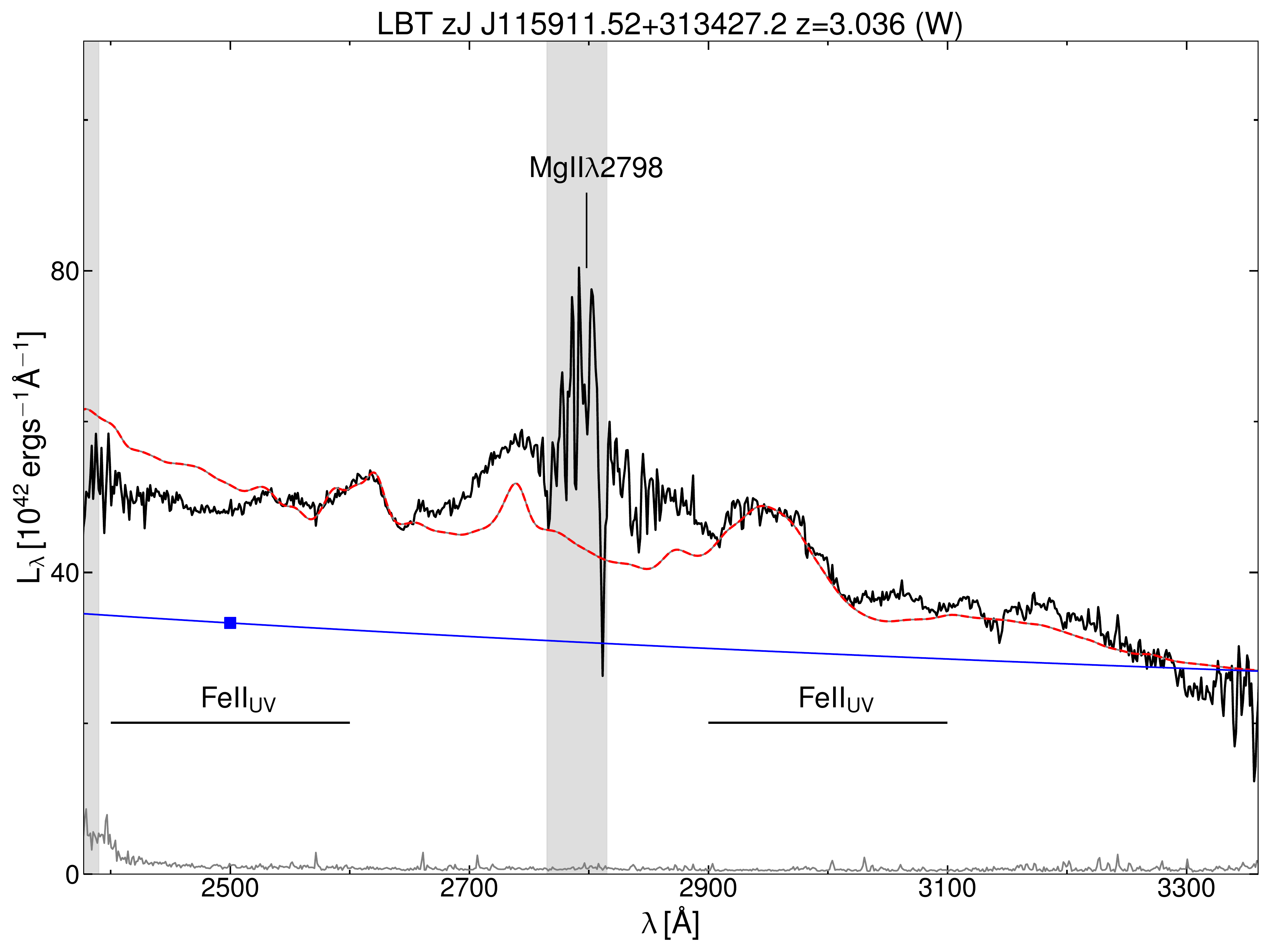}\hfill
    \includegraphics[width=.33\textwidth]{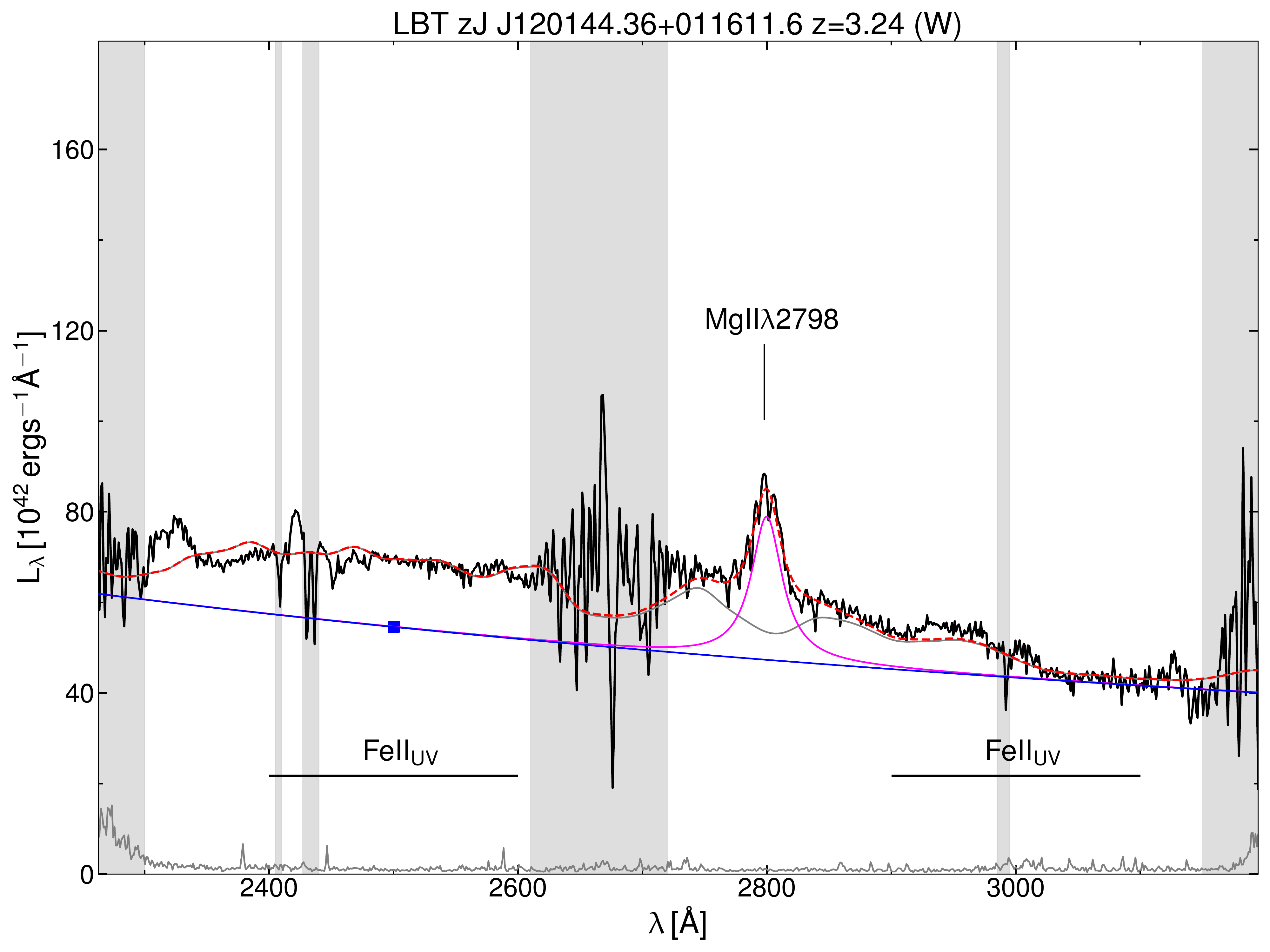}

    \includegraphics[width=.33\textwidth]{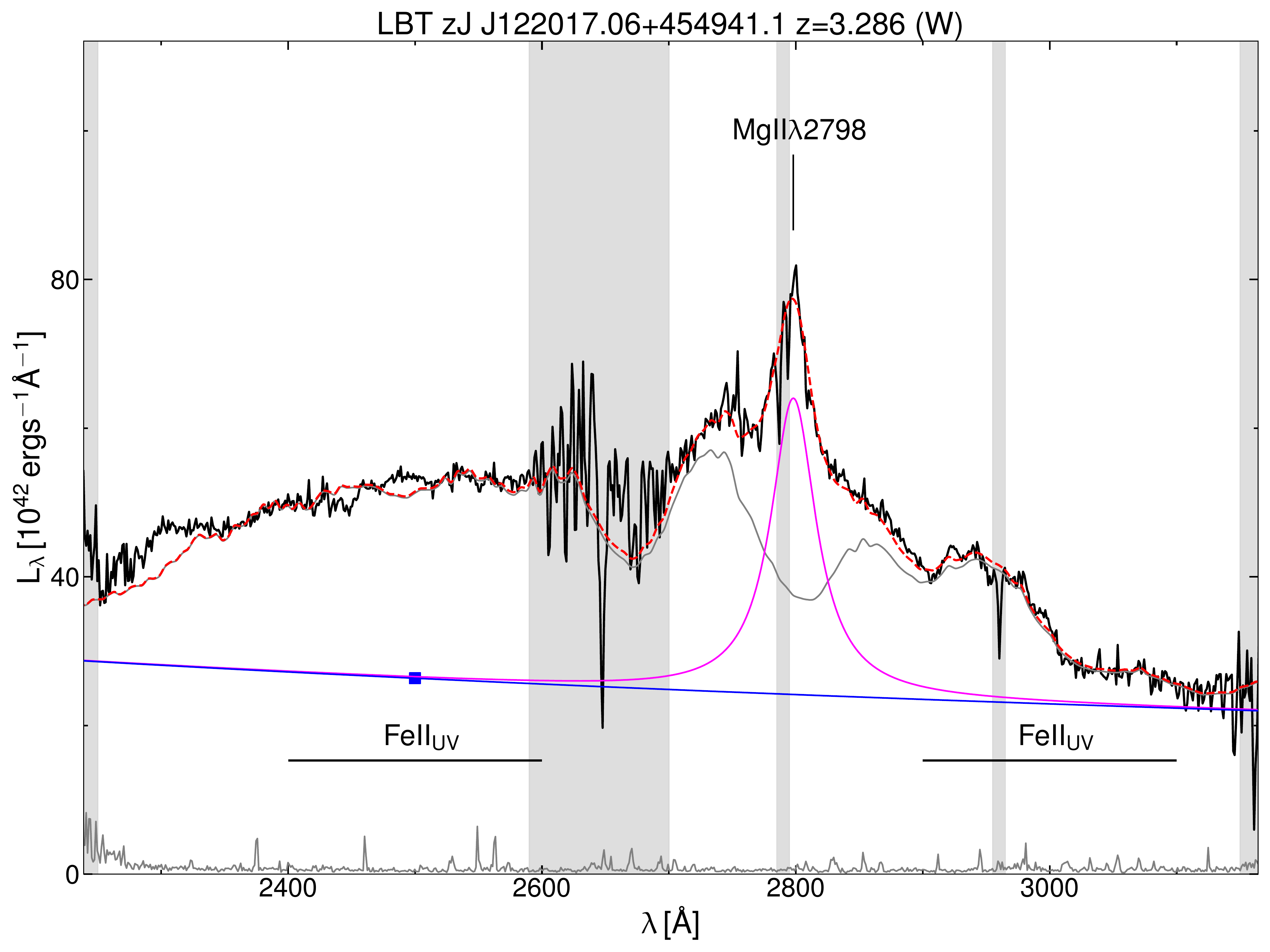}\hfill
    \includegraphics[width=.33\textwidth]{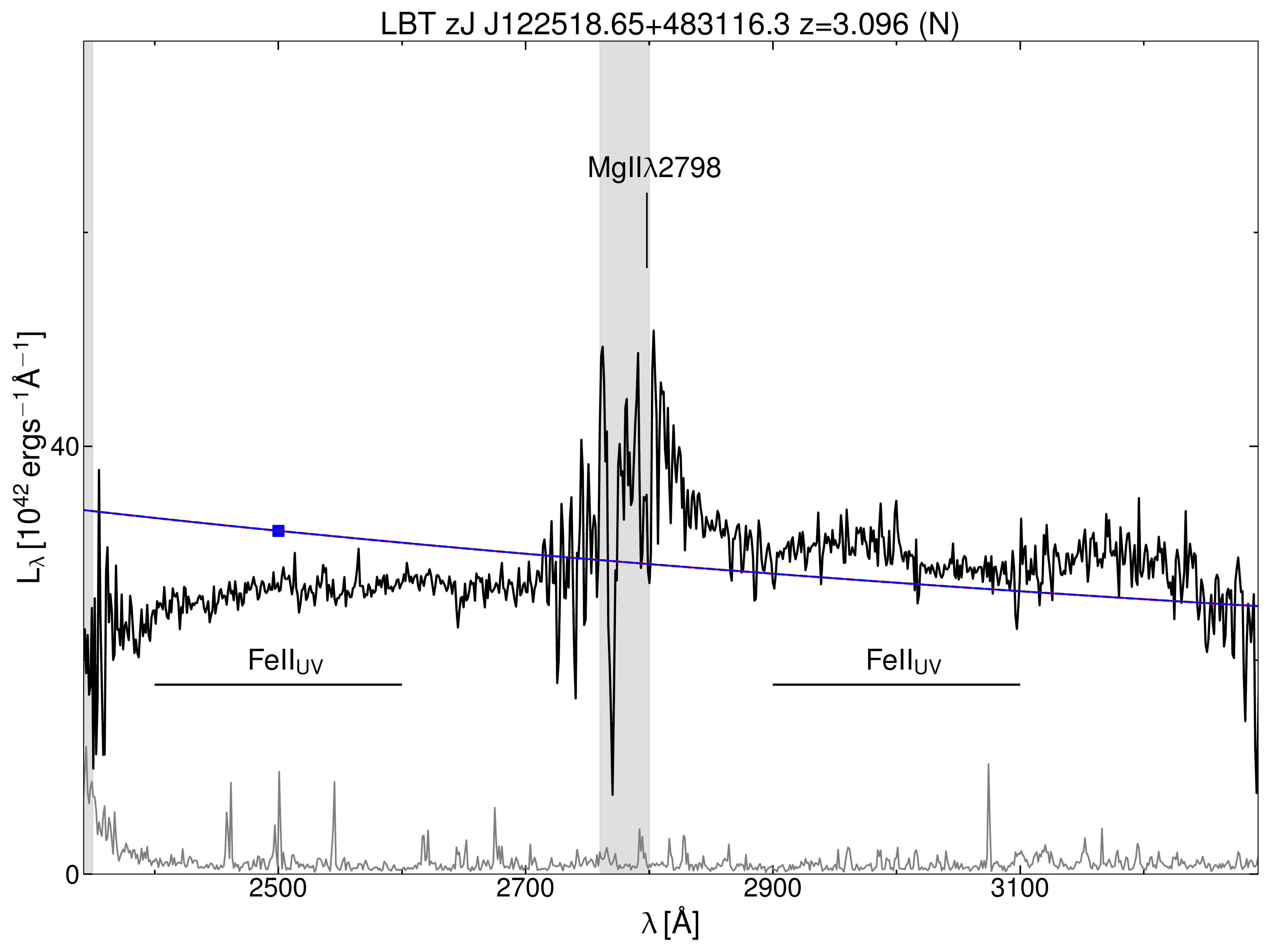}\hfill
    \includegraphics[width=.33\textwidth]{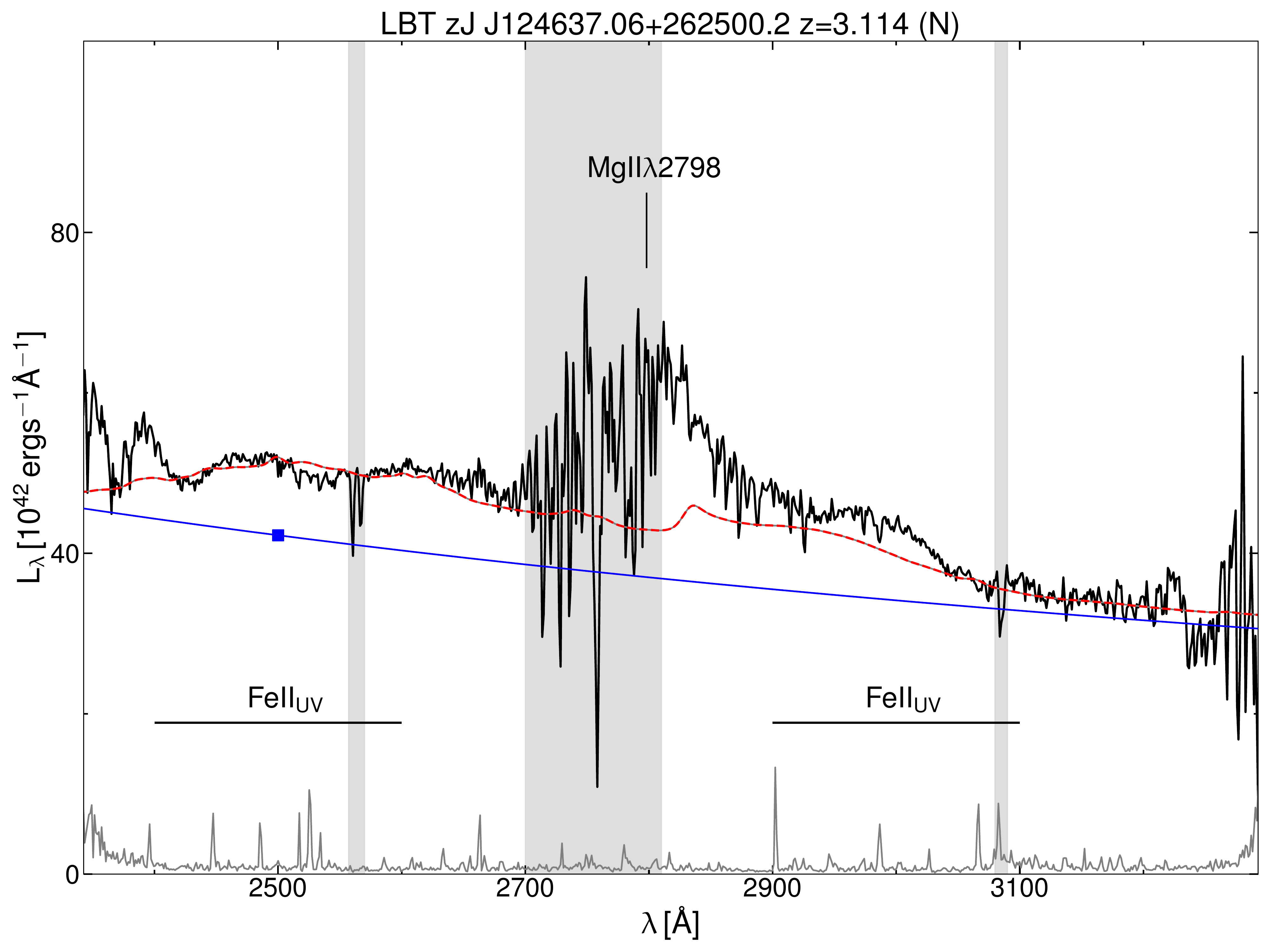}

    \includegraphics[width=.33\textwidth]{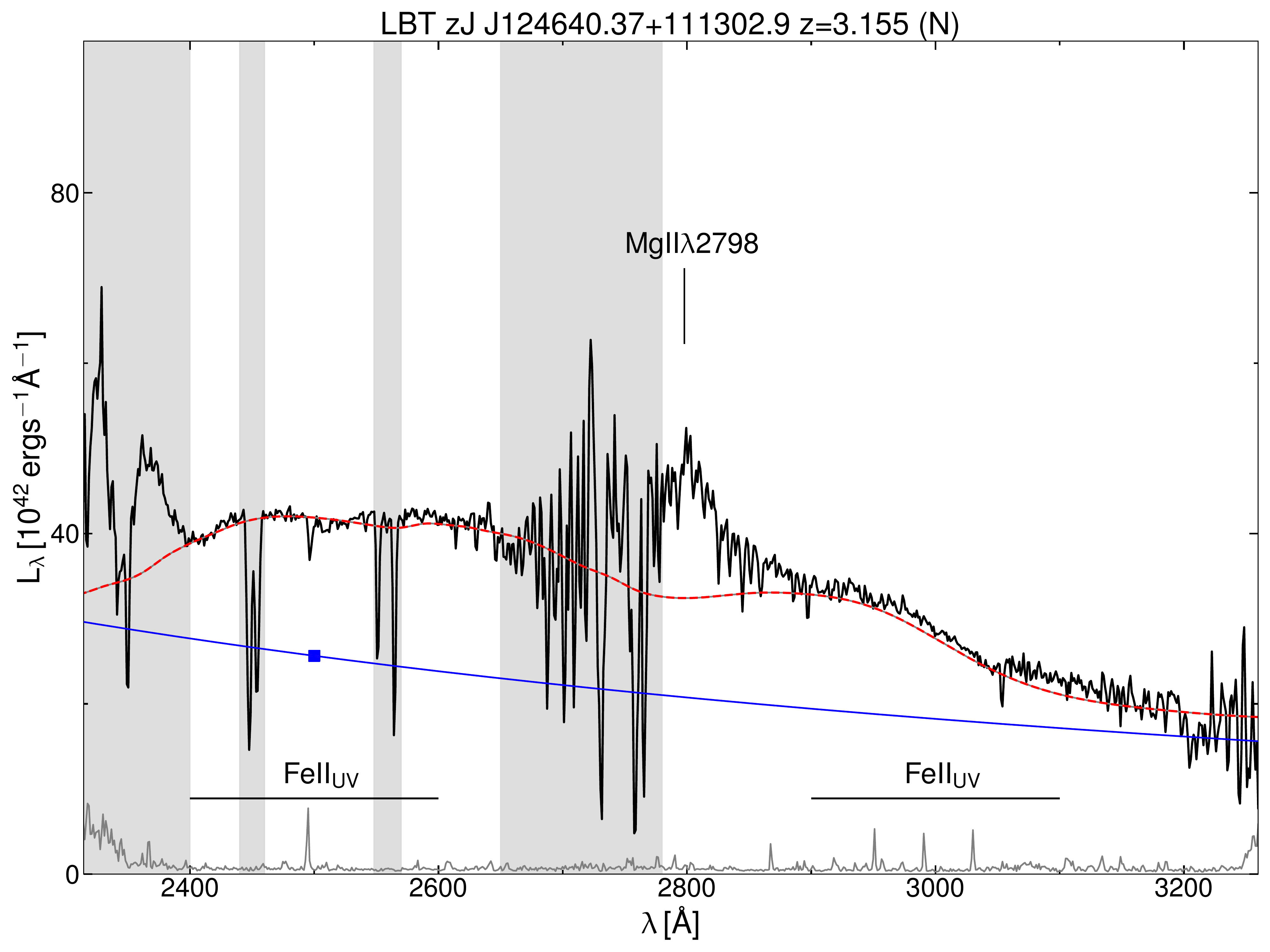}\hfill
    \includegraphics[width=.33\textwidth]{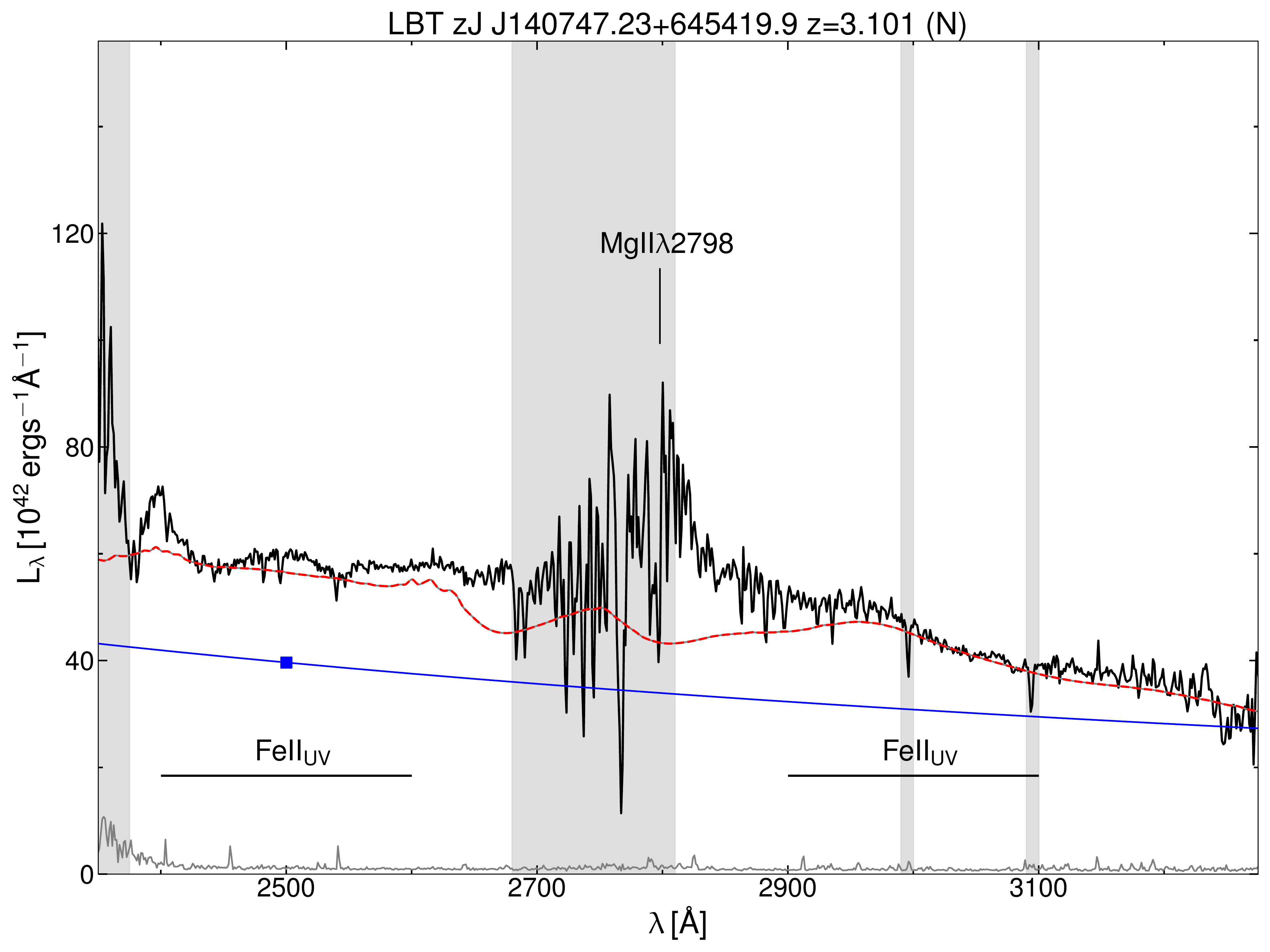}\hfill
    \includegraphics[width=.33\textwidth]{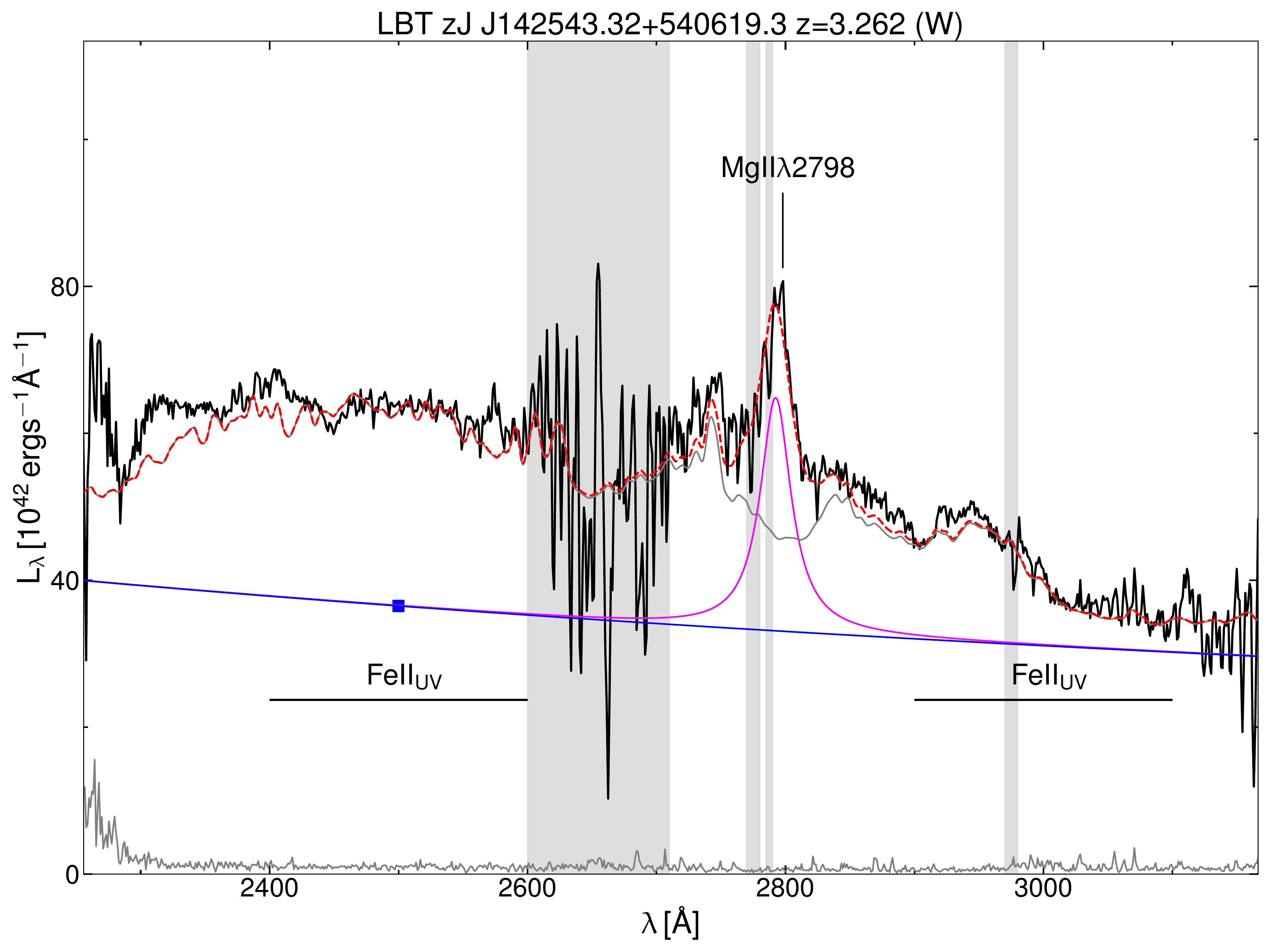}

    \includegraphics[width=.33\textwidth]{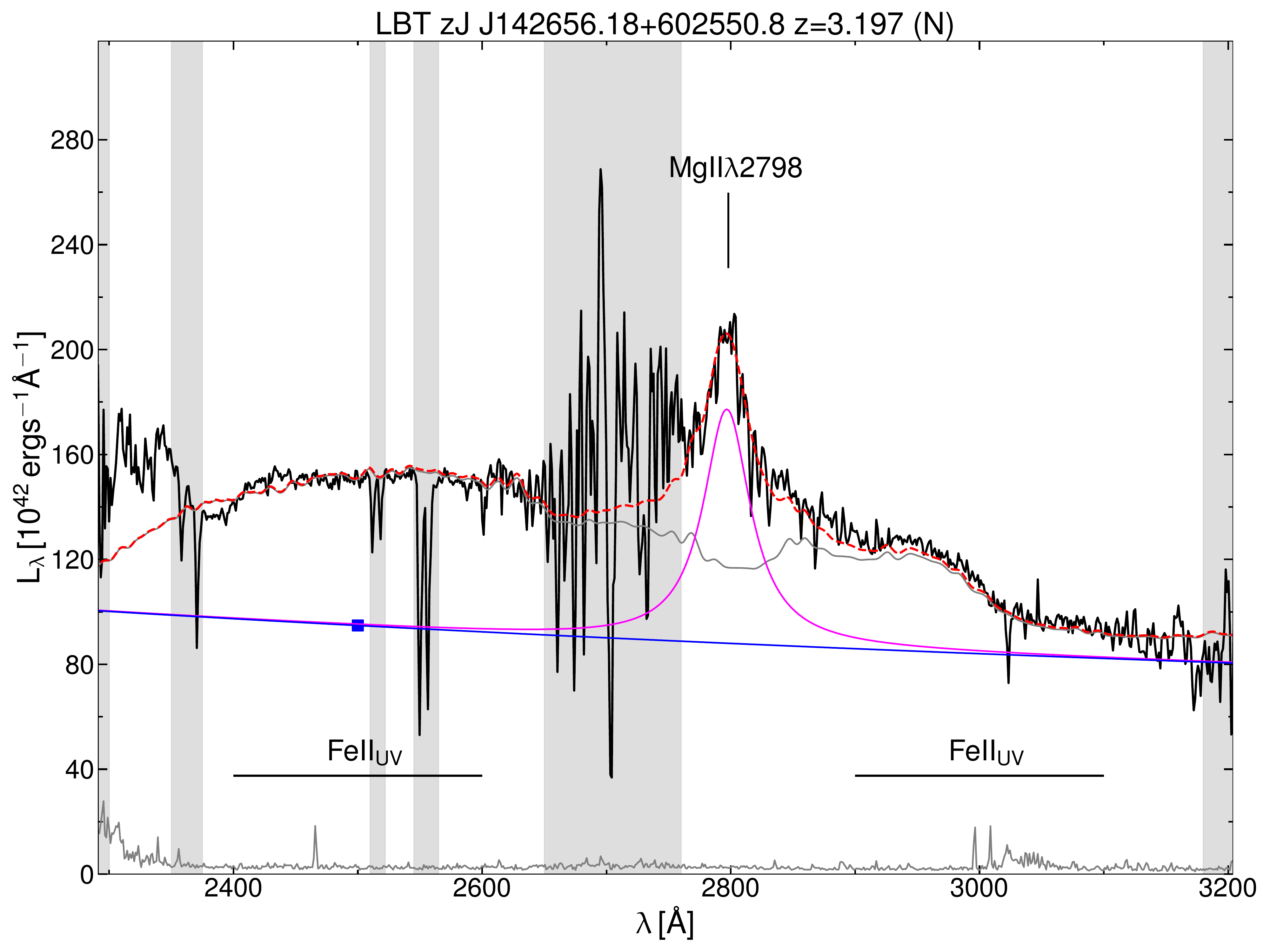}\hfill
    \includegraphics[width=.33\textwidth]{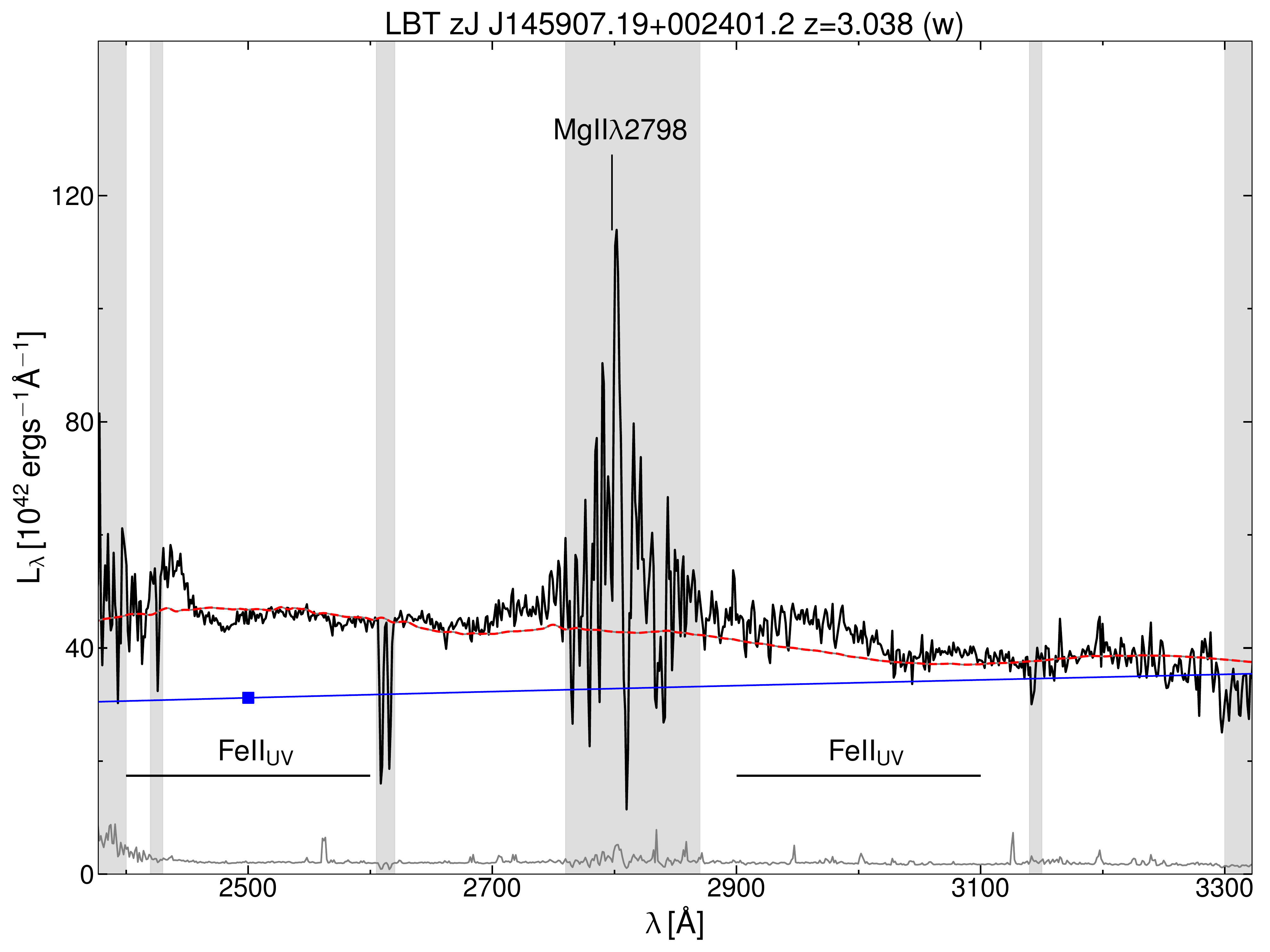}\hfill
    \includegraphics[width=.33\textwidth]{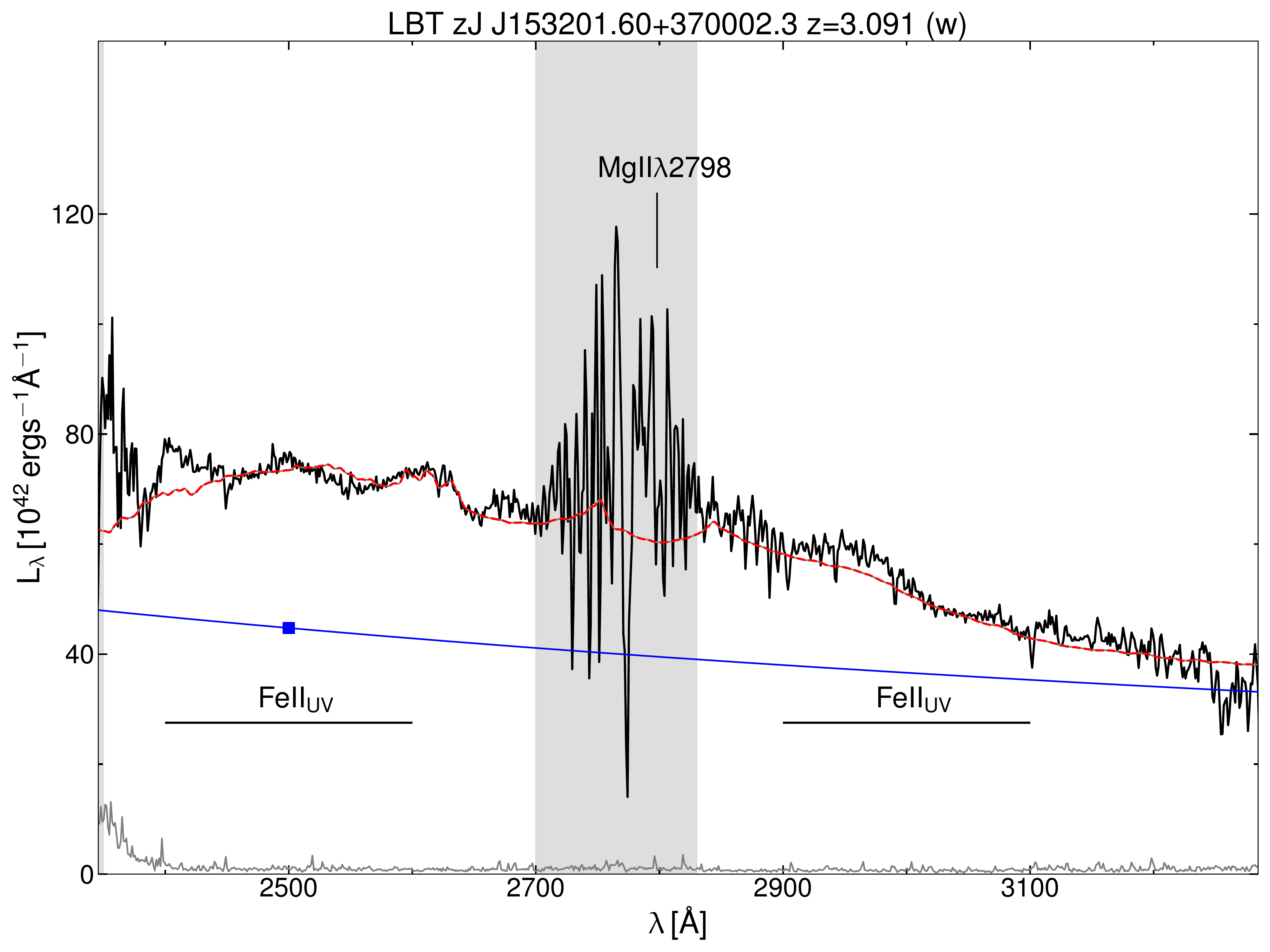}

    \includegraphics[width=.33\textwidth]{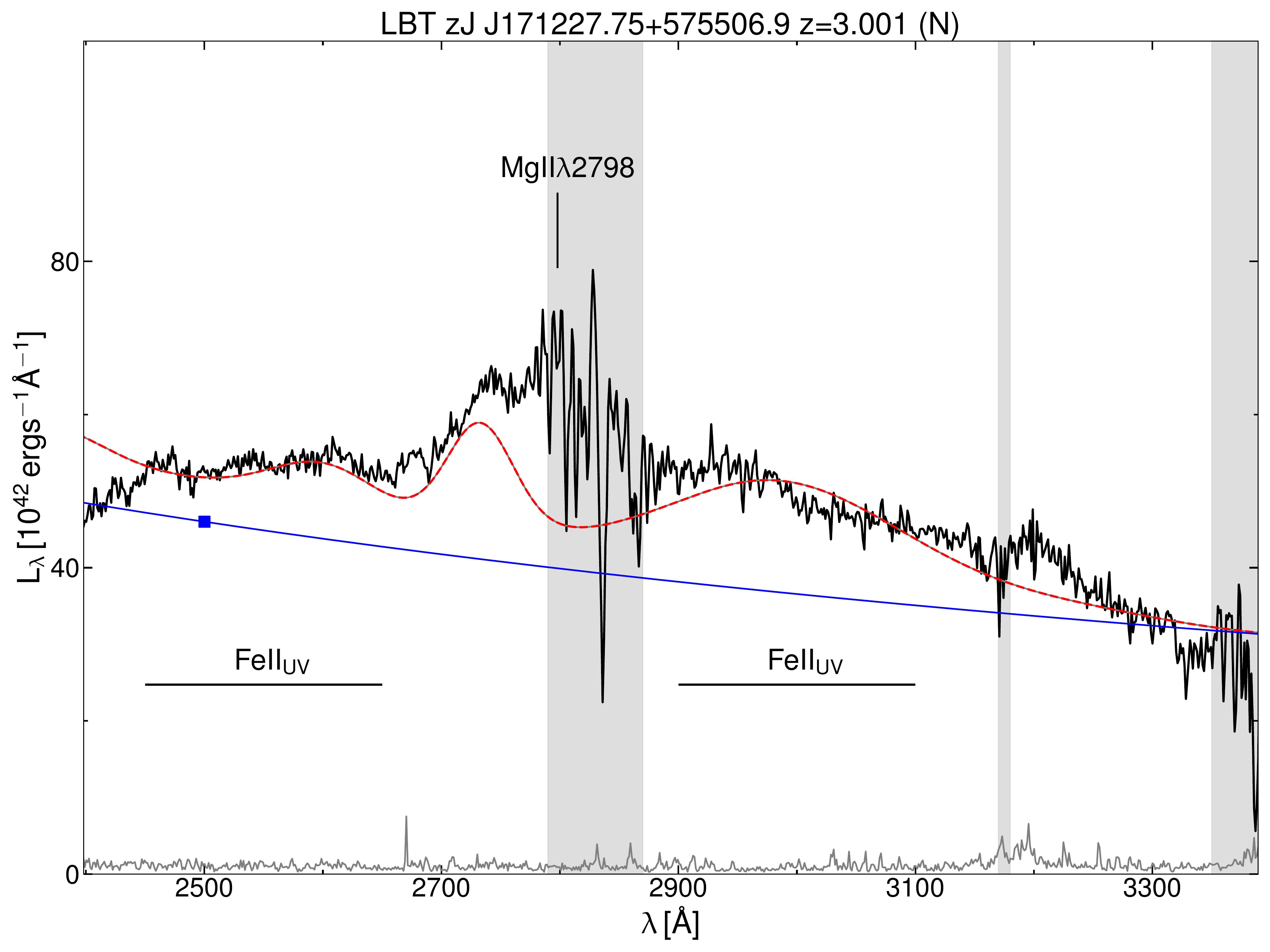}
    \includegraphics[width=.33\textwidth]{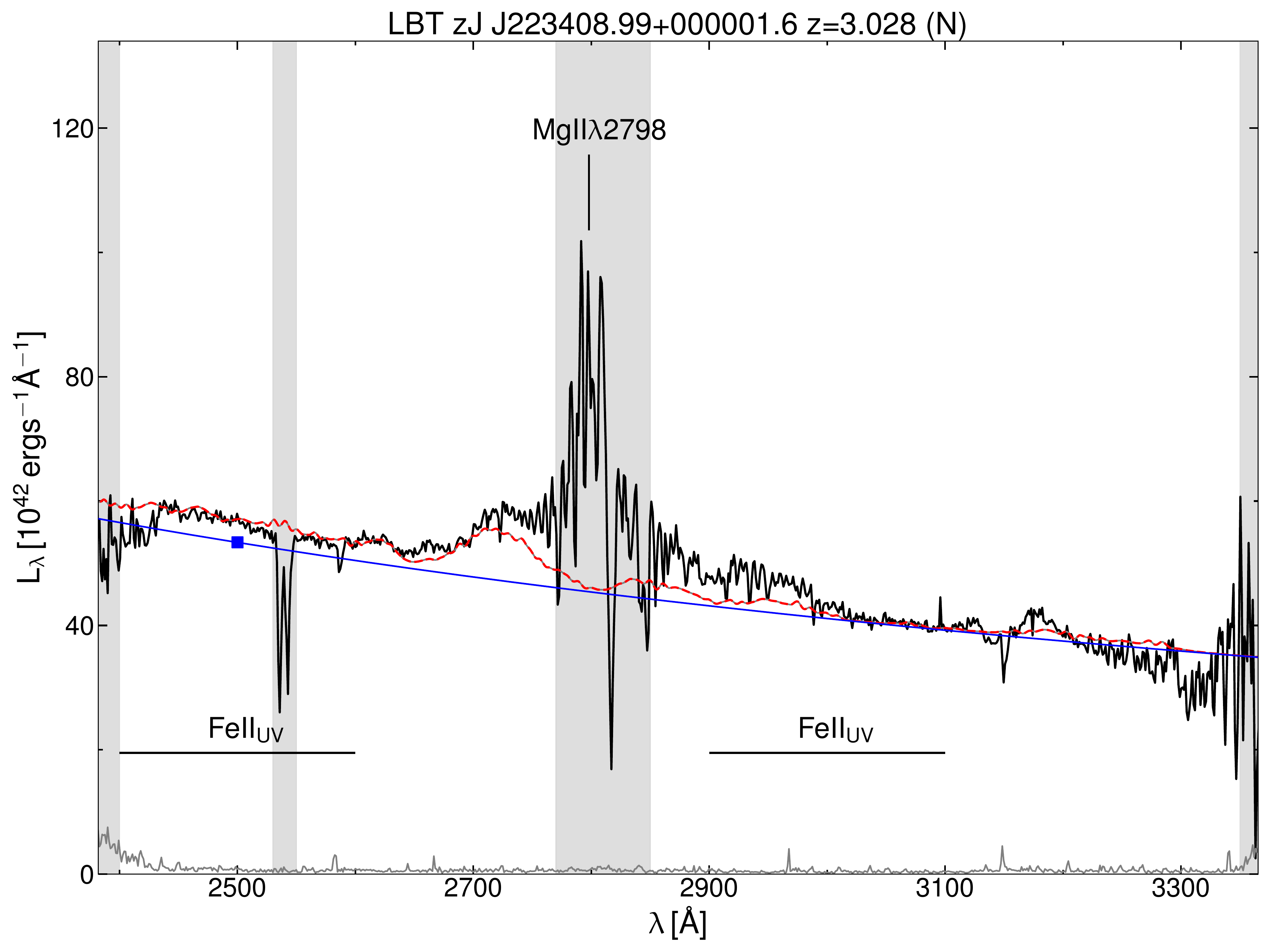}\hfill

\label{fig:A1}    
\captionsetup{labelformat=empty}
\caption{Fig. A1 continued: LBT $zJ$ spectra}
\addtocounter{figure}{-1}
\end{figure}

\newpage

\begin{figure}
    \includegraphics[width=.33\textwidth]{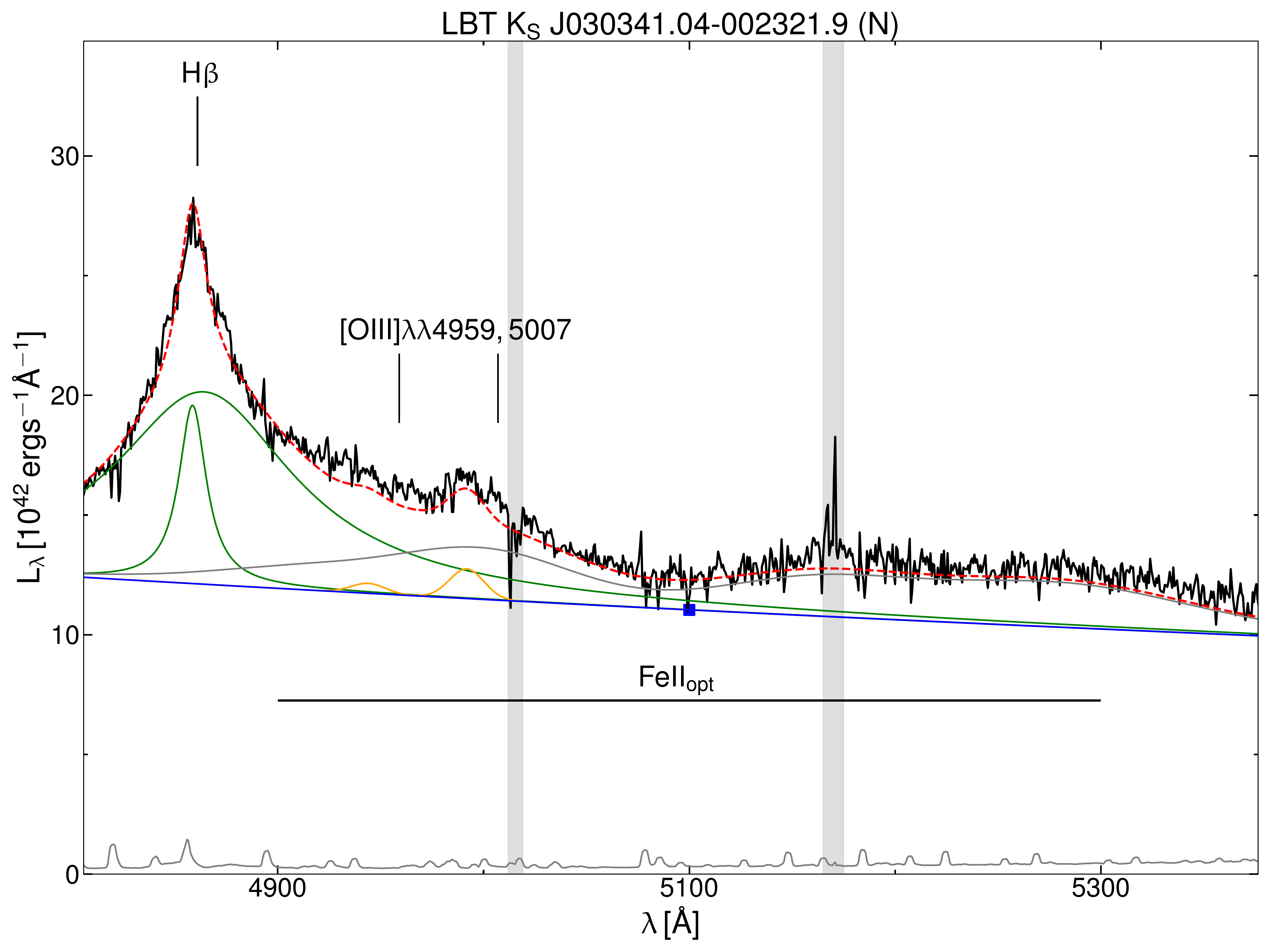}\hfill
    \includegraphics[width=.33\textwidth]{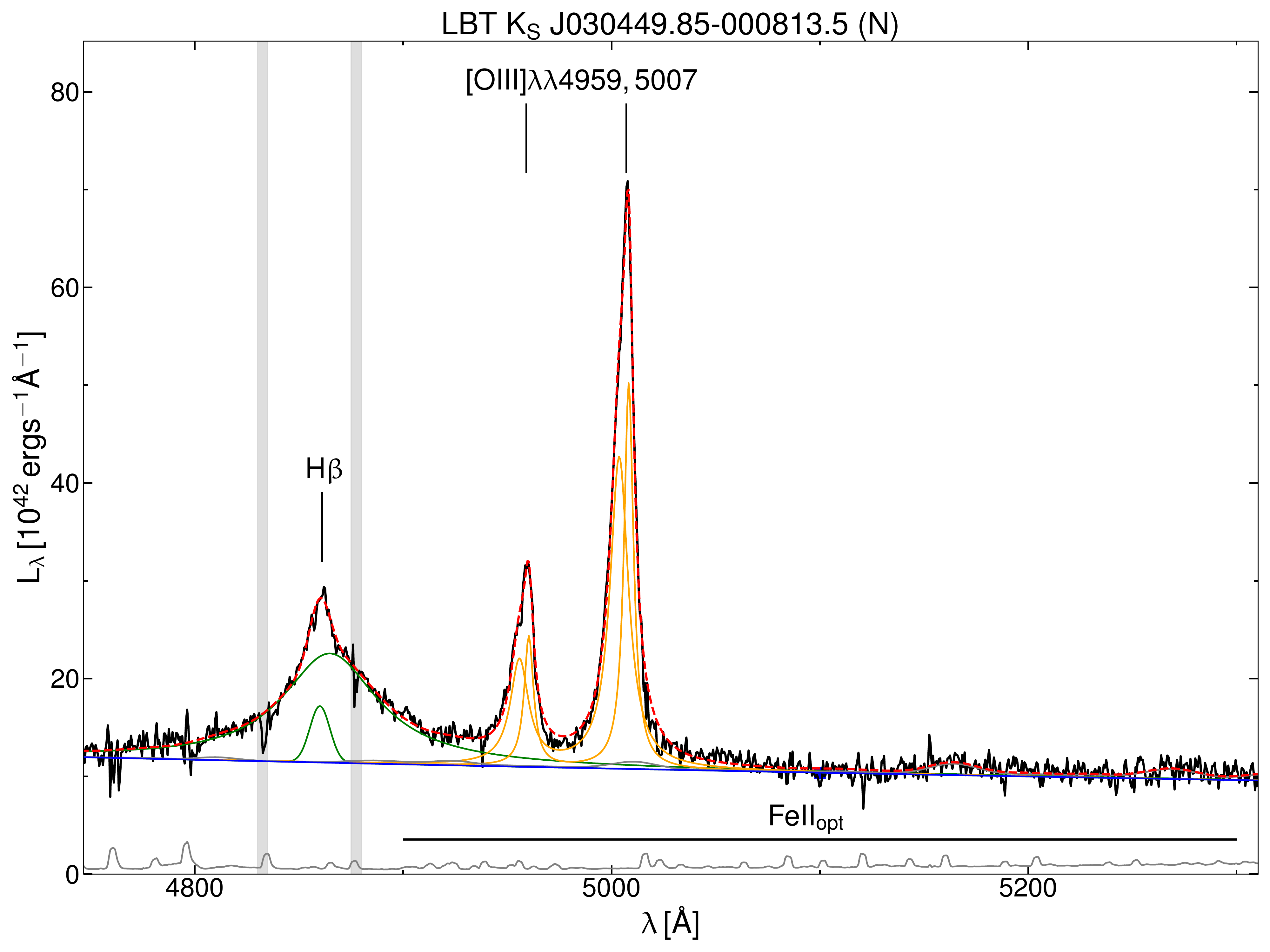}\hfill
    \includegraphics[width=.33\textwidth]{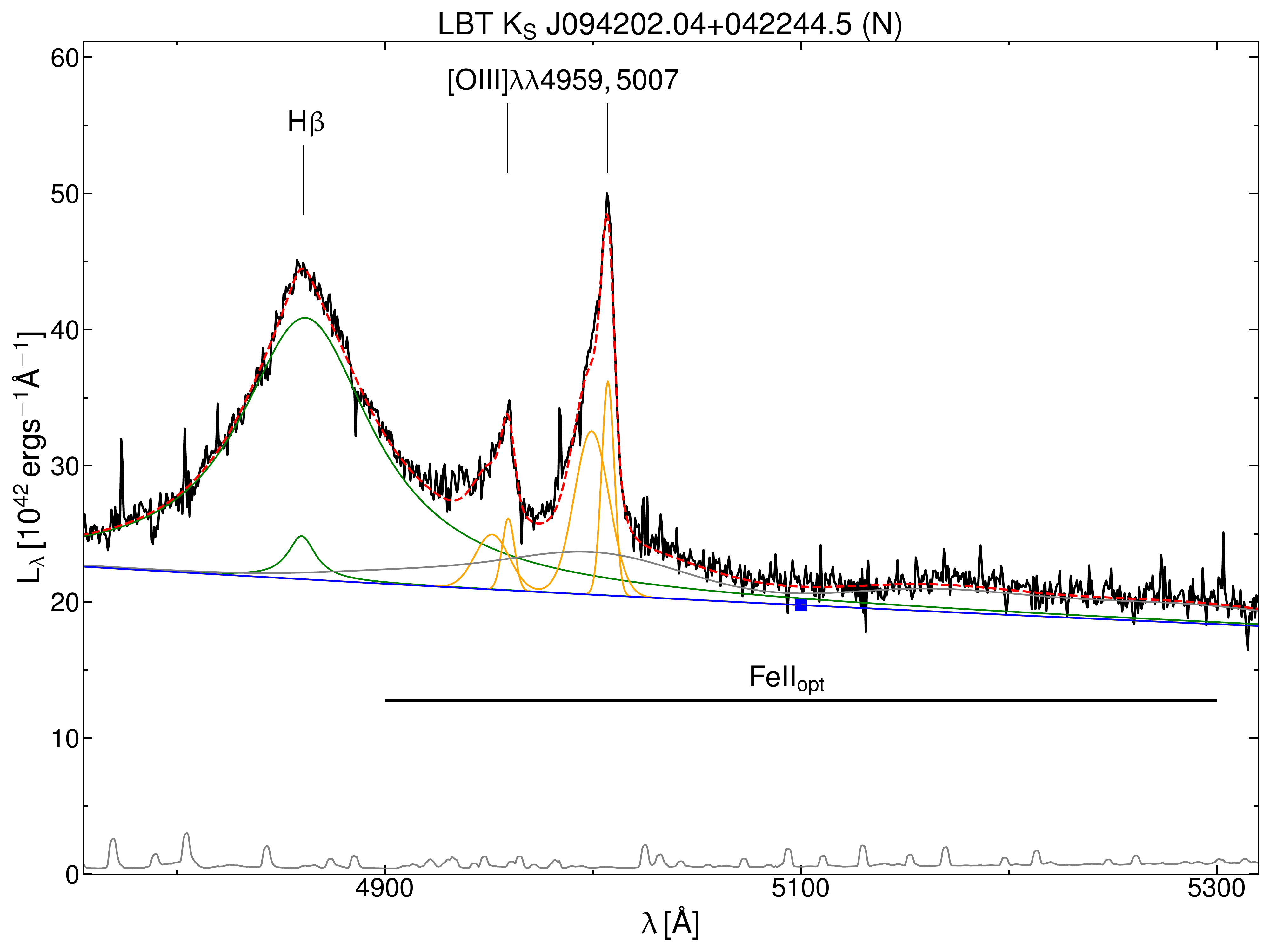}
    \\[\smallskipamount]

    \includegraphics[width=.33\textwidth]{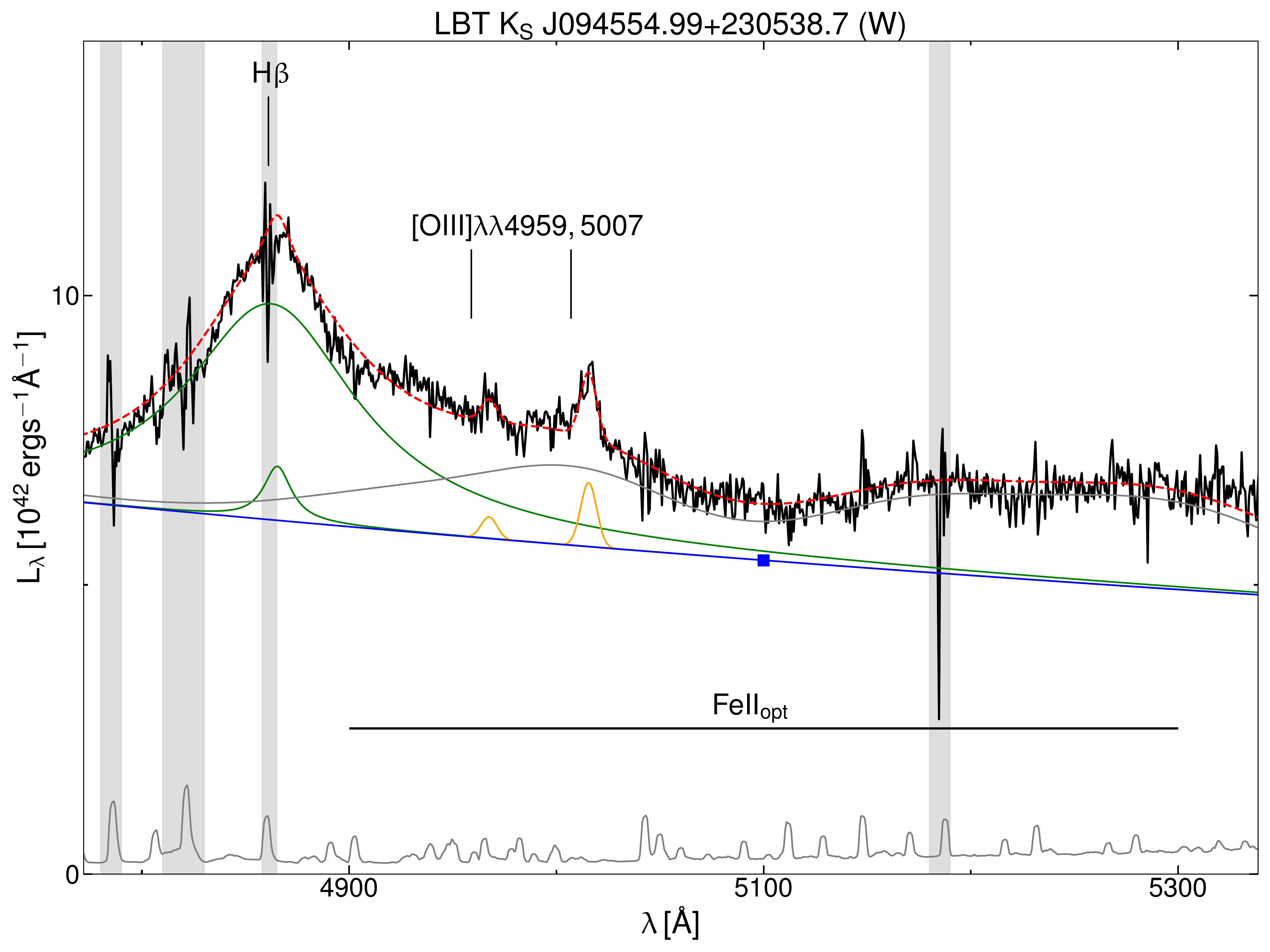}\hfill
    \includegraphics[width=.33\textwidth]{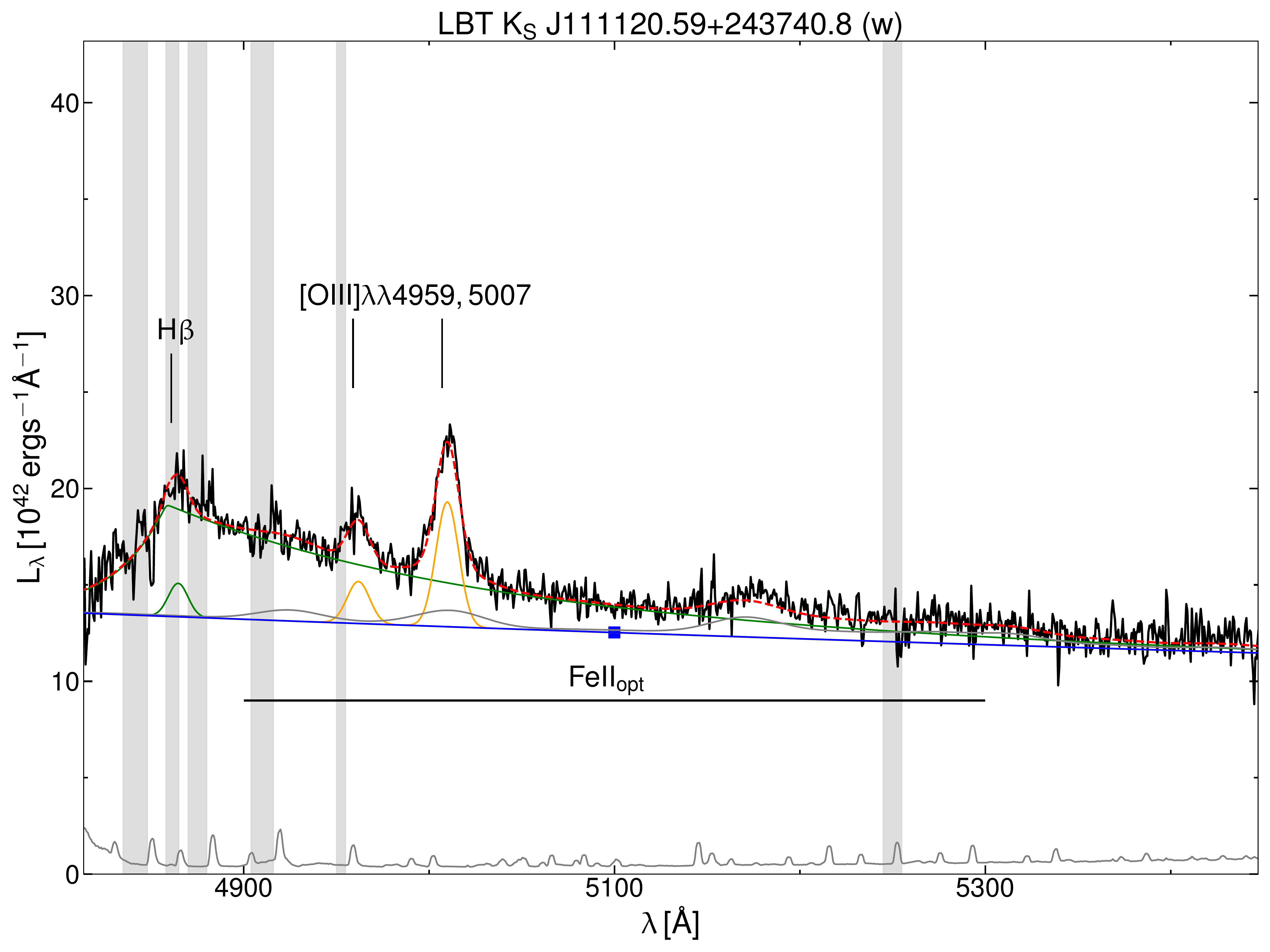}\hfill
    \includegraphics[width=.33\textwidth]{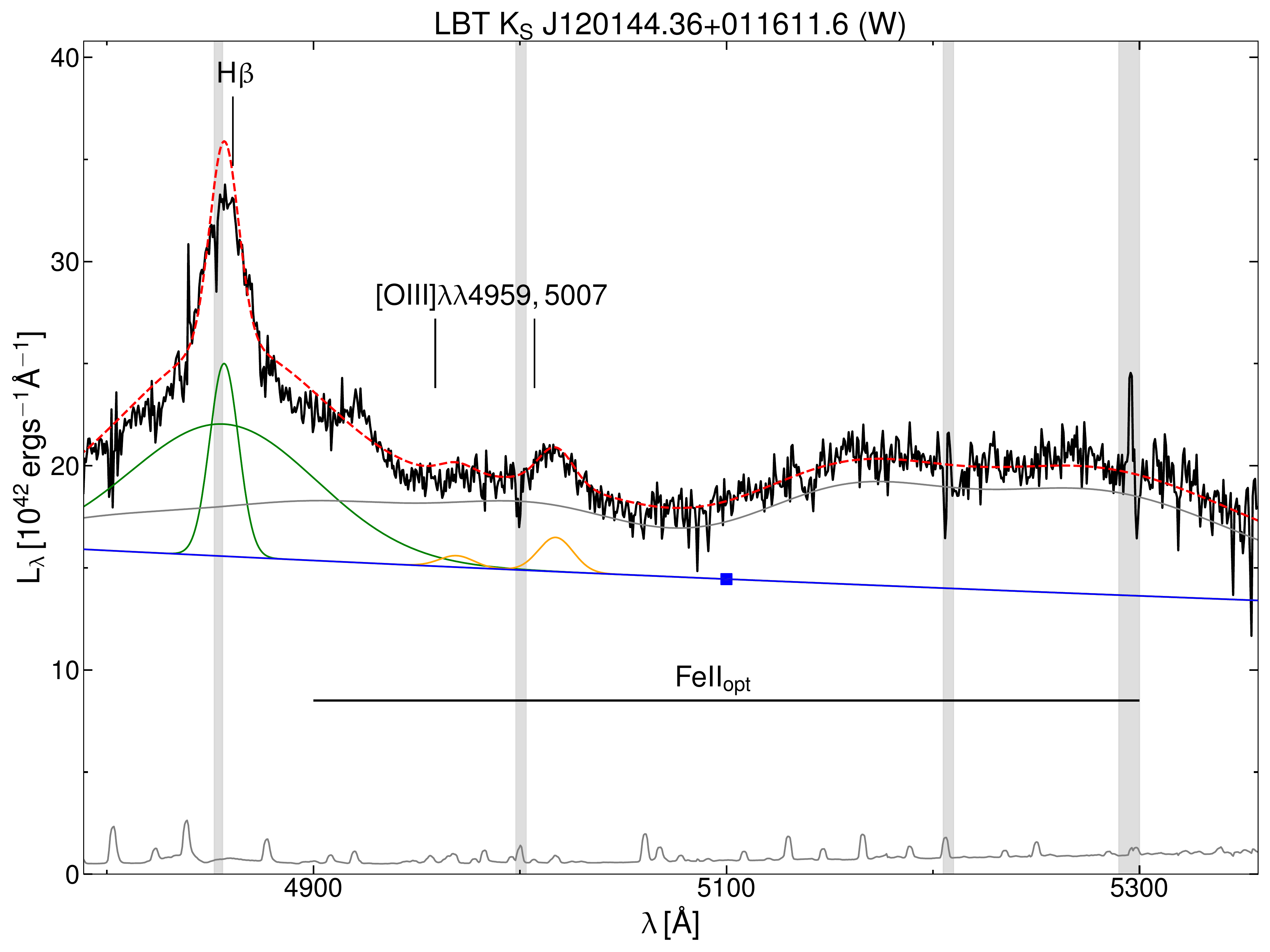}
    \\[\smallskipamount]

    \includegraphics[width=.33\textwidth]{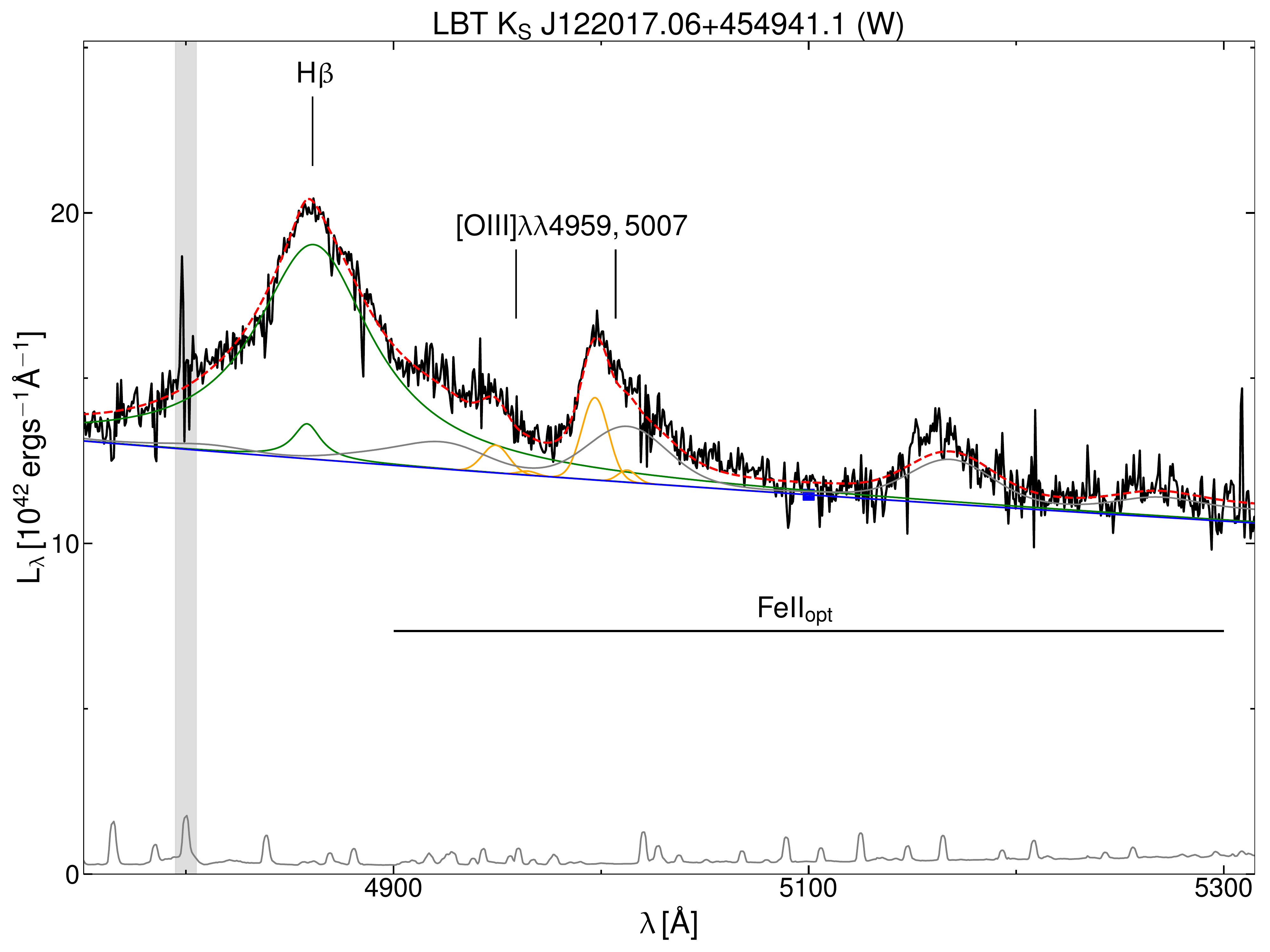}\hfill
    \includegraphics[width=.33\textwidth]{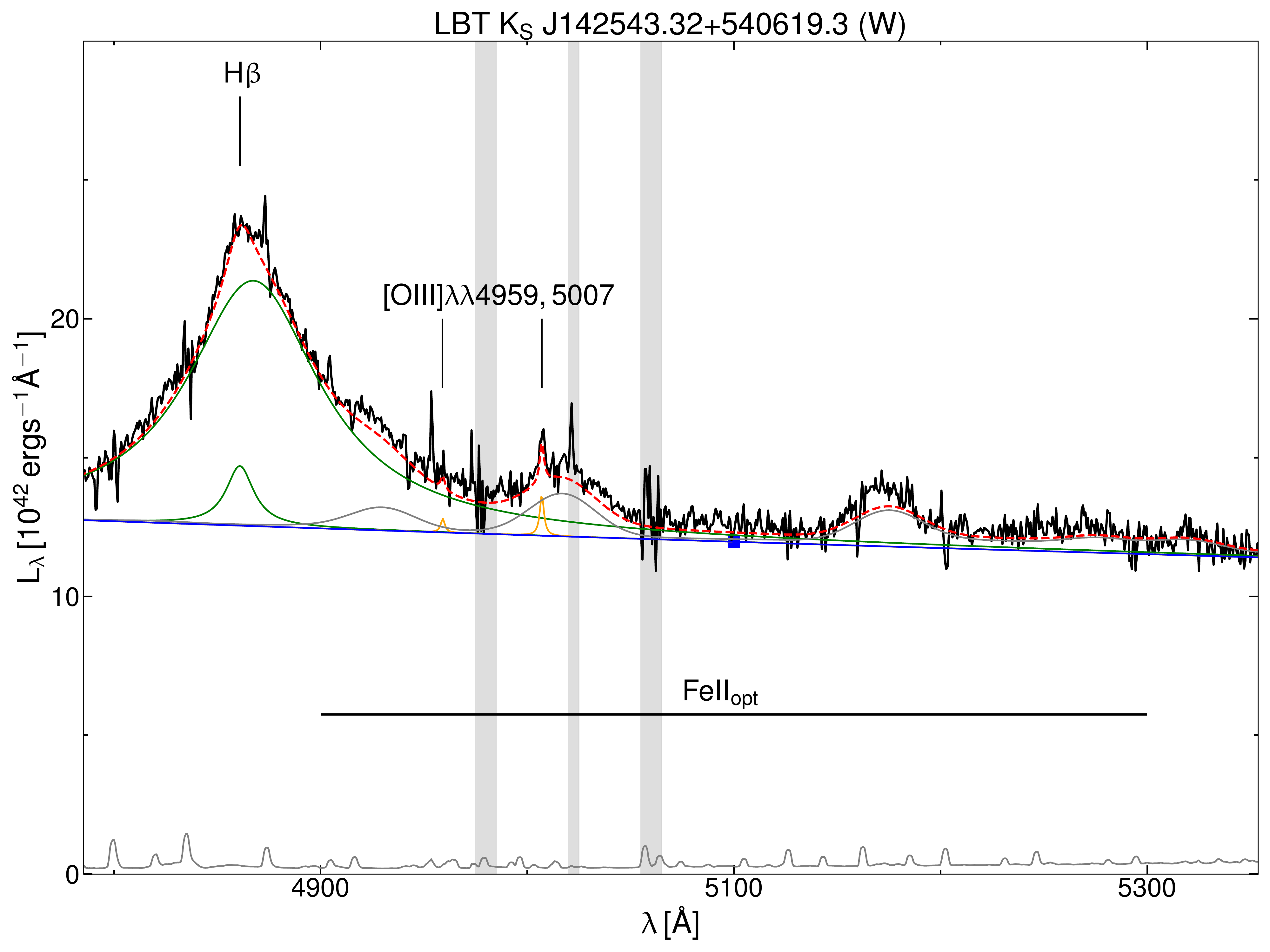}\hfill
    \includegraphics[width=.33\textwidth]{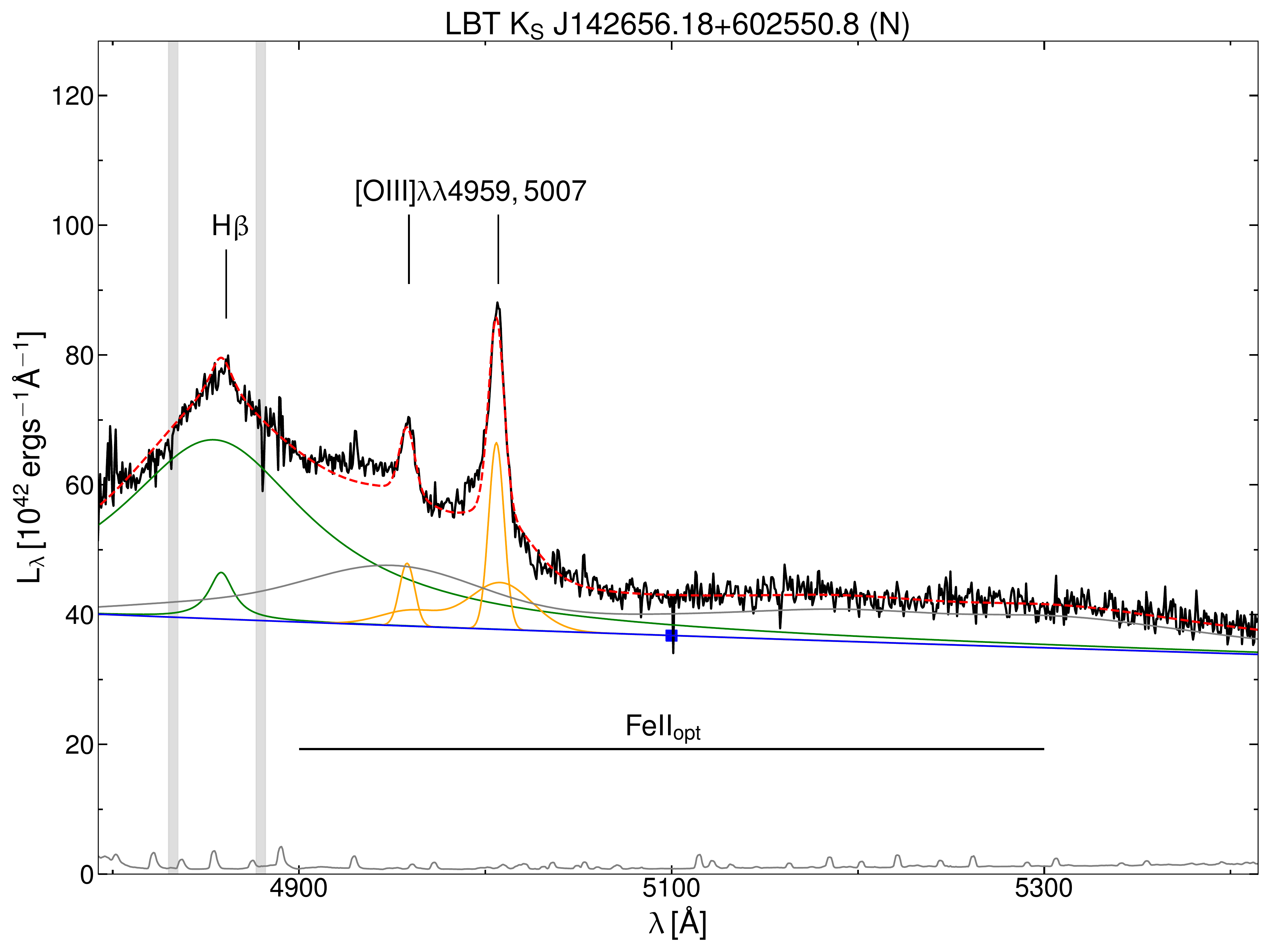}
    \\[\smallskipamount]

\captionsetup{labelformat=empty}
\caption{Fig. A2: LBT $\ks$ spectra. The colour code is described in Appendix \ref{app:appendix_b}.}
\label{fig:appendix2}    

\end{figure}

\newpage

\clearpage
\onecolumn
\begin{longtable}{lccccccc}

\caption*{Table A1: UV spectral properties of the $z$\,$\sim$\,3 sample.}\\
\label{tbl:TA1}\\

\hline\hline
Name & X-ray class & EW\,\ion{Mg}{ii} & EW\,\ion{Fe}{ii}\textsubscript{UV} & \ion{Fe}{ii}/\ion{Mg}{ii} & EW\,H$\beta$ & EW\,[\ion{O}{iii}] &  EW\,\ion{Fe}{ii}\textsubscript{opt}\tablefootmark{a}\\

  &  &  [\AA]  &  [\AA]  &  &  [\AA]  &  [\AA]  &  [\AA] \\

\hline

J0301-0035 & $N$ &  82.8$\pm$9.5  & 373.4$\pm$36.9  & 4.5$\pm$0.5  & 109.9$\pm$4.1  & 2.3$\pm$0.2  & 53.6$\pm$2.5   \\
J0304-0008 & $N$ &  35.3$\pm$2.6  & 196.7$\pm$49.7  & 5.6$\pm$1.0  & 77.2$\pm$2.5   & 77.0$\pm$2.1 & 7.3$\pm$5.8  \\
J0826+3148 & $N$ &  -             & 127.6$\pm$74.8  & -            & -              & -            & -             \\
J0835+2122 & $N$ &  -             & 135.6$\pm$8.6   & -            & -              & -            & -             \\
J0900+4215 & -   &  40.9$\pm$0.8  & 131.9$\pm$14.6  & 3.2$\pm$0.3  & -              & -            & -             \\
J0901+3549 & $N$ &  -             & 299.7$\pm$87.2  & -            & -              & -            & -             \\
J0905+3057 & $N$ &  -             & 695.5$\pm$144.9 & -            & -              & -            & -             \\
J0942+0422 & $N$ &  62.0$\pm$2.1  & 315.0$\pm$14.2  & 5.1$\pm$0.5  & 108.0$\pm$3.7  & 18.1$\pm$3.1 & 32.1$\pm$8.4  \\
J0945+2305 & $W$ &  72.4$\pm$4.0  & 511.9$\pm$70.2  & 7.1$\pm$1.5  & 95.5$\pm$8.9   & 1.9$\pm$0.3  & 77.4$\pm$7.5  \\
J0947+1421 & $N$ &  -             & 312.2$\pm$150.5 & -            & -              & -            & -             \\
J1014+4300 & $N$ &  -             & 318.3$\pm$21.0  & -            & -              & -            & -             \\
J1027+3543 & $N$ &  -             & 341.6$\pm$48.1  & -            & -              & -            & -             \\
J1111-1505 & $w$ & -              & 32.2 $\pm$2.9   & -            & -              & -            & -             \\
J1111+2437 & $w$ &  59.2$\pm$14.3 & 801.3$\pm$70.1  & 13.5$\pm$7.4 & 81.4$\pm$41.1  & 7.8$\pm$0.6  & 13.3$\pm$2.0  \\
J1143+3452 & $N$ &  36.5$\pm$0.5  & 77.2 $\pm$11.8  & 2.1$\pm$0.3  & -              & -            & -             \\
J1148+2313 & $W$ & -              & 650.4$\pm$123.9 & -            & -              & -            & -             \\
J1159+3134 & $W$ &  -             & 547.4$\pm$49.6  & -            & -              & -            & -             \\
J1201+0116 & $W$ &  34.3$\pm$1.7  & 178.7$\pm$8.3   & 5.2$\pm$0.5  & 47.0$\pm$8.6   & 2.4$\pm$1.0  & 96.6$\pm$13.8 \\
J1220+4549 & $W$ &  118.5$\pm$12.7& 679.3$\pm$31.3  & 5.7$\pm$1.1  & 55.1$\pm$1.1   & 3.7$\pm$0.3  & 20.0$\pm$1.5  \\
J1225+4831 & $N$ &  -             & -               & -            & -              & -            & -             \\
J1246+2625 & $N$ &  -             & 151.1$\pm$30.5  & -            & -              & -            & -             \\
J1246+1113 & $N$ &  -             & 584.3$\pm$183.6 & -            & -              & -            & -             \\
J1407+6454 & $N$ &  -             & 394.6$\pm$104.6 & -            & -              & -            & -             \\
J1425+5406 & $W$ &  46.4$\pm$2.8  & 486.4$\pm$52.2  & 10.5$\pm$1.2 & 85.8$\pm$1.4   & 0.5$\pm$0.6  & 17.7$\pm$1.3  \\
J1426+6025 & $N$ &  76.7$\pm$5.5  & 364.5$\pm$56.8  & 4.8$\pm$0.6  & 136.6$\pm$55.3 & 15.9$\pm$2.6 & 51.2$\pm$25.5 \\
J1459+0024 & $w$ &  -             & 324.7$\pm$25.7  & -            & -              & -            & -             \\
J1532+3700 & $w$ & -              & 472.5$\pm$102.9 & -            & -              & -            & -             \\
J1712+5755 & $N$ &  -             & 222.1$\pm$17.3  & -            & -              & -            & -             \\
J2234+0000 & $N$ &  -             & 54.0 $\pm$0.5   & -            & -              & -            & -             \\
\hline

\end{longtable}
\tablefoottext{a}{EW \ion{Fe}{ii}\textsubscript{opt} evaluated between 4900--5300 \AA.}

\newpage

\clearpage
\onecolumn
\begin{longtable}{lcccccccccc}
\caption*{Table A1 continued}\\

\hline\hline
FWHM \ion{Mg}{ii} & FWHM H$\beta$ & L\textsubscript{[\ion{O}{iii}]$\lambda$5007} & Log L$_{3000 \AA}$ & Log L$_{5100 \AA}$ & Log M$\rm{_{BH}}$ & $\lambda_{\rm{Edd}}$   & $\alpha_{\rm OX}$ & $\Delta \alpha_{\rm OX}$ & $\Gamma$  \\

  [km s$^{\rm -1}$]  &  [km s$^{-1}$]  & [$10^{42}$ erg s$^{\rm -1}$] & [erg s$^{\rm -1}$]  & [erg s$^{\rm -1}$] & [M$_{\odot}$]  &  &  &  \\

\hline

4693$\pm$25      & 6509$\pm$134  & 25.8$\pm$2.3      & 46.94  & 46.75 & 9.8$\pm$0.2 & 0.8 & -1.68 & 0.02   & 1.87$_{-0.07}^{+0.08}$ \\
2105$\pm$15      & 3946$\pm$180  & 839.0$\pm$13.4    & 46.95  & 46.73 & 9.2$\pm$0.2 & 4.6  & -1.64 & 0.07  & 1.99$_{-0.06}^{+0.05}$ \\
\hspace{0.6cm} - & -             & -                 & 46.90  & -     & 9.7$\pm$0.4 & 0.7  & -1.87 & -0.17 & 1.56$_{-0.16}^{+0.17}$ \\
\hspace{0.6cm} - & -             & -                 & 46.95  & -     & 9.8$\pm$0.4 & 0.4  & -1.67 & 0.03  & 1.77$_{-0.06}^{+0.07}$ \\
3214$\pm$34      & -             & -                 & 47.34  & -     & 9.9$\pm$0.2 & 1.0  & -1.56 & 0.20  & 1.83$_{-0.03}^{+0.03}$ \\
\hspace{0.6cm} - & -             & -                 & 46.94  & -     & 9.7$\pm$0.4 & 0.9  & -1.82 & -0.12 & 1.60$_{-0.08}^{+0.07}$ \\
\hspace{0.6cm} - & -             & -                 & 46.81  & -     & 9.1$\pm$0.4 & 4.0  & -1.60 & 0.11  & 2.12$_{-0.05}^{+0.06}$ \\
3830$\pm$52      & 4779$\pm$75   & 369.5$\pm$64.9    & 47.06  & 47.00 & 9.6$\pm$0.2 & 1.2  & -1.70 & 0.02  & 2.11$_{-0.10}^{+0.11}$ \\
4487$\pm$39      & 6156$\pm$223  & 10.7$\pm$1.5      & 46.56  & 46.44 & 9.7$\pm$0.2 & 0.1  & -1.97 & -0.3  & 1.80(f) \\
\hspace{0.6cm} - & -             & -                 & 47.07  & -     & 10.0$\pm$0.4 & 0.5 & -1.71 & 0.03  & 1.88$_{-0.05}^{+0.06}$ \\
\hspace{0.6cm} - & -             & -                 & 47.32  & -     & 10.0$\pm$0.4 & 1.1 & -1.83 & -0.05 & 2.21$_{-0.09}^{+0.08}$ \\
\hspace{0.6cm} - & -             & -                 & 47.22  & -     & 10.0$\pm$0.4 & 0.9 & -1.71 & 0.05  & 1.91$_{-0.05}^{+0.06}$ \\
\hspace{0.6cm} - & -             & -                 & 46.82  & -     & 10.0$\pm$0.4 & 0.3 & -1.88 & -0.20 & 1.71$_{-0.13}^{+0.13}$ \\
3481$\pm$150     & 8187$\rev{>}$ & 99.5$\pm$4.7   & 46.86  & 46.81 & 9.6$\pm$0.2 & 1.2  & -1.91 & -0.19 & 1.77$_{-0.12}^{+0.13}$ \\
3177$\pm$39      & -             & -                 & 47.09  & -     & 9.7$\pm$0.2 & 1.4  & -1.68 & 0.04  & 1.94$_{-0.05}^{+0.06}$ \\
\hspace{0.6cm} - & -             & -                 & 46.94  & -     & 10.0$\pm$0.4 & 0.3 & -2.12 & -0.35 & 1.16$_{-0.11}^{+0.11}$  \\
\hspace{0.6cm} - & -             & -                 & 46.94  & -     & 10.1$\pm$0.4 & 0.3 & -2.14 & -0.42 & 1.8(f)  \\
3208$\pm$27      & 6530$\pm$177  & 35.3$\pm$15.2     & 47.11  & 46.86 & 9.7$\pm$0.2 & 1.0  & -2.00 & -0.30 & 1.60$_{-0.14}^{+0.14}$ \\
4646$\pm$30      & 4183$\pm$117  & 44.1$\pm$3.9      & 46.84  & 46.77 & 9.5$\pm$0.2 & 0.4  & -2.04 & -0.34 & 1.70$_{-0.28}^{+0.31}$ \\
\hspace{0.6cm} - & -             & -                 & 46.91  & -     & 9.9$\pm$0.4 & 0.4  & -1.59 & 0.12  & 1.89$_{-0.04}^{+0.05}$ \\
\hspace{0.6cm} - & -             & -                 & 47.01  & -     & 10.4$\pm$0.4 & 0.2 & -1.75 & -0.04 & 2.00$_{-0.07}^{+0.07}$ \\
\hspace{0.6cm} - & -             & -                 & 46.74  & -     & 9.6$\pm$0.4 & 0.9  & -1.70 & -0.02 & 2.14$_{-0.28}^{+0.31}$ \\
\hspace{0.6cm} - & -             & -                 & 46.97  & -     & 9.9$\pm$0.4 & 0.8  & -1.71 & 0.02  & 2.07$_{-0.07}^{+0.08}$ \\
3110$\pm$15      & 4778$\pm$70   & 5.5$\pm$7.2       & 46.97  & 46.79 & 9.5$\pm$0.2 & 0.9  & -1.99 & -0.28 & 1.80(f) \\
4932$\pm$52      & 7645$\pm$275  & 597.8$\pm$102.2   & 47.40  & 47.27 & 10.1$\pm$0.2 & 1.1 & -1.74 & 0.04  & 1.81$_{-0.04}^{+0.05}$ \\
\hspace{0.6cm} - & -             & -                 & 47.01  & -     & 9.6$\pm$0.4 & 0.3  & -1.91 & -0.20 & 1.72$_{-0.24}^{+0.27}$ \\
\hspace{0.6cm} - & -             & -                 & 47.04  & -     & 10.1$\pm$0.5 & 0.4 & -1.92 & -0.18 & 1.69$_{-0.11}^{+0.11}$ \\
\hspace{0.6cm} - & -             & -                 & 47.04  & -     & 9.0$\pm$0.4 & 4.7  & -1.62 & 0.09  & 1.68$_{-0.05}^{+0.04}$ \\
\hspace{0.6cm} - & -             & -                 & 47.09  & -     & 9.0$\pm$0.4 & 5.0  & -1.68 & 0.05  & 1.86$_{-0.05}^{+0.05}$ \\
\hline
\end{longtable}

\end{document}